\documentclass[]{jfm}

\usepackage{graphicx}
\usepackage{newtxtext}
\usepackage{newtxmath}
\usepackage{natbib}
\usepackage{hyperref}
\hypersetup{
    colorlinks = true,
    urlcolor   = red,
    citecolor  = blue,
}
\usepackage{subfigure}
\usepackage{tikz}
\usepackage{xcolor}
\usepackage{bm}
\usepackage{diagbox}
\usepackage{lineno}

\newcommand{\RomanNumeralCaps}[1]

\title{Causal analysis of inner and outer motions in near-wall turbulence}

\author{Jingxuan Zhang\aff{1},
  Zhengping Zhu\aff{2},
  Limin Wang\aff{2},
  Wei Gao\aff{3,4}
 \and Ruifeng Hu\aff{1}
  \corresp{\email{hurf@lzu.edu.cn}}}

\affiliation{
\aff{1}Center for Particle-Laden Turbulence, Key Laboratory of Mechanics on Disaster and Environment in Western China, The Ministry of Education, College of Civil Engineering and Mechanics, Lanzhou University, Lanzhou 730000, PR China
\aff{2}Zhejiang Laboratory, Hangzhou 310000, PR China
\aff{3}State Key Laboratory of Water Resources Engineering and Management, Wuhan University, Wuhan, Hubei 430072, PR China
\aff{4}Mechanical Engineering, Physical Science and Engineering Division, King Abdullah University of Science and Technology, Thuwal 23955-6900, Saudi Arabia
}

\begin{document}
\maketitle

\begin{abstract}
In this work, we study the causality of near-wall inner and outer turbulent motions. The inner motions are defined as the self-sustained near-wall cycle, and the outer motions as those living in the logarithmic layer exhibiting footprints on the near-wall region. Causal inference with three typical methods is performed, \emph{i.e.} transfer entropy, information flow, and SURD (synergistic--unique--redundant decomposition of causality). The causal inference methods are first applied to several canonical problems to illustrate their abilities, including a linear problem, a non-linear problem, and a low-dimensional model of near-wall turbulence. It is demonstrated that all three methods can produce consistent causal findings.
Furthermore, we study the causalities between the inner and outer turbulent motions in a channel flow using the three methods with an improved inner-outer decomposition method. 
It is revealed that both the inner and outer motions are causally self-sustained, 
supporting the self-sustaining mechanism of turbulent motions at all scales.
{We also find that there are top-down and bottom-up causalities between the outer motions and their near-wall footprints owing to the dynamical coherence of inclined large-scale turbulent motions.}
More interestingly, pressure is identified to play an active role in the inner-outer causalities and may act as a bridge in linking the inner and outer turbulent motions.
\end{abstract}


{\bf MSC Codes }  {\it(Optional)} Please enter your MSC Codes here

\section{Introduction}
Wall-bounded turbulence is ubiquitous in both natural phenomena and engineering applications, such as atmospheric boundary layers and internal/external flows of aircraft. Due to its critical role in the determination of friction drag and heat transfer, wall-bounded turbulence has long been an active topic in turbulence research \citep{Robinson1991,Adrian2000,marusicWallboundedTurbulentFlows2010,Smits2011,McKeon2017engine,Jimenez2018coherent,liuLargescaleStructuresWallbounded2021}. Unlike turbulent flows in free space, wall-bounded turbulence is strongly influenced by boundary constraints and shear effects, leading to significant inhomogeneity and anisotropy. The multiscale interactions of turbulent motions further complicate its dynamics, posing great challenges for theoretical, numerical, and experimental studies. Therefore, a deep understanding of multiscale interactions in wall-bounded turbulence should be critical and beneficial for practical applications such as flow control and drag reduction \citep{choi2008control,marusic2021energy,fukagataTurbulentDragReduction2024}.

\subsection{Relationship between inner and outer turbulent motions}

It is well known that wall-bounded turbulent flows are comprised of coherent motions spanning a wide range of spatial and temporal scales \citep{Robinson1991, Adrian2007, Jimenez2018coherent,leeFlowStructuresTransitional2019a,wuNewInsightsTurbulent2023,cremadesIdentifyingRegionsImportance2024}. The prevalent coherent motions include near-wall streamwise streaks \citep{klineStructureTurbulentBoundary1967,smithCharacteristicsLowspeedStreaks1983,baeLifeCycleStreaks2021}, quasi-streamwise vortices \citep{jeongCoherentStructuresWall1997,schoppaCoherentStructureGeneration2002,Schlatter2014near}, hairpin vortices \citep{Head1981,wuDirectNumericalSimulation2009a,wangHairpinVortexGeneration2015}, hairpin packets \citep{zhouMechanismsGeneratingCoherent1999,Adrian2000,christensenStatisticalEvidenceHairpin2001,dennisExperimentalMeasurementLargescale2011,wangExperimentalStudyDominant2019a}, large-scale motions (LSMs) \citep{kovasznayLargescaleMotionIntermittent1970,raoBurstingPhenomenonTurbulent1971a,brownLargeStructureTurbulent1977a}, very-large-scale motions (VLSMs) \citep{Kim1999,leeVerylargescaleMotionsTurbulent2011,wangVeryLargeScale2016}, and superstructures \citep{Hutchins2007a}, among others. However, the definitions of these coherent structures are usually qualitative, and there are no clear boundaries among them. 

A wall-bounded turbulent flow can be roughly divided into inner and outer regions/layers in the wall-normal direction, traditionally in the sense of the mean velocity profile \citep{Pope2000}. For turbulent motions, a self-sustaining near-wall regeneration cycle has been discovered \citep{jimenez1991minimal,Hamilton1995,Waleffe1997}, which is free of the influence of outer turbulent motions \citep{Jimenez1999,Hwang2013near}.
In this cycle, near-wall velocity streaks can be profoundly amplified by quasi-streamwise vortices through the transfer of energy from the mean shear to streamwise velocity fluctuations via the lift-up effect. Then the amplified streaks rapidly oscillate and break down due to instability or transient growth, which in turn leads to the generation of new quasi-streamwise vortices. Therefore, the near-wall vortex-streak cycle was considered universal (independent of Reynolds number) as a cornerstone of the inner-outer interaction model (IOIM) \citep{Marusic2010,Mathis2011,Baars2016}. The increasing influence of the outer motions on the inner motions with Reynolds number is modelled by the superposition and modulation effects \citep{hutchinsLargescaleInfluencesNearwall2007,mathisLargescaleAmplitudeModulation2009,zhangQuasisteadyQuasihomogeneousDescription2016,doganQuantificationAmplitudeModulation2019,chernyshenko2021extension,andreolliSeparatingLargescaleSuperposition2023,sunUncoveringParameterVariations2026}. 
{It should be stressed that the inner motions can only be unconditionally universal (independent of Reynolds number) once the modulation effect has been fully removed. Otherwise, the inner motions conditionally depend on the Reynolds number due to the modulation effect.}
{The quasi-steady-quasi-homogeneous (QSQH) theory proposed by Chernyshenko and coworkers \citep{zhangQuasisteadyQuasihomogeneousDescription2016,chernyshenko2021extension} is a fundamental framework for understanding the inner-outer interactions, in which it is assumed that the inner motions vary much faster than the outer motions, therefore it should be universal if scaled by the local large-scale wall-shear stress rather than the mean one.}
\citet{Wang2021A} found that $y^+=100$ is a proper height to divide inner and outer regions for turbulent motions in the IOIM framework, as the extracted inner motions are truly Reynolds-number independent at $Re_\tau > 1000$, where $y^+=yu_\tau/\nu$ is the normalised wall-normal height with viscous units, $u_\tau$ is the friction velocity, $\nu$ is the fluid kinematic viscosity, and $Re_\tau=\delta u_\tau/\nu$ is the friction Reynolds number ($\delta$ is outer length scale, such as half-channel height, pipe radius or boundary-layer thickness).

A significant issue of long-term controversy is the relationship between inner and outer motions, which is the focus of this study.
There are several different views, such as the bottom-up view \citep{Adrian2000,Adrian2007}, the top-down view \citep{Hunt2000}, the co-supporting view \citep{Toh2005Interaction}, and the self-sustaining view  \citep{Cossu2017Selfsustaininga}.
In the bottom-up view, the hairpin packets in the outer region are believed to be generated via the mechanism proposed by \citet{zhouMechanismsGeneratingCoherent1999}, \emph{i.e.} new hairpins are induced by existing hairpins, and a hairpin packet is formed by a group of hairpins with a similar propagation velocity \citep{Adrian2000,Adrian2007}. Furthermore, \citet{Kim1999} hypothesised that VLSMs are a consequence of spatial coherence between hairpin packets. Since hairpin vortices are usually regarded as near-wall structures, the hypothesis of the subsequent generation of hairpin packets and VLSMs in the outer region is termed the `bottom-up' mechanism. 
In the top-down view, it is hypothesised that the prominent outer energetic turbulent motions impinge downward on the wall and generate near-wall motions at high Reynolds numbers \citep{Hunt2000}.  In addition, \citet{morrisonInteractionInnerOuter2007} underscored that the top-down interaction should be nonlinear. 
In the co-supporting view, the large-scale outer motions are suggested to be generated by the collective behaviour of near-wall motions, and the generation of the near-wall motions is enhanced by the large-scale outer motions. The numerical studies by \citet{Zhou2022Interaction,zhouInfluenceOuterLargescale2024} partly support this view by showing that the spanwise drift velocity of the near-wall streaks is affected by the outer LSMs, while the merger of the near-wall streaks is only weakly correlated with the generation of the LSMs. In the self-sustaining view, it is suggested that a whole family of self-sustaining motions exists in wall-bounded turbulent flows, ranging from the near-wall buffer-layer streak-vortex cycle to the outer LSMs and VLSMs \citep{Cossu2017Selfsustaininga}.  

The complex multiscale and nonlinear nature of turbulence hinders clear identification and clarification of the generation mechanisms and relationship of the inner and outer motions. 
Energy budget analysis is a crucial method to understand how turbulence is produced, transported, and dissipated. It can provide certain insights into the direction of energy transfer in space and scale, thus providing a useful tool for revealing the interaction between inner and outer motions. 
In particular, spectral energy budget analyses of wall-bounded turbulent flows in several studies have shed light on this issue. 
\citet{mizunoSpectraEnergyTransport2016} uncovered both downward and upward energy fluxes in turbulent channel flows with moderate Reynolds numbers. In the inner layer, the downward energy fluxes are found to originate in the outer layer and emerge at large scales, and the upward fluxes remove energy from wider scales to narrower scales as a forward energy cascade. \citet{choScaleInteractionsSpectral2018} found that there exist downward energy fluxes from small to large scales (inverse cascade) in the near-wall region, and the relation of turbulence production and pressure-strain supports the self-sustaining mechanism of turbulent motions at all scales.
\citet{leeSpectralAnalysisBudget2019} also highlighted the downward and upward energy transfers, the former of which is dominated by the VLSMs. In the near-wall region, they reported an inverse energy transfer from small to large scales in the streamwise velocity fluctuations.
It is further noted that several studies utilised the generalised Kolmogorov equations for the second-order structure function to conduct budget analysis in space and scale, and similar findings of the upward/downward and forward/inverse energy transfers were discovered \citep{maratiEnergyCascadeSpatial2004,cimarelliSourcesFluxesScale2015,cimarelliCascadesWallnormalFluxes2016,chiariniAscendingdescendingDirectinverseCascades2022}. Despite these advances, a definitive consensus of the problem remains elusive, as previous studies did not effectively separate the inner and outer motions in the near-wall region, both of which may contribute to the spectral energy at a single scale. Besides, novel diagnostic tools beyond the traditional energy transfer analysis may provide new insights into this issue. 

\subsection{Causal inference studies}

A causal relationship describes a mechanism in which one event directly affects another, which forms a fundamental basis for understanding complex systems. 
Causal inference can provide an elegant way to reveal the inherent causal relationship in a physical system. 
Essentially, causal inference methods can be categorised into intrusive and non-intrusive methods.
In intrusive methods, the system is modified directly, allowing observation of intervention-induced consequences. 
Indeed, quite a number of studies on turbulence have fallen into this category \citep{Jimenez1999,jimenezMachineaidedTurbulenceTheory2018,jimenezMonteCarloScience2020,lozano-duranCauseandeffectLinearMechanisms2021,jimenezStreaksWallboundedTurbulence2022,encinarIdentifyingCausallySignificant2023,andreolliSeparatingLargescaleSuperposition2023,osawa2024causal}. However, intrusive methods need to introduce perturbations to the flow field, which brings additional complexities of numerical simulations. In addition, intrusive methods can hardly be implemented to deal with measurement or monitoring data in practice, or in cases where the governing equations are completely or partly unknown.    

Due to the limitations of intrusive approaches, non-intrusive causal inference methods have seen significant developments and applications in recent years \citep{Pearl2000,Pearl2009Causala,liang2013liang,moraffahCausalInferenceTime2021,yaoSurveyCausalInference2021,assaadSurveyEvaluationCausal2022,camps2023discovering,Runge2023Causal}. 
As comprehensively summarised in these overviews, various causal inference methods have been proposed.
\citet{Granger1969Investigating} pioneeringly proposed the Granger causality (GC), which is a statistical test for whether a variable can be effectively predicted by the past of another variable with a linear autoregressive model. 
In information theory, causal relationships between variables can be quantified by measuring the information transfer from one variable to another. The foundation of this approach lies in the Shannon entropy \citep{shannon1948mathematical}, with early causal measures introduced by \cite{massey1990causality}, who used the conditional Shannon entropy to capture directional dependencies. 
Later, \cite{schreiber2000measuring} formalised this concept by proposing transfer entropy (TE), which builds on the conditional Shannon entropy. The key intuition of TE is that a cause variable should reduce the uncertainty of a result variable. This framework enables a rigorous quantification of causal relationships. TE offers several distinct advantages over alternative methods. As demonstrated by \cite{barnett2009granger}, TE is equivalent to linear GC for Gaussian variables. 
Parallel to TE, \cite{liang2005information} developed an alternative approach named information flow (IF) based on dynamical systems, later rigorously derived and validated by \cite{liang2013liang}. Unlike TE, IF quantifies causality through the time derivative of Shannon entropy, which is derived from first principles through the dynamic equations. 
More recently, \cite{martinez2024decomposing} developed a novel framework that effectively decomposes causality into synergistic, unique, and redundant (SURD) components, which can provide a deeper insight into the causal relationship in a system.
\citet{martinez2025observational} further extended the SURD to account for the effects of states.
Other causal inference methods can be found in the aforementioned overviews.

These causal inference methods have found extensive applications across diverse disciplines, including fluid dynamics and especially turbulent flows \citep{wang2026causal}. 
\cite{tissot2014granger} employed GC to investigate the self-sustaining processes in a turbulent channel flow. They demonstrated that GC can identify the causal links between streak breakdown and wall-normal activity transferred by shear deformation. 
\citet{materassi2014information} applied TE to the Gledzer-Ohkitani-Yamada shell model of turbulence, showing the existence of distinct energy cascades and the locality of the scale interactions. 
\citet{liang2016preliminary} adopted IF to study the near-wall self-sustained turbulence cycle, and revealed that there may be two subcycles in it.
Through TE, \cite{lozano2020causality} revealed that the energetic turbulent motions exhibit similar causal characteristics in both the buffer and logarithmic layers. 
This approach was subsequently extended by \cite{Wang2021Informationa, Wang2022Spatial} to investigate the causal interactions in turbulent flow over porous media. 
\cite{Lozano-Duran2022Informationtheoretic} employed information flux between variables to quantify causality, and applied it to the problems of energy cascade in isotropic turbulence, subgrid modelling of large-eddy simulation, as well as optimal drag reduction in wall-bounded turbulence using opposition control.
\cite{martinez2023causality} analysed the formation mechanism of large-scale coherent structures in the flow around a wall-mounted square cylinder using TE.
\citet{arranzInformativeNoninformativeDecomposition2024} proposed a method that can decompose a source field into its informative and residual non-informative components relative to a target field in wall-bounded turbulent flows. 
\cite{ling2024causality} investigated the streak-burst causality in wall turbulence using information flux. 
\citet{lopez-dorigaLinearNonlinearGranger2024a} leveraged linear and nonlinear GC to identify the causalities of coherent structures in turbulent flows in square and rectangular ducts. It was found that the secondary flow fluctuations are the primary cause of both the presence and movement of near-wall streaks towards and away from the duct corners.
\citet{tanogamiInformationthermodynamicBoundInformation2024} demonstrated that information of turbulent fluctuations is transferred from large to small scales along with the energy cascade in fully-developed three-dimensional turbulence.
\citet{martinez2024decomposing} applied the SURD method to the DNS data of a three-dimensional isotropic turbulence and the experimental data of a turbulent boundary layer, revealing more detailed causal relationships in the energy cascade and inner-outer interactions. \citet{wangCausalAnalysisTurbulent2026} proposed a Galerkin-based Granger causal inference method and applied it in the causal analysis of a low-dimensional near-wall turbulence model \citep{moehlis2004low}.
\citet{zhao2026causal} leveraged the information flow to identify the causal structures for skin friction generation.

\subsection{The present study}

The objective of this study is to reveal the causal relationship between inner and outer turbulent motions in wall-bounded turbulence. Three typical causal inference methods are leveraged, \emph{i.e.} TE, IF and SURD. First, we assess and compare these methods in some simple systems, including a linear problem, a nonlinear problem and a low-order nonlinear model of near-wall turbulence. Then, we introduce an improved inner-outer decomposition method based on the IOIM and identify the causal relationship between inner and outer motions.

This paper is organised as follows. The fundamental methodologies of the adopted causal inference methods are introduced in \S \ref{section2}. In \S \ref{section3}, the applications of these methods in several simplified model systems are given. An improved inner-outer decomposition method is described in \S \ref{iod}. 
The causal relationship between the inner and outer motions is thoroughly examined in \S \ref{section5}. 
Finally, the conclusions of the study are drawn in \S \ref{section6}. 

\section{Causal inference methods}\label{section2}

Recent years have witnessed a growing interest in causal discovery and inference as a powerful approach to uncovering the fundamental causal structures governing physical systems. Since the pioneering work of \citet{Granger1969Investigating}, an expanding array of methodologies has been proposed and developed to operate under various theoretical assumptions, as comprehensively reviewed in \citet{Pearl2000,Pearl2009Causala,liang2013liang,moraffahCausalInferenceTime2021,yaoSurveyCausalInference2021,assaadSurveyEvaluationCausal2022,camps2023discovering,Runge2023Causal}. 
In the present investigation, we focus on three well-established non-intrusive causal inference techniques, \emph{i.e.} transfer entropy \citep{schreiber2000measuring}, information flow \citep{liang2005information}, and SURD \citep{martinez2023causality}. 

\subsection{Transfer entropy}
\label{subsection2.1}

Quantification of causal relationships between time-dependent variables can be achieved effectively using the framework of information theory \citep{shannon1948mathematical}. 
The core of this approach lies in Shannon entropy, which serves as the fundamental measure for assessing causal interactions.
For a random variable $X$ with probability distribution function (p.d.f) $p(x)$, where $x$ is any measurement of $X$, its Shannon entropy is defined as 
\begin{equation}
    \label{shannonentropy}
    S(X) = -\sum_{x}p(x)\log p(x),
\end{equation}
which characterises the uncertainty inherent in $X$. The joint entropy for a pair of variables $(X, Y)$ extends this measure to quantify their combined uncertainty: 
\begin{equation}
    \label{jointshannonentropy}
    S(X,Y) = -\sum_{x} \sum_{y} p(x,y)\log p(x,y),
\end{equation}
where $Y$ represents another random variable, and $p(x,y)$ denotes the joint p.d.f of $X$ and $Y$. 

The conditional Shannon entropy is  defined as
\begin{equation}
    \label{conditionalshannonentropy}
    S(X|Y) = S(X,Y)-S(Y),
\end{equation}
which quantifies the remaining uncertainty in $X$ given knowledge of $Y$. 

Building upon conditional entropy and leveraging the temporal asymmetry inherent in causal relationships, transfer entropy is formally defined as \citep{schreiber2000measuring} 
\begin{equation}
    \label{transferentropy}
        T_{X \to Y}(\Delta t)=S(Y(t)|Y(t-\Delta t))-S(Y(t)|X(t-\Delta t),Y(t-\Delta t)),
\end{equation}
where $\Delta t$ denotes time lag. The term $S(Y(t)|Y(t-\Delta t))$ quantifies the uncertainty in $Y$ that is conditioned on the past state of $Y$. Similarly, $S(Y(t)|X(t-\Delta t),Y(t-\Delta t))$ represents the residual uncertainty in $Y$ when conditioned on the historical states of $X$ and $Y$ simultaneously. Therefore, transfer entropy fundamentally measures the reduction in the uncertainty of $Y$ obtained by observing the historical evolution of $X$. A non-zero positive transfer entropy indicates that there exists causality. This formulation naturally captures a crucial property of time series, \emph{i.e.}, inherent directionality (asymmetry) in the causal relationship between variables, which distinguishes it from other conventional analysis methods. 
Within this information-theoretic framework, transfer entropy emerges as a useful measure of causality that has found a number of applications in recent turbulence research \citep{materassi2014information,lozano2020causality,Wang2021Informationa,Wang2022Spatial,martinez2023causality}.

For a multivariate system, the transfer entropy from variable $V_j$ to $V_i$ can be computed as \citep{lizier2014jidt,lozano2020causality}
\begin{equation} \label{eq4}
    T_{j \to i}(\Delta t) = S(V_i(t)|\boldsymbol{V}^{\not{j}}(t-\Delta t)) - S(V_i(t)|\boldsymbol{V}(t-\Delta t)), 
\end{equation}
where $\boldsymbol{V}^{\not{j}}$ denotes the vector $\boldsymbol{V}$ excluding the $j$th component.

\subsection{Information flow}
\label{subsection2.2}

The framework of the information flow was derived from first principles \citep{liang2005information,liang2013liang}, considering a minimal two-dimensional dynamical system capable of sustaining information transfer, as
\begin{equation}
    \frac{{\rm d}x_1}{{\rm d}t}=F_1(x_1,x_2,t), \quad
    \frac{{\rm d}x_2}{{\rm d}t}=F_2(x_1,x_2,t),
\end{equation}
where $F_1$ and $F_2$ represent general functions. This minimal dimensionality of the system is necessary and sufficient for demonstration. 
As the dynamical system evolves temporally, the joint p.d.f $p(x_1,x_2)$ undergoes corresponding evolution governed by the Liouville equation \citep{lasota2013chaos} as
\begin{equation}\label{eqLiouv}
    \frac{\partial p(x_1,x_2)}{\partial t}+\frac{\partial}{\partial x_1}(F_1(x_1,x_2,t)p(x_1,x_2))+\frac{\partial}{\partial x_2}(F_2(x_1,x_2,t)p(x_1,x_2))=0.
\end{equation}

The temporal evolution rate of the Shannon entropy ${\rm d}S_1/{\rm d}t$ plays a fundamental role in quantifying the information flow from $x_2$ to $x_1$. To derive ${\rm d}S_1/{\rm d}t$, the Liouville equation (\ref{eqLiouv}) is integrated over $x_2$: 
\begin{equation}\label{eqinteLiouv}
    \frac{\partial p(x_1)}{\partial t}+\frac{\partial}{\partial x_1}\int F_1(x_1,x_2,t)p(x_1,x_2) {\rm d}x_2 = 0, 
\end{equation}
in which, the other terms vanish due to the compact support assumption for $p(x_1,x_2)$. 
Multiplying equation~(\ref{eqinteLiouv}) by $-(1 + \log p(x_1))$ and integrating it over $x_1$ gives: 
\begin{equation}\label{eqds1dt}
    \begin{aligned}
    \frac{{\rm d}S_1}{{\rm d}t} &=\iint \left[\log p(x_1)\frac{\partial(p(x_1,x_2) F_1(x_1,x_2,t))}{\partial x_1}\right]{\rm d}x_1{\rm d}x_2\\
    &=-E\left(\frac{F_1(x_1,x_2,t)}{p(x_1)}\frac{\partial p(x_1)}{\partial x_1}\right),
    \end{aligned}
\end{equation}
where $E(\cdot)$ denotes the mathematical expectation. 

\cite{liang2005information} suggested that in evolving two-dimensional systems, the marginal entropy change of $x_1$ comes from two distinct mechanisms: 
\begin{equation}
    \frac{{\rm d} S_1}{{\rm d}t}=\frac{{\rm d}S_1^*}{{\rm d}t}+L_{2 \to 1}, 
\end{equation}
where ${\rm d}S_1^*/{\rm d}t$ represents the autonomous entropy production from $x_1$ itself, and $L_{2 \to 1}$ quantifies the information flow rate from $x_2$ to $x_1$. The latter term $L_{2 \to 1}$ serves as the fundamental causality measure in information flow theory. 
The total entropy rate ${\rm d} S_1/{\rm d}t$ can be calculated using (\ref{eqds1dt}), but determining the information flow $L_{2 \to 1}$ requires knowledge of ${\rm d}S_1^*/{\rm d}t$. By multiplying the Liouville equation (\ref{eqLiouv}) by $-(1 + \log p(x_1,x_2))$ and integrating over $x_1$ and $x_2$ (assuming compact support), the joint Shannon entropy evolution can be derived as 
\begin{equation}\label{eqdsdt}
    \frac{{\rm d} S}{{\rm d}t} = E(\nabla\cdot \bm{F}), 
\end{equation}
where $\nabla$ denotes the divergence operator and $\bm{F}=[F_1(x_1,x_2,t),F_2(x_1,x_2,t)]^T$ represents the vector field of the coupled system. Building upon (\ref{eqdsdt}), \cite{liang2005information} proposed that the autonomous contribution to the marginal entropy change of $x_1$ can be determined by 
\begin{equation}
    \frac{{\rm d} S_1^*}{{\rm d}t} = E\left(\frac{\partial F_1(x_1,x_2,t)}{\partial x_1}\right). 
\end{equation}
Therefore, the information flow from $x_2$ to $x_1$ is consequently obtained as
\begin{equation}
    L_{2 \to 1} = \frac{{\rm d} S_1}{{\rm d}t} - \frac{{\rm d}S_1^*}{{\rm d}t} = -E\left(\frac{F_1(x_1,x_2,t)}{p(x_1)}\frac{\partial p(x_1)}{\partial x_1}\right) - E\left(\frac{\partial F_1(x_1,x_2,t)}{\partial x_1}\right). 
\end{equation}

\citet{Liang2008} extended the formalism to stochastic dynamical systems.
The theoretical foundation of the information flow was further rigorously established by \cite{liang2016information} for both deterministic and stochastic dynamical systems. 
For practical applications, \cite{liang2014unraveling} developed a maximum likelihood estimator that can express the information flow purely in terms of covariances between two variables in the linear sense. For the directional information flow from $x_2$ to $x_1$, the estimator takes the form: 
\begin{equation}
    L_{2\to 1}=\frac{C_{11}C_{12}C_{2,{\rm d}1}-C_{12}^{2}C_{1,{\rm d}1}}{C_{11}^{2}C_{22}-C_{11}C_{12}^{2}},
\end{equation}
where $C_{ij}={\rm cov}(V_i,V_j)$ denotes the covariance between system components $V_i$ and $V_j$, $C_{i,{\rm d}j}={\rm cov}(V_i,\dot{V}_j)$ represents the covariance between $V_i$ and the temporal derivative $\dot{V}_j$. 

The multivariate form of the information flow was given by \citet{liang2021normalized} similarly to the bivariate case, as 
\begin{equation}\label{eq5} 
    L_{j \to i}=\frac{{\rm d}S_i}{{\rm d}t}-\frac{{\rm d}S_{i \not j}}{{\rm d}t} \approx \frac{1}{{\rm det} (\mathsfbi{C})}\cdot \sum\limits_{m=1}^d \Delta_{jm}C_{m,{\rm d}i} \cdot \frac{C_{ij}}{C_{ii}},
\end{equation}
where $\mathsfbi{C}$ represents the covariance matrix, ${\rm det}(\cdot)$ denotes the determinant of a matrix, $\Delta_{ij}$ indicates the cofactors of $\mathsfbi{C}$, and $d$ is the system dimensionality. ${\rm d}S_i/{\rm d}t$ is the temporal variation rate of $S(V_i)$, and ${\rm d}S_{i \not j}/{\rm d}t$ is the temporal variation rate of $S(\boldsymbol{V})$ with the effect of $V_j$ excluded, thus making $L_{j \to i}$ a direct measure of the information transfer rate from $V_j$ to $V_i$.

\subsection{Synergistic-unique-redundant decomposition (SURD)}
\label{subsection2.3}

A novel approach to causal analysis, termed synergistic unique-redundant decomposition of causality (SURD), was recently proposed by \cite{martinez2024decomposing}. SURD provides a framework for causal quantification by decomposing information into redundant, unique, and synergistic components, and can also account for unobserved confounders. 

The SURD approach still operates within the information-theoretic framework, with mutual information serving as the fundamental measure for causality quantification. Within the framework of information theory, mutual information between two variables is formally defined as
\begin{equation}
    \label{mutualinformation}
    I(X;Y) = S(X)-S(X|Y)=S(Y)-S(Y|X)=S(X)+S(Y)-S(X,Y).
\end{equation}
Here, $S(X)$ and $S(X|Y)$ represent the Shannon entropy and the conditional Shannon entropy, as defined in eqs.~(\ref{shannonentropy}) and ~(\ref{conditionalshannonentropy}), respectively. Mutual information quantifies the decrease in uncertainty associated with one random variable when another random variable is known. This reduction in uncertainty can be understood as the information jointly known by both variables, \emph{i.e.}, their shared information. Therefore, mutual information constitutes a symmetric metric, \emph{i.e.} $I(X;Y)=I(Y;X)$. 
It is noted that $X$ and $Y$ may denote a collection of variables. 

Consider a system with $N$ observable variables at time $t$, represented by the vector $\bm{\mathit{Q}}(t)=[Q_1(t),Q_2(t),...,Q_N(t)]$, where each $Q_i(t)$ denotes a variable of interest. Information about the future state of the variable $Q_j$ is quantified by $S(Q_j(t+\Delta t))$. According to the definition of mutual information, information about the future state of the variable $Q_j$ can be decomposed into two components based on all observable variables, as 
\begin{equation}
    S(Q_j(t+\Delta t)) = I(Q_j(t+\Delta t);\bm{\mathit{Q}}(t))+S(Q_j(t+\Delta t)|\bm{\mathit{Q}}(t)),
\end{equation}
where $I(Q_j(t+\Delta t);\bm{\mathit{Q}}(t))$ represents the reduction in uncertainty about the future state of $Q_j$ given the current state of the variables $\bm{\mathit{Q}}$, which can be interpreted as the information transferred from $\bm{\mathit{Q}}(t)$ to $Q_j(t+\Delta t)$. On the other hand, $S(Q_j(t+\Delta t)|\bm{\mathit{Q}}(t))$ denotes the residual uncertainty in the future state of $Q_j$ conditioned on the current state of the observable variables $\bm{\mathit{Q}}$. This term represents information transferred to $Q_j$ from unobserved sources beyond the system $\bm{\mathit{Q}}$, which can be characterised as information leakage $\Delta I_{leak \to j}$.


Based on the specific mutual information \citep{deweese1999measure}, the mutual information can be decomposed as:
\begin{equation}
    I(Q_j(t+\Delta t);\bm{\mathit{Q}}(t))=\sum_{i=1}^{N} \Delta I^U_{i \to j} + \sum_{\bm{\mathit{i}}} \Delta I^R_{\bm{\mathit{i}} \to j} + \sum_{\bm{\mathit{i}}} \Delta I^S_{\bm{\mathit{i}} \to j},
\end{equation}
where $\Delta I^U_{i \to j}$, $\Delta I^R_{\bm{\mathit{i}} \to j}$, and $\Delta I^S_{\bm{\mathit{i}} \to j}$ are unique, redundant, and synergistic causalities, respectively. The symbol $\bm{\mathit{i}}$ denotes a higher-order combination of variables. 
The redundant causality represents the shared causal information common to all components of $\bm{\mathit{Q_i}}$. This condition occurs when all variables in $\bm{\mathit{Q_i}}$ provide identical predictive information about $Q_j(t+\Delta t)$. 
The unique causality quantifies the distinct causal information that is exclusively attributable to $Q_i$ and cannot be derived from any other individual variable. This form of causality emerges when the observation of $Q_i$ provides greater informational gain outcomes of $Q_j(t+\Delta t)$ compared to the observation of any other isolated variable in the system. 
The synergistic causality refers to the causal influence arising from the joint interaction of variables in $\bm{\mathit{Q_i}}$. This form of causality is characterised by an informational gain in predicting $Q_j(t+\Delta t)$ when the entire set $\bm{\mathit{Q_i}}$ is observed simultaneously, exceeding the information obtained from observing each variable in isolation. 
These components are derived through the computation of the expectations of specific mutual information terms. 

Finally, the information on the future state of variable $Q_j$ can be ultimately decomposed into four distinct contributions, \emph{i.e.} unique, redundant, synergistic causalities and information leaks \citep{martinez2024decomposing}: 
\begin{equation}
    S(Q_j(t+\Delta t)) = \sum_{i=1}^{N} \Delta I^U_{i \to j} + \sum_{\bm{\mathit{i}}} \Delta I^R_{\bm{\mathit{i}} \to j} + \sum_{\bm{\mathit{i}}} \Delta I^S_{\bm{\mathit{i}} \to j}+\Delta I_{leak \to j}.
\end{equation}

\subsection{Implementation details}
In this work, we employ the histogram method to calculate the p.d.f. in transfer entropy. The number of bins in each dimension is usually set to 10 if the number of variables is less than 9. 
Otherwise, 5 bins are used per dimension due to the excessive storage requirement. This reduction in the number of bins does not affect the conclusion of causal inference, according to our tests in simple problems. 
Furthermore, we calculate the SURD using the multivariate compatible procedure introduced by \cite{martinez2024decomposing}. This method involves discretising the variables into 50 equally spaced bins per dimension and computing the joint and marginal p.d.f. directly from the resulting histograms. 

We normalise the transfer entropy and information flow by the sum of the causal contributions from all source variables to the target variable. Similarly, we use total mutual information to normalise the redundant, synergistic, and unique information components in SURD. This can yield the proportional contribution of each component to the overall information of the target variable. 
Consistent with \cite{martinez2024decomposing}, we normalise the information leak associated with a variable by the total information encompassed in its future states. This procedure quantifies the relative contribution of unobserved variables to the information contained in the future states of the target variable. 

In this study, our main interest lies in the causal interactions between variables. In the subsequent analysis, the self-information transfer (self-causality) is omitted by setting it to zero after normalisation. 
Moreover, following \cite{martinez2024decomposing}, we restrict the presentation of our SURD results to third-order synergy. Higher-order synergistic effects are consequently subsumed within the information leak term. Finally, all components are sorted in descending order on the basis of their magnitudes in SURD.

\section{Causal analysis of simple systems}\label{section3}
In this section, we apply the causal inference methods of transfer entropy, information flow, and SURD in three simple systems, \emph{i.e.}, a linear problem, a nonlinear problem, and a low-dimensional model of near-wall turbulence. As causalities can be explicitly inferred from their mathematical formulations, we can directly evaluate the performance of the three methods in these simple systems in a preliminary manner. 

\subsection{The linear problem}

\begin{figure}
    \centering
    \subfigure{
        \begin{tikzpicture}
            \node[anchor=north west, inner sep=0] (image) at (0,0) {
                \includegraphics[width=0.48\textwidth]{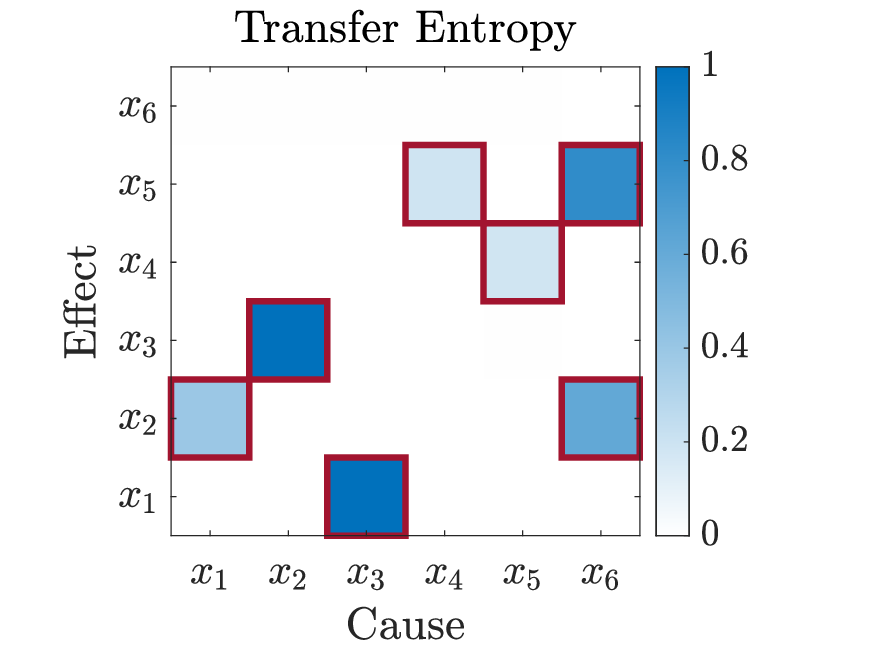}
            };
            \node[anchor=north west] at (image.north west) {{\rmfamily\fontsize{12}{13}\fontseries{l}\selectfont(a)}};
        \end{tikzpicture}
        \label{figlineara}} 
    \subfigure{    
        \begin{tikzpicture}
            \node[anchor=north west, inner sep=0] (image) at (0,0) {
                \includegraphics[width=0.48\textwidth]{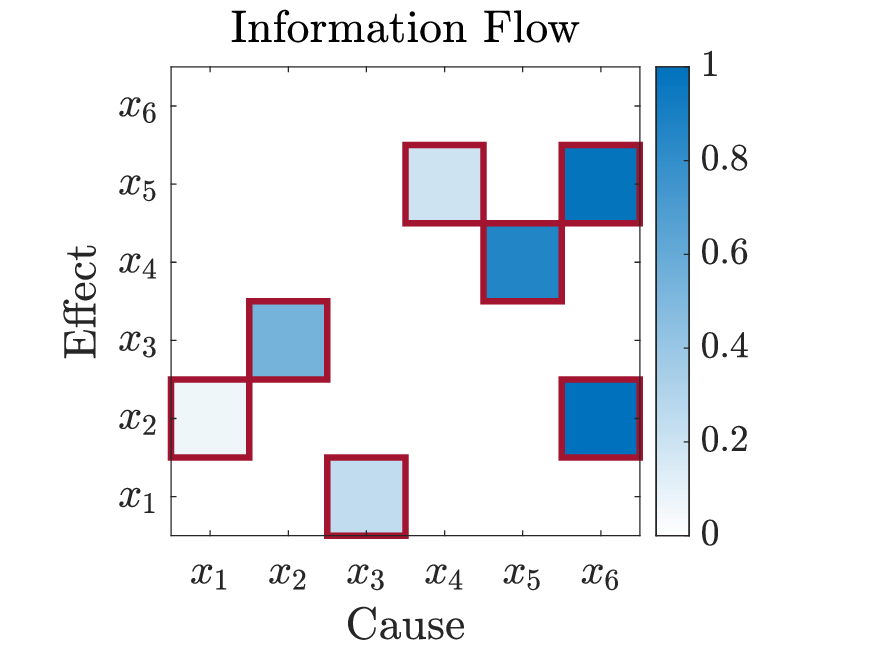}
            };
            \node[anchor=north west] at (image.north west) {{\rmfamily\fontsize{12}{13}\fontseries{l}\selectfont(b)}};
        \end{tikzpicture}
        \label{figlinearb}}
    \subfigure{
        \begin{tikzpicture}
            \node[anchor=north west, inner sep=0] (image) at (0,0) {
                \includegraphics[width=0.85\textwidth]{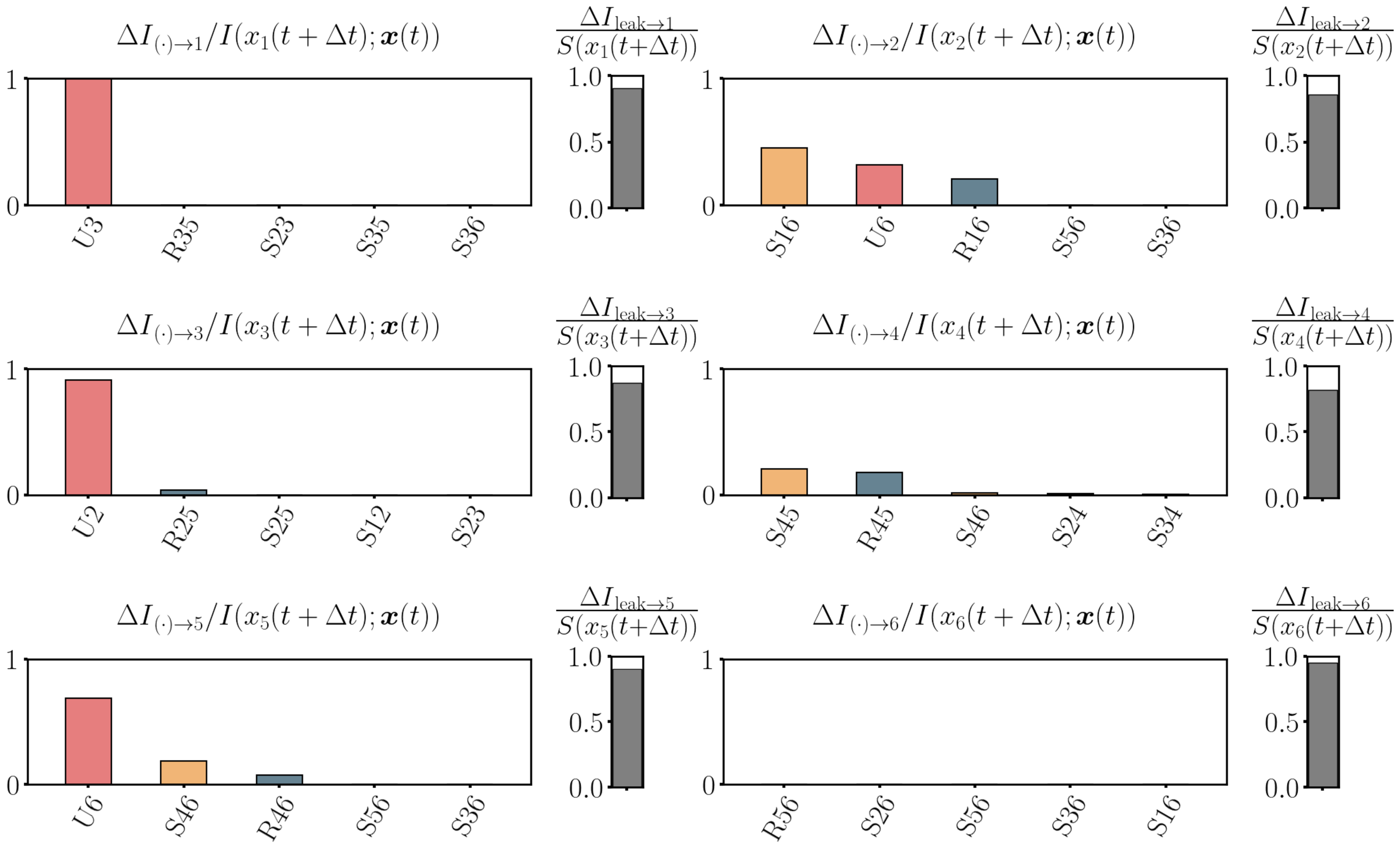}
            };
            \node[anchor=north west] at (image.north west) {{\rmfamily\fontsize{12}{13}\fontseries{l}\selectfont(c)}};
        \end{tikzpicture}
        \label{figlinearc}}
    \caption{Causal maps of the linear problem. (a) Transfer entropy; (b) Information flow; (c) SURD. Red boxes: the ground-truth causal links in the system. }
    \label{figlinear}
\end{figure}

We first examine causal relationships in a six-dimensional vector autoregressive process $\boldsymbol{x}=(x_1,...,x_6)^{\rm{T}}$ as
\begin{equation}\label{lineareq}
    {\boldsymbol{x}(n+1)=\boldsymbol{\alpha}+\mathsfbi{A}\boldsymbol{x}(n)+\mathsfbi{B}\boldsymbol{e}(n+1)},
\end{equation}
where $\boldsymbol{\alpha}$ is a constant vector and $\mathsfbi{A}$ denotes the coefficient matrix. The perturbation vector $\boldsymbol{e}=(e_1,...,e_6)^{\rm{T}}$ consists of independent standard normal random variables, and each $b_{ii}$ in the diagonal perturbation matrix $\mathsfbi{B}$ quantifies the noise amplitude of the $i$th component. In the present study of a strong perturbation, we set uniformly high noise levels of $b_{ii}=100$ $(i=1,...,6)$. Following \cite{liang2021normalized}, the other parameters are specified as: 
\begin{equation}
    {\boldsymbol{\alpha}=(0.1,0.7,0.5,0.2,0.8,0.3)^{\rm{T}}},
\end{equation}
and 
\begin{equation}
\renewcommand{\arraystretch}{1.3}
    {\mathsfbi{A}=\left[
            \begin{array}{cccccc}
               0 & 0 & -0.6 & 0 & 0 & 0 \\
               -0.5 & 0 & 0 & 0 & 0 & 0.8 \\
               0 & 0.7 & 0 & 0 & 0 & 0 \\
               0 & 0 & 0 & 0.7 & 0.4 & 0 \\
               0 & 0 & 0 & 0.2 & 0 & 0.7 \\
               0 & 0 & 0 & 0 & 0 & -0.5 
            \end{array}
               \right]}.
\label{linearma}
\end{equation}
By numerically solving the system~(\ref{lineareq}) with uniformly distributed random initial conditions, 
the numerical simulation produced a total of 500 realisations. Each realisation comprises six distinct time series, each spanning 500,000 time steps. To ensure that the system had reached its attractor, the first 100,000 time steps of each time series were discarded as transient. Consequently, the final dataset of each realisation consists of 400,000 time steps per series. 
In this study, the causal analysis is performed with a lag of one time step.

Figure~\ref{figlinear} presents the results of the causal analysis of the linear system. In this problem, if $x_j$ exists on the right-hand side of the equation of $x_i$, $x_j \to x_i$ is regarded as a ground-truth causal link. All the ground-truth causal links in the system~(\ref{lineareq}) are highlighted by the red bounding boxes in figure~\ref{figlinear}. 
It is seen that both transfer entropy and information flow demonstrate consistent identifications of the causal links of the linear system. The detected non-zero causalities, \emph{i.e.}, $x_1 \to x_2$, $x_2 \to x_3$, $x_3 \to x_1$, $x_4 \to x_5$, $x_5 \to x_4$, $x_6 \to x_2$ and $x_6 \to x_5$, are in perfect agreement with the non-zero elements in the coefficient matrix~(\ref{linearma}). 

The identification result by SURD is presented in figure~\ref{figlinearc}, in which $\bm{\mathit{x}}$ denotes the set of all variables in the linear model. Higher-than-second-order redundant contributions are not explicitly shown in the figure due to limited space, and we mainly focus on low-order causal interactions, as the dynamics of the linear model dictate that any variable can be influenced by at most two other variables. 
It can be found that the inference result of SURD is in close agreement with the ground-truth causal links of the linear model. The link $x_3 \to x_1$ is expressed through unique information transfer $\Delta I_{3 \to 1}^U$. The causalities of $x_1$ and $x_6$ to $x_2$ are manifested by a combination of unique ($\Delta I_{6 \to 2}^U$), synergistic ($\Delta I_{16 \to 2}^S$), and redundant ($\Delta I_{16 \to 2}^R$) contributions, respectively, with the synergistic component being the most substantial. The link $x_2 \to x_3$ is identified as a predominant unique contribution $\Delta I_{2 \to 3}^U$. The link $x_4 \to x_5$ involves a combination of synergistic ($\Delta I_{45 \to 4}^S$) and redundant ($\Delta I_{45 \to 4}^R$) contributions, while the causalities of $x_4$ and $x_6$ to $x_5$ are jointly governed by unique ($\Delta I_{6 \to 5}^U$), synergistic ($\Delta I_{46 \to 5}^S$), and redundant ($\Delta I_{46 \to 5}^R$) mechanisms, in which the unique one is the most important. 

From the above results, we can see that all three methods can successfully identify the causal links embedded in the linear system with strong noise. The SURD method can further distinguish contributions from unique, synergistic, and redundant mechanisms, providing more detailed causal information.

\subsection{The nonlinear problem} 
Next, we examine a nonlinearly coupled system to evaluate the performance of the causal inference methods in more complex dynamical systems. 

\begin{figure}
    \centering
    \subfigure{
        \begin{tikzpicture}
            \node[anchor=north west, inner sep=0] (image) at (0,0) {
                \includegraphics[width=0.48\textwidth]{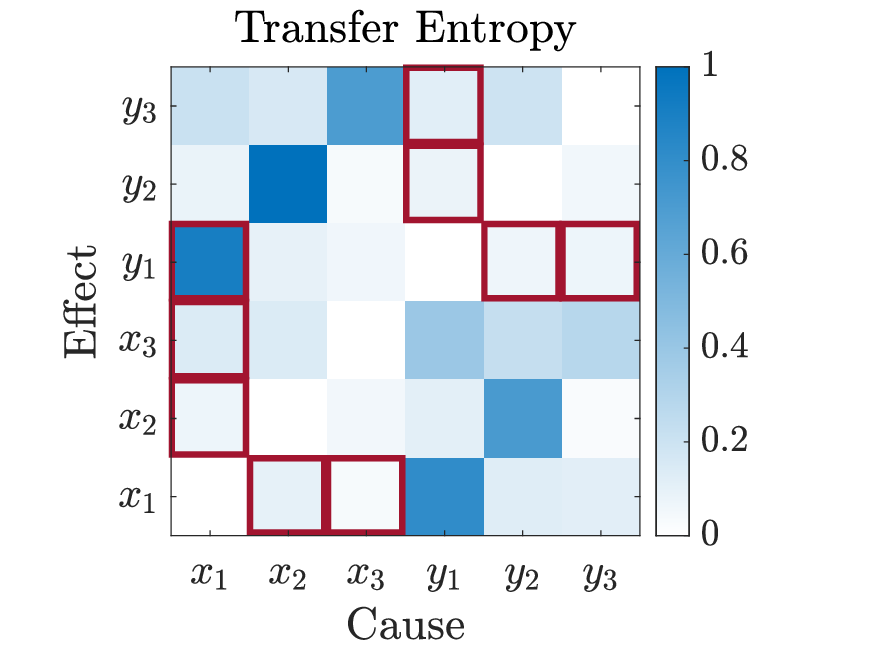}
            };
            \node[anchor=north west] at (image.north west) {{\rmfamily\fontsize{12}{13}\fontseries{l}\selectfont(a)}};
        \end{tikzpicture}
        \label{fignonlineara}}
    \subfigure{    
        \begin{tikzpicture}
            \node[anchor=north west, inner sep=0] (image) at (0,0) {
                \includegraphics[width=0.48\textwidth]{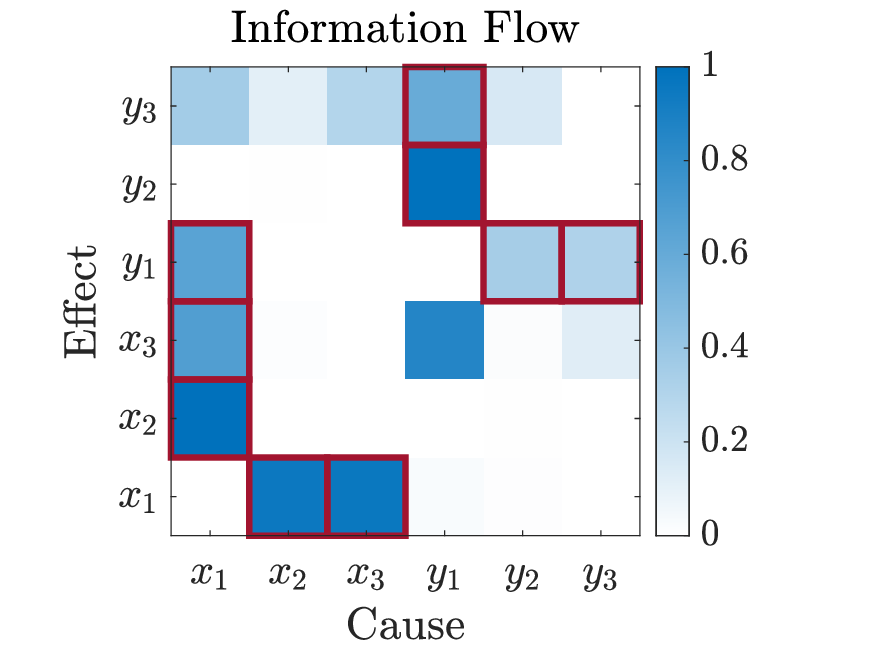}
            };
            \node[anchor=north west] at (image.north west) {{\rmfamily\fontsize{12}{13}\fontseries{l}\selectfont(b)}};
        \end{tikzpicture}
        \label{fignonlinearb}}
    \subfigure{
        \begin{tikzpicture}
            \node[anchor=north west, inner sep=0] (image) at (0,0) {
                \includegraphics[width=0.85\textwidth]{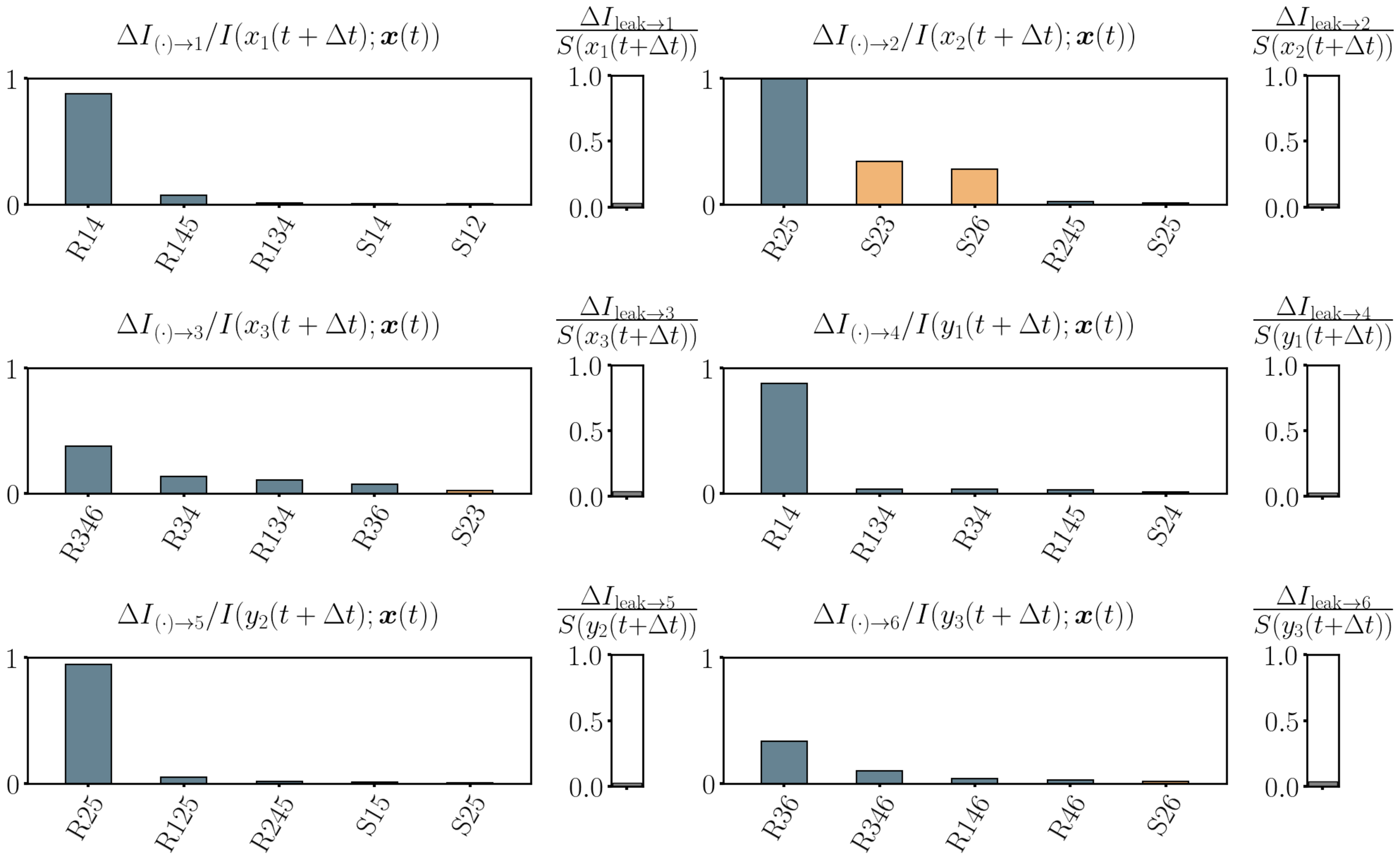}
            };
            \node[anchor=north west] at (image.north west) {{\rmfamily\fontsize{12}{13}\fontseries{l}\selectfont(c)}};
        \end{tikzpicture}
        \label{fignonlinearc}}
    \caption{Causal maps of the nonlinear problem. (a) Transfer entropy; (b) Information flow; (c) SURD. Red boxes: the causal links in the system. }
    \label{fignonlinear}
\end{figure}

Following \citet{paluvs2018causality}, we consider the following nonlinear system: 
\begin{equation}\label{nonlineareq}
    \begin{aligned}
    &\mathcal{X}=\left\{
    \begin{array}{ll}
    {\rm d}x_1/{\rm d}t = -\omega_1x_2(t) - x_3(t), \\
    {\rm d}x_2/{\rm d}t = \omega_1x_1(t) + 0.15x_2(t), \\
    {\rm d}x_3/{\rm d}t = 0.2 + x_3(t)\left[x_1(t)-10\right],
    \end{array}
    \right. \\
    &\mathcal{Y}=\left\{
    \begin{array}{ll}
    {\rm d}y_1/{\rm d}t = -\omega_2y_2(t) - y_3(t) + \epsilon\left[x_1(t)-y_1(t)\right], \\
    {\rm d}y_2/{\rm d}t = \omega_2y_1(t) + 0.15y_2(t), \\
    {\rm d}y_3/{\rm d}t = 0.2 + y_3(t)\left[y_1(t)-10\right],
    \end{array}
    \right.
    \end{aligned}
\end{equation}
in which, $\omega_1 = 1.015$, $\omega_2 = 0.985$, and $\epsilon$ is the coupling strength parameter, which is set to $0.25$ in this study. Eq.~(\ref{nonlineareq}) is a unidirectionally coupled R{\"o}ssler system, \emph{i.e.}, $\mathcal{X}$ is the driving or master system and $\mathcal{Y}$ is the driven system. 

The nonlinear system is numerically solved using a variable-order numerical differentiation formula (NDF) method (the `ode15s' function in MATLAB) with a fixed time step of $\Delta t=0.001$. The numerical solution generated a total of 500 realisations. Each realisation begins from a random initial condition and evolves for $N=500,000$ time steps, with the first 100,000 steps discarded as transient to ensure that the solution reaches its attractor. Causal analysis is performed with a time lag of $2\Delta t$ to resolve the characteristic timescale of coupled dynamics \citep{paluvs2018causality}. 

Figure~\ref{fignonlinear} presents the causal inference results of the nonlinear system. It is seen that both transfer entropy and information flow correctly identify the primary driver-response causal relationships specified in eq.~(\ref{nonlineareq}), \emph{i.e.} the link $x_1 \to y_1$. 
However, when examining other causal links, some differences emerge between the two methods. It is seen that both methods can detect the causal links of $x_1 \leftrightarrow x_2$ and $y_1 \leftrightarrow y_2$, but the information flow can produce more pronounced values. 
In particular, the information flow can clearly identify $x_1 \leftrightarrow x_3$ and $y_1 \leftrightarrow y_3$, while the values in the transfer entropy result are quite small. In addition, much richer causalities are identified by transfer entropy than by information flow. 
In figure~\ref{fignonlinearc}, the variables are indexed as $(1,2,3) \to (x_1,x_2,x_3)$ and $(4,5,6) \to (y_1,y_2,y_3)$ to illustrate the SURD result. Similar causal links are found with the primary causalities identified by transfer entropy, \emph{e.g.}, the primary driver-response causal link $x_1 \to y_1$ manifested as the redundant causality $\Delta I_{14 \to 4}^R$ in SURD. 
The link $y_1 \to x_1$ identified by transfer entropy is manifested by the redundant causality $\Delta I_{14 \to 1}^R$.
The links $y_2 \leftrightarrow x_2$ are manifested by the redundant causalities $\Delta I_{25 \to 2}^R$ and $\Delta I_{25 \to 5}^R$. The link $x_3 \to y_3$ is by the redundant one $\Delta I_{36 \to 6}^R$. 

\subsection{The low-dimensional model of near-wall turbulence}

\begin{figure}
    \centering
    \subfigure{
        \begin{tikzpicture}
            \node[anchor=north west, inner sep=0] (image) at (0,0) {
                \includegraphics[width=0.48\textwidth]{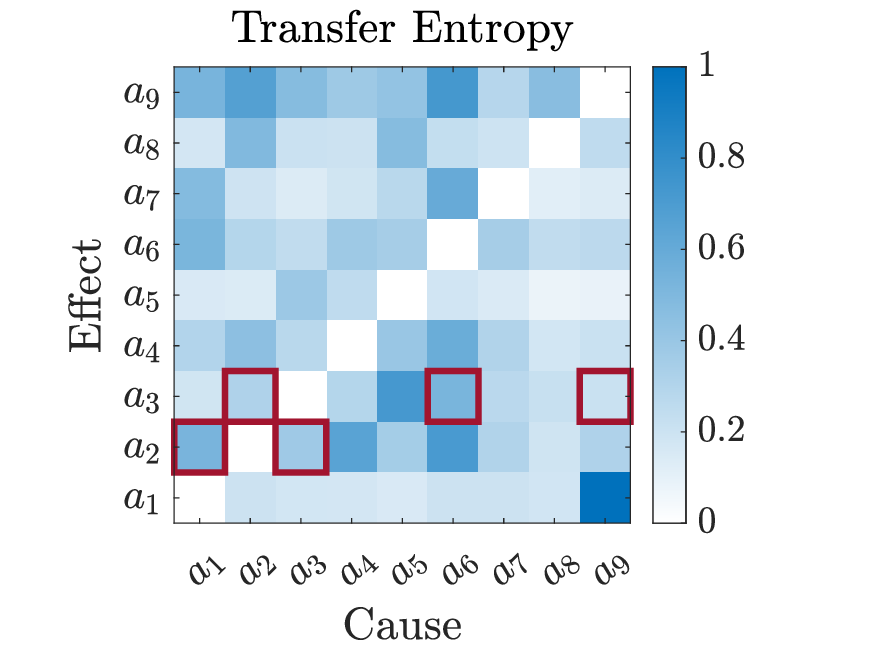}
            };
            \node[anchor=north west] at (image.north west) {{\rmfamily\fontsize{12}{13}\fontseries{l}\selectfont(a)}};
        \end{tikzpicture}
        \label{figloworderta}}
    \subfigure{    
        \begin{tikzpicture}
            \node[anchor=north west, inner sep=0] (image) at (0,0) {
                \includegraphics[width=0.48\textwidth]{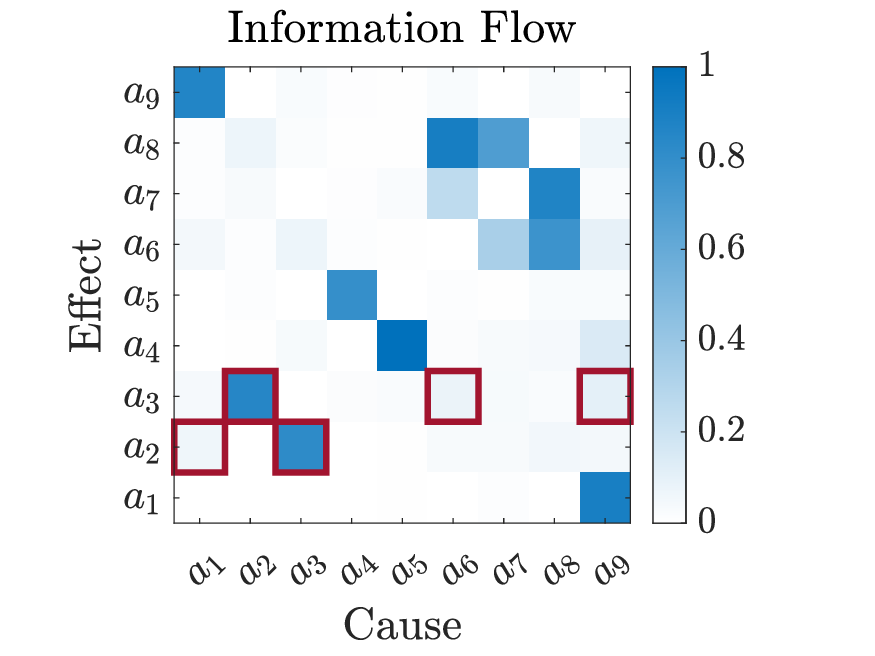}
            };
            \node[anchor=north west] at (image.north west) {{\rmfamily\fontsize{12}{13}\fontseries{l}\selectfont(b)}};
        \end{tikzpicture}
        \label{figlowordertb}}
    \subfigure{
        \begin{tikzpicture}
            \node[anchor=north west, inner sep=0] (image) at (0,0) {
                \includegraphics[width=0.85\textwidth]{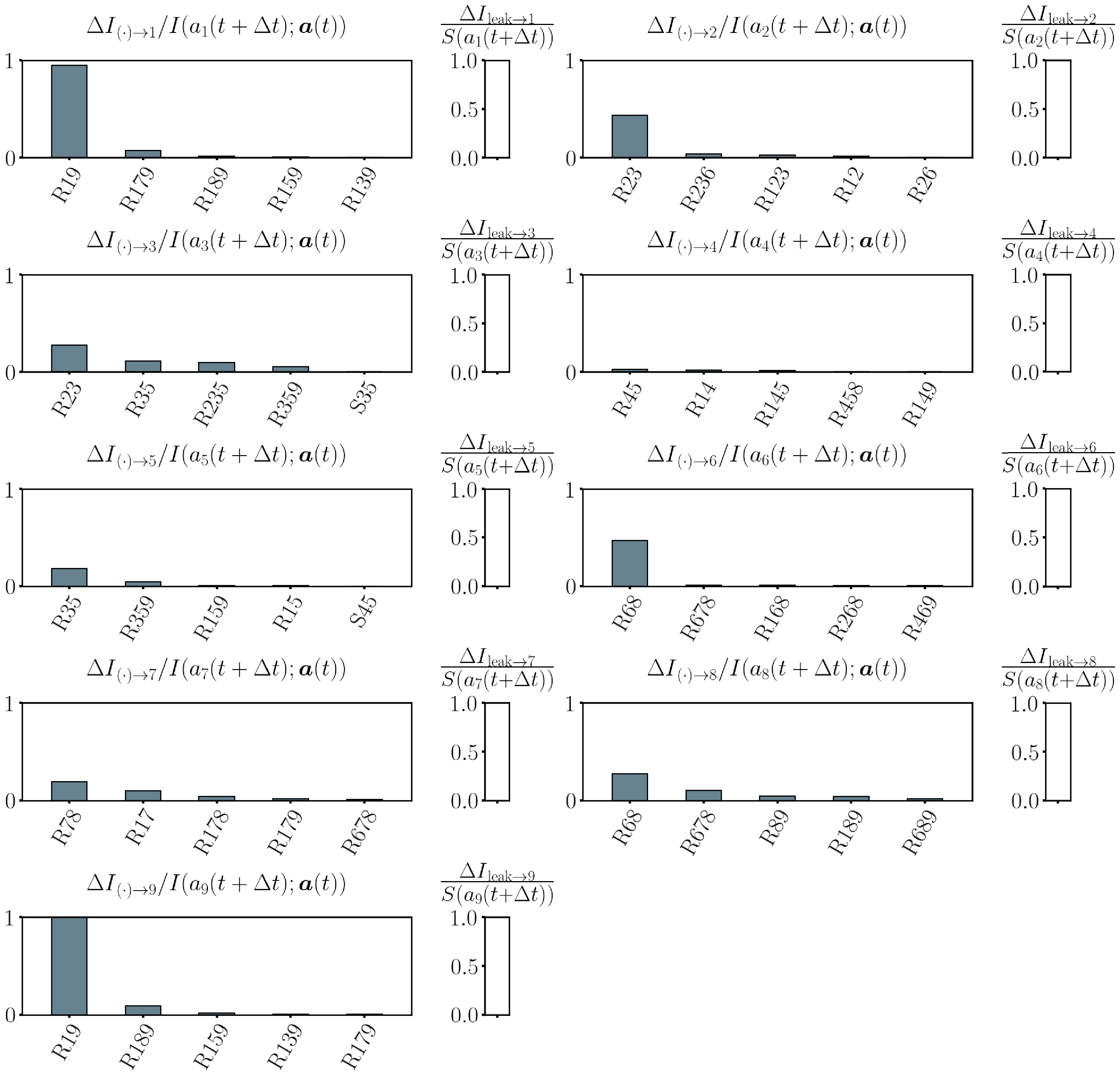}
            };
            \node[anchor=north west] at (image.north west) {{\rmfamily\fontsize{12}{13}\fontseries{l}\selectfont(c)}};
        \end{tikzpicture}
        \label{figlowordertc}}
    \caption{Causal maps of the low-dimensional model of near-wall turbulence. (a) Transfer entropy; (b) Information flow; (c) SURD. Red boxes: the causal links associated with the self-sustaining cycle of streaks and vortex. }
    \label{figlowordert}
\end{figure}


Next, we analyse the causal relations in a low-dimensional model of the plane Couette flow with a sinusoidal body force \citep{moehlis2004low}. It presents an improvement of the eight-mode model of \citet{Waleffe1997}. The self-sustaining process of near-wall turbulence, \emph{i.e.} the streak-vortex cycle, can be well represented by the low-dimensional model. 
There are nine modes of the model, which are (1) the basic profile mode, (2) the streak mode, (3) the downstream vortex mode, (4, 5) the spanwise flow modes, (6, 7) the normal vortex modes, (8) a fully three-dimensional mode, and (9) the mode of modification of the basic profile. The first eight modes are the same as those in \citet{Waleffe1997}.
By applying the Galerkin projection to the Navier-Stokes equations without the pressure term, a set of nine ordinary differential equations is obtained for the temporal amplitude coefficients $a_i(t)$ ($i=1$ to 9), featuring quadratic nonlinearities. 

Here we adopt the same parameters as \cite{martinez2023causality}. The system is numerically solved to generate 500 realisations, each spanning 4,000 time units with a fixed time step of 0.01 (400,000 steps per realisation). 
The solution data are generated by applying a {uniformly distributed} random perturbation to initialise $a_4$. The initial conditions are $(1, 0.07066, -0.07076, 0, 0, 0, 0, 0, 0)$ for the nine equations, which are solved using the variable-order differentiation method. In causal analysis, the time lag is set to one time step ($\Delta t=0.01$). 

Figure~\ref{figlowordert} shows the causal inference results of the low-dimensional model, which reveals distinct sensitivities of methodologies to capture the self-sustaining mechanism of near-wall turbulence. 
First, it can be seen that both transfer entropy and information flow can successfully identify the core interactions, including the lift-up mechanism ($a_1 \to a_2$ and $a_3 \to a_2$) and streak-to-vortex mechanism ($a_2 \to a_3$). 
In figure~\ref{figlowordertc}, these two mechanisms are manifested through the redundant causalities of $\Delta I_{23 \to 2}^R$ and $\Delta I_{23 \to 3}^R$. 
Second, figure~\ref{figlowordert} also shows that transfer entropy identifies a wider spectrum of causalities, while information flow detects a more compact one. 
It indicates that information flow can detect primary energy transfers, while the transfer entropy shows a higher sensitivity to higher-order interactions, particularly spanwise flow effects ($a_4 \to a_2$, $a_5 \to a_3$) and three-dimensional mode couplings ($a_6 \to a_2$). 
Third, the causalities identified by SURD exhibit a certain similarity with information flow. 
The causal links $a_9 \leftrightarrow a_1$ identified by information flow manifest themselves as redundant causalities $\Delta I_{19 \to 1}^R$ and $\Delta I_{19 \to 9}^R$ in the SURD result. The causal links $a_6 \leftrightarrow a_8$ correspond to the redundant causalities $\Delta I_{68 \to 6}^R$ and $\Delta I_{68 \to 8}^R$. The causal links $a_7 \leftrightarrow a_8$ resemble the redundant causalities $\Delta I_{78 \to 7}^R$ and $\Delta I_{678 \to 8}^R$.

\subsection{Discussions}

{The results of the nonlinear problem and low-dimensional model of near-wall turbulence problem imply that different methods may have different sensitivities to nonlinearity. It should be beneficial to identify the actual causal links and evaluate causal inference methods if the ground truth is known. However, it is often not possible for a nonlinear system, as nonlinear couplings can propagate causal influences across variables over finite time lags, making a ground truth solely based on dynamical equations infeasible.}

{Taking the low-dimensional model of near-wall turbulence as an example, the dynamical equations of the system are given in Appendix~\ref{appA}. First, it is observed that $a_9$ does not appear on the right-hand side of ${\rm d}a_1/{\rm d}t$, while all the three methods identify the causal link of $a_9 \to a_1$. By subtracting equation~(\ref{eqn:a9}) from (\ref{eqn:a1}), one can obtain a differential equation that contains only $a_1$ and $a_9$ (and their derivatives), \emph{i.e.} (\ref{eqn:a1_a9}).
This indicates that $a_1$ and $a_9$ are closely coupled with each other, hence the identified causal link has a theoretical foundation. 
The second is the identified causal link of $a_2 \to a_3$. It is seen that $a_2$ is absent in the dynamical equation of $a_3$, \emph{i.e.} (\ref{eqn:a3}). 
By multiplying the equations of each mode by their respective coefficients and summing them, one can completely eliminate all crossing terms and obtain (\ref{eqn:sum_ai}), in which all coefficients are coupled. As a result, the reformulation of the dynamical equations reveals the possible existence of the causal link from $a_2$ to $a_3$. 
The above two examples illustrate the complexity of nonlinear problems, in which the underlying causal relationships may not be explicit in dynamical systems. 
}

{Nevertheless, the key point is that all three methods successfully identify the primary driver–response relationship ($x_1 \to y_1$) in the nonlinear system (\S 3.2) and the core streak–vortex self-sustaining cycle in the low-dimensional turbulence model (\S 3.3). These are the most physically important causal links in each system. The fact that the methods capture these essential mechanisms, despite their differences in secondary links, demonstrates their applicability in the analyses of complex nonlinear problems.}

{To further strengthen the reliability of the causal analysis of the turbulent channel flow, we adopt a conservative interpretation strategy in the following sections. The causal links that are consistently identified by at least two of the three methods are kept, which significantly reduces the uncertainty and risk of spurious detections.}
{Moreover, sensitivity analyses with respect to the data length and histogram bin number for the verification cases are provided in the supplementary material, confirming the robustness of the methods. }

\section {Inner-outer decomposition of near-wall turbulence}\label{iod}

The components of the turbulent flow velocity of near-wall inner and outer motions are decomposed using a scaling-improved inner-outer decomposition method based on the predictive inner-outer model (PIOM) \citep{Marusic2010, Mathis2011, Baars2016}. 
In the PIOM framework, the velocity of turbulent flow near the wall can be expressed as 
\begin{equation}
  u_{i} = u_{i,S} + u_{i,L} = {u_i^*\left(1+ \Gamma_{i} u_{i,L}\right)} + u_{i,L},  
\end{equation} 
where $u_{i}$ is the velocity fluctuation in the $i$th direction ($i=1,2,3$ corresponding to the streamwise, wall-normal, and spanwise directions, respectively), $u_{i, S}=u_i^*\left(1+ \Gamma_{i} u_{i,L}\right)$ denotes the velocity fluctuation of inner motions, $u_i^*$ is the unmodulated velocity fluctuation of inner motions or the universal near-wall turbulence cycle, $\Gamma_{i}$ is the modulation coefficient, and $u_{i,L}$ indicates the velocity fluctuation of outer motions or the footprint of large-scale turbulent motions in the outer layer. Here, lowercase letters consistently indicate velocity fluctuations. The near-wall region is restricted to $y^+ \le 100$ \citep{Wang2021A}. The "+" superscript indicates the normalisation in wall units by the friction velocity $u_\tau$ and fluid kinematic viscosity $\nu$. 

The velocity of outer motions is calculated by \citep{Baars2016,Wang2021A}: 
\begin{equation}\label{ls} 
    u_{i,L}(x,y,z,t)= \\ 
    F_x^{-1} \left\{\hat{H}_{i,L}(\lambda_x,y)F_x \left[u_i(x,y_o,z,t) \right] \right\},
\end{equation}
where $F_x$ and $F_x^{-1}$ represent the streamwise Fourier transform and the inverse Fourier transform, respectively; $\lambda_x$ denotes the streamwise wavelength; $x,y,z$ correspond to the streamwise, wall-normal, and spanwise coordinates; $y_o$ specifies the outer reference height for the calculation of the outer footprint near the wall, which is set to $y^+_o=100$ according to \citet{Wang2021A}; $\hat{H}_{i,L}$ is the scale-dependent complex-valued kernel function derived from linear stochastic spectral estimation \citep{Baars2016}, which is defined as: 
\begin{equation}\label{kf} 
    \hat{H}_{i,L}(\lambda_x,y)=\frac{\langle \hat{u}_i(\lambda_x,y,z,t) \overline{\hat{u}_i(\lambda_x,y_o,z,t)} \rangle_{z,t}}{\langle \hat{u}_i(\lambda_x,y_o,z,t) \overline{\hat{u}_i(\lambda_x,y_o,z,t)} \rangle_{z,t}},
\end{equation}
where $\hat{(\cdot)}$ represents variables in Fourier space, the overbar represents complex conjugation, and $\langle \cdot \rangle_{z,t}$ denotes the averaging in spanwise and time. 

Based on the IOIM and $y^+_o=100$, \cite{Wang2021A} successfully extracted universal (Reynolds number independent) inner motions from DNS data of turbulent channel flows at friction Reynolds numbers $Re_\tau = 1000,2000$ and $5200$, where $Re_\tau=u_{\tau}\delta/\nu$, and $\delta$ is the half height of the channel.
However, \citet{Wang2021A} only showed the autocorrelations of each velocity component itself, but did not show the cross-correlation between different components, \emph{e.g.} the Reynolds shear stress (RSS).
Since inner motions are generated by an autonomous and self-sustaining regeneration cycle which is {independent} of outer motions, the correlations of cross-velocity components between inner and outer motions should also be zero.

In the inner-outer decomposition framework, the total RSS can be expressed as: 
\begin{equation}\label{rsd} 
    \langle uv \rangle = \langle u_Sv_S \rangle + \langle u_Lv_L \rangle + \langle u_Sv_L \rangle + \langle u_Lv_S \rangle, 
\end{equation}
where the first two terms $\langle u_S v_S \rangle$ and $\langle u_L v_L \rangle$ on the right-hand side represent the RSS components contributed by the inner and outer motions themselves, respectively. The remaining two terms $\langle u_S v_L \rangle$ and $\langle u_L v_S \rangle$ arise from the cross-correlations between the inner and outer motions, both of which are expected to be zero. 

\begin{figure}  \centerline{\includegraphics[width=0.8\textwidth]{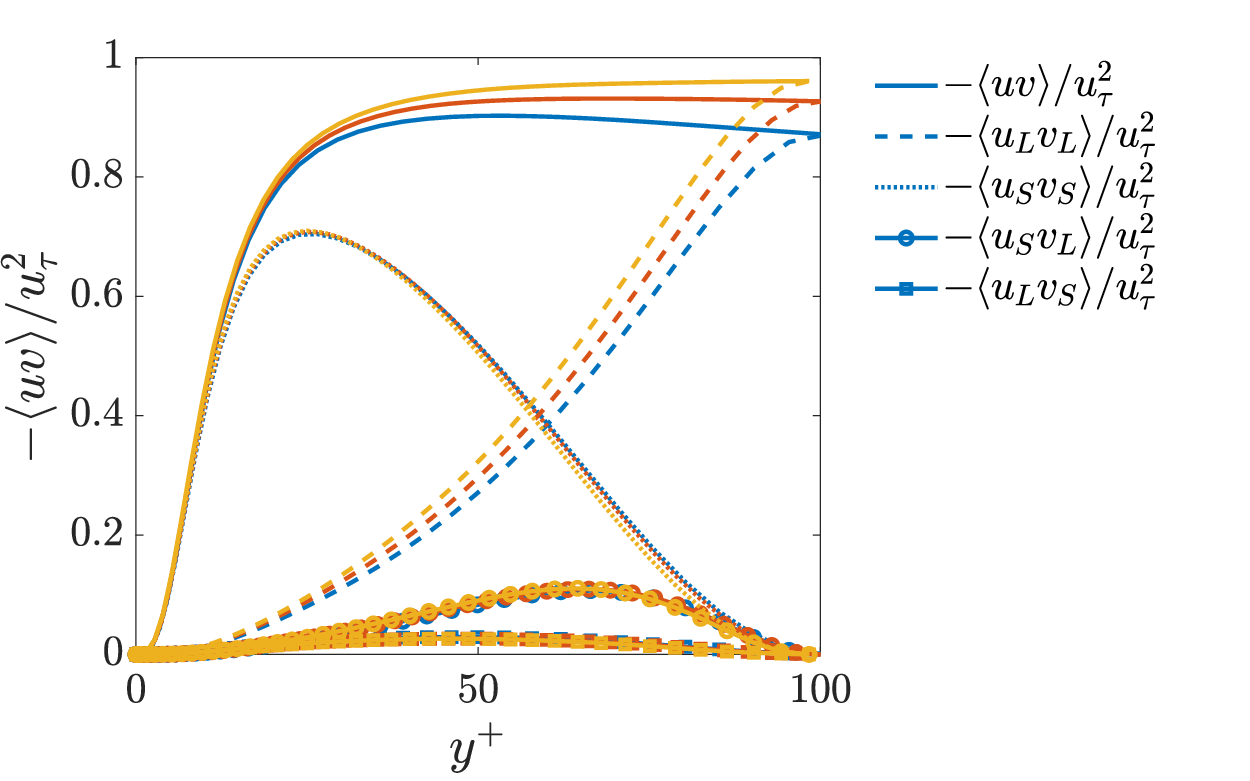}}
  \caption{Wall-normal profiles of the Reynolds shear stress and its decomposed components at $Re_\tau=1000,2000$, and $5200$. The line colours correspond to the cases listed in table~\ref{detailinfor}. }
\label{figrsd}
\end{figure}

We will evaluate whether $\langle u_S v_L \rangle$ and $\langle u_L v_S \rangle$ are zero using high-fidelity DNS data of turbulent channel flows. The DNS datasets at $Re_{\tau}=1000$ and $5200$ were generated by \citet{Lee2015} and obtained from the Johns Hopkins Turbulence Database (JHTDB) \citep{graham2016web}, while the $Re_{\tau}=2000$ dataset is from \citet{hoyas2006scaling}. 
Detailed parameters for the DNS data are provided in table~\ref{detailinfor}. 
Figure~\ref{figrsd} presents the wall-normal profiles of the RSS and its decomposed components in (\ref{rsd}). 
It can be seen that the total RSS $\langle uv \rangle$ exhibits a notable Reynolds number dependence, which increases with $Re_\tau$, indicating the growing influence of outer motions on the near-wall RSS. 
For the decomposed components, the inner RSS $\langle u_Sv_S \rangle$ appears to be independent of the Reynolds number, and the cross RSS components $\langle u_Sv_L \rangle$ and $\langle u_Lv_S \rangle$ exhibit a very weak Reynolds number dependence. However, due to the Reynolds number effect of the outer motions, the outer RSS $\langle u_Lv_L \rangle$ maintains a visible Reynolds number effect. As evident in figure~\ref{figrsd}, both of the cross RSS components $\langle u_Sv_L \rangle$ and $\langle u_Lv_S \rangle$ are nonzero, which indicates that some correlations still exist between the inner and outer motions. This contradicts the {independent} nature between the inner and outer motions, thus calling for a further improvement of the inner-outer decomposition method. 

\begin{table}
    \centering
    \begin{tabular}{c c c c c c c c c}
        $Re_{\tau}$ & $L_x/\delta$ & $L_z/\delta$ & $\Delta x^+$ & $\Delta z^+$ &
        $\Delta y^+_w$ & $\Delta y^+_c$ & $Tu_\tau/\delta$ & Line Color \\
          1000        &    $8\pi$    &    $3\pi$    &     12.3     &     6.14     &
          0.017        &       6.15     &        12.5      &  
          \begin{tabular}[c]{@{}c@{}}
                    \vspace{0.05cm}
                    {\color[rgb]{0,0.4470,0.7410}\rule{1cm}{0.5mm}}
                    \vspace{0.1cm}
                    \end{tabular} \\
          2003         &    $8\pi$    &    $3\pi$    &     8.2      &     4.10     &
          0.323        &       8.89     &        10.3      &  
          \begin{tabular}[c]{@{}c@{}}
                    \vspace{0.05cm}
                    {\color[rgb]{0.8500,0.3250,0.0980}\rule{1cm}{0.5mm}}
                    \vspace{0.1cm}
                    \end{tabular} \\
          5200         &    $8\pi$    &    $3\pi$    &     12.7     &     6.38     &
          0.07         &       10.3     &        7.80      &  
          \begin{tabular}[c]{@{}c@{}}
                    \vspace{0.05cm}
                    {\color[rgb]{0.9290,0.6940,0.1250}\rule{1cm}{0.5mm}}
                    \vspace{0.1cm}
                    \end{tabular} \\
    \end{tabular}
    \caption{Summary of the DNS datasets. The computational domain sizes in the streamwise and spanwise directions are denoted by $L_x$ and $L_z$, respectively, scaled by the outer length scale $\delta$. The viscous-scaled grid resolutions are represented by $\Delta x^+$ and $\Delta z^+$ for the streamwise and spanwise directions, with $\Delta y^+_w$ and $\Delta y^+_c$ indicating the wall-normal grid spacings at the wall and channel centre, respectively. The time span of the data is normalised as $Tu_{\tau}/\delta$. }
    \label{detailinfor}
\end{table}

\subsection{An improved inner-outer decomposition method}

The reason for the existence of the nonzero cross RSS components lies in the fact that equation~(\ref{ls}) does not take into account the nonzero contributions of the outer footprints from the cross velocity component. 
For example, in the estimation of $u_L$, only the contribution of $u$ itself is included in (\ref{ls}), while that of $v$ is neglected. And so is $v_L$.
Therefore, we develop an improved inner-outer decomposition method that can include the effects of the cross velocity components in calculating the outer footprint velocity $u_{i,L}$. More specifically, we consider cross contributions between $u$ and $v$ as their product constitutes the non-zero RSS components.

First, we apply the spectral linear stochastic estimation (SLSE) to calculate the correlated component of the streamwise and wall-normal velocity fluctuations at $y_o^+=100$, as 
\begin{equation}\label{acu} 
    u^{c}(x,y_{o},z,t)= \\ 
    F_x^{-1} \left\{\hat{H}_{v}^{c}(\lambda_x)F_x \left[v(x,y_o,z,t) \right] \right\}, 
\end{equation}
\begin{equation}\label{acv} 
    v^{c}(x,y_{o},z,t)= \\ 
    F_x^{-1} \left\{\hat{H}_{u}^{c}(\lambda_x)F_x \left[u(x,y_o,z,t) \right] \right\}, 
\end{equation}
where
\begin{equation}\label{kfacu} 
    \hat{H}_{u}^c(\lambda_x)=\frac{\langle \hat{v}(\lambda_x,y_o,z,t) \overline{\hat{u}(\lambda_x,y_o,z,t)} \rangle_{z,t}}{\langle \hat{u}(\lambda_x,y_o,z,t) \overline{\hat{u}(\lambda_x,y_o,z,t)} \rangle_{z,t}}, 
\end{equation}
\begin{equation}\label{kfacv} 
    \hat{H}_{v}^c(\lambda_x)=\frac{\langle \hat{u}(\lambda_x,y_o,z,t) \overline{\hat{v}(\lambda_x,y_o,z,t)} \rangle_{z,t}}{\langle \hat{v}(\lambda_x,y_o,z,t) \overline{\hat{v}(\lambda_x,y_o,z,t)} \rangle_{z,t}},
\end{equation}
are the scale-dependent kernel functions. 

Next, the uncorrelated components of the streamwise and wall-normal velocity fluctuations at the reference height can be obtained by 
\begin{equation}\label{uiac} 
    u^{uc}(x,y_{o},z,t)= \\
    u(x,y_o,z,t)-u^{c}(x,y_o,z,t),
\end{equation}
\begin{equation}\label{viac} 
    v^{uc}(x,y_{o},z,t)= \\
    v(x,y_o,z,t)-v^{c}(x,y_o,z,t).
\end{equation}

In the following, we will use $v^{uc}$ and $u^{uc}$ to calculate the outer footprint velocity components in $u$ and $v$ from the cross-velocity contributions, respectively, as the contributions from $v^c$ and $u^c$ have been included in $u$ and $v$ themselves.
Consequently, the improved estimation for the outer footprint velocity components can be written as 
\begin{equation}\label{nlsu} 
    \begin{split}
    u_{L}(x,y,z,t) = & F_x^{-1} \left\{\hat{H}_{u,L}(\lambda_x,y)F_x \left[u(x,y_o,z,t) \right] \right\} \\
    &+F_x^{-1} \left\{\hat{H}_{v,L}^{uc}(\lambda_x,y)F_x \left[v^{uc}(x,y_o,z,t) \right] \right\},
    \end{split}
\end{equation}
\begin{equation}\label{nlsv} 
    \begin{split}
    v_{L}(x,y,z,t) = & F_x^{-1} \left\{\hat{H}_{v,L}(\lambda_x,y)F_x \left[v(x,y_o,z,t) \right] \right\} \\
    &+F_x^{-1} \left\{\hat{H}_{u,L}^{uc}(\lambda_x,y)F_x \left[u^{uc}(x,y_o,z,t) \right] \right\}.
    \end{split}
\end{equation}
In this formulation, the first term on the right-hand side remains identical to that in equation~(\ref{ls}), and the second term is from the cross-velocity contribution. The terms $\hat{H}_{u,L}^{uc}$ and $\hat{H}_{v,L}^{uc}$ represent scale-dependent kernel functions for calculating the cross velocity components, which are  
\begin{equation}\label{kffacu} 
    \hat{H}_{u,L}^{uc}(\lambda_x,y)=\frac{\langle \hat{v}(\lambda_x,y,z,t) \overline{\hat{u}^{uc}(\lambda_x,y_o,z,t)} \rangle_{z,t}}{\langle \hat{u}^{uc}(\lambda_x,y_o,z,t) \overline{\hat{u}^{uc}(\lambda_x,y_o,z,t)} \rangle_{z,t}},
\end{equation}
\begin{equation}\label{kffacv} 
    \hat{H}_{v,L}^{uc}(\lambda_x,y)=\frac{\langle \hat{u}(\lambda_x,y,z,t) \overline{\hat{v}^{uc}(\lambda_x,y_o,z,t)} \rangle_{z,t}}{\langle \hat{v}^{uc}(\lambda_x,y_o,z,t) \overline{\hat{v}^{uc}(\lambda_x,y_o,z,t)} \rangle_{z,t}}.
\end{equation}

Then the demodulated inner motions are obtained as 
\begin{equation}\label{demoduu}
    u^*(x, y, z, t) = \frac{u_S(x, y, z, t)}{1 + \Gamma_{uu}(y) u_L(x, y, z, t)}, 
\end{equation}
\begin{equation}\label{demoduv}
    v^*(x, y, z, t) = \frac{v_S(x, y, z, t)}{1 + \Gamma_{uv}(y) u_L(x, y, z, t)}, 
\end{equation}
\begin{equation}\label{demoduw}
    w^*(x, y, z, t) = \frac{w_S(x, y, z, t)}{1 + \Gamma_{uw}(y) u_L(x, y, z, t)}, 
\end{equation}
where $u^*$, $v^*$, and $w^*$ represent the demodulated inner motions, while $\Gamma_{uu}$, $\Gamma_{uv}$, and $\Gamma_{uw}$ denote the modulation coefficient. Both the demodulated inner motions and modulation coefficients are obtained iteratively. First, an initially prescribed $\Gamma_{uu}$, $\Gamma_{uv}$, and $\Gamma_{uw}$ are used to compute $u^*$, $v^*$, and $w^*$ via equation~(\ref{demoduu}), (\ref{demoduv}), and (\ref{demoduw}). 

Subsequently, the modulation coefficients are determined by
\begin{equation}\label{itermoducoeffu}
    M_{A}(u^{*}) = \frac{ \langle E_{L}(u^{*}) u_{L} \rangle }{ \sqrt{ \langle E_{L}(u^{*})^{2} \rangle } \sqrt{ \langle u_{L}^{2} \rangle } } 
\end{equation}
\begin{equation}\label{itermoducoeffv}
    M_{A}(v^{*}) = \frac{ \langle E_{L}(v^{*}) u_{L} \rangle }{ \sqrt{ \langle E_{L}(v^{*})^{2} \rangle } \sqrt{ \langle u_{L}^{2} \rangle } } 
\end{equation}
\begin{equation}\label{itermoducoeffw}
    M_{A}(w^{*}) = \frac{ \langle E_{L}(w^{*}) u_{L} \rangle }{ \sqrt{ \langle E_{L}(w^{*})^{2} \rangle } \sqrt{ \langle u_{L}^{2} \rangle } } 
\end{equation}
where $E_{L}$ denotes the signal envelope obtained by the Hilbert transform. When the modulation coefficients meet $M_{A}(u^{*})=0$, $M_{A}(v^{*})=0$, and $M_{A}(w^{*})=0$, the correct solutions of $\Gamma_{uu}$, $\Gamma_{uv}$, and $\Gamma_{uw}$ and $u^*$, $v^*$, and $w^*$ are obtained. Otherwise, the $\Gamma_{uu}$, $\Gamma_{uv}$, and $\Gamma_{uw}$ values must be adjusted iteratively until $M_{A}(u^{*})=0$, $M_{A}(v^{*})=0$, and $M_{A}(w^{*})=0$ are satisfied. 

\begin{figure}
    \centering
    \subfigure{
        \begin{tikzpicture}
            \node[anchor=north west, inner sep=0] (image) at (0,0) {
                \includegraphics[width=0.48\textwidth]{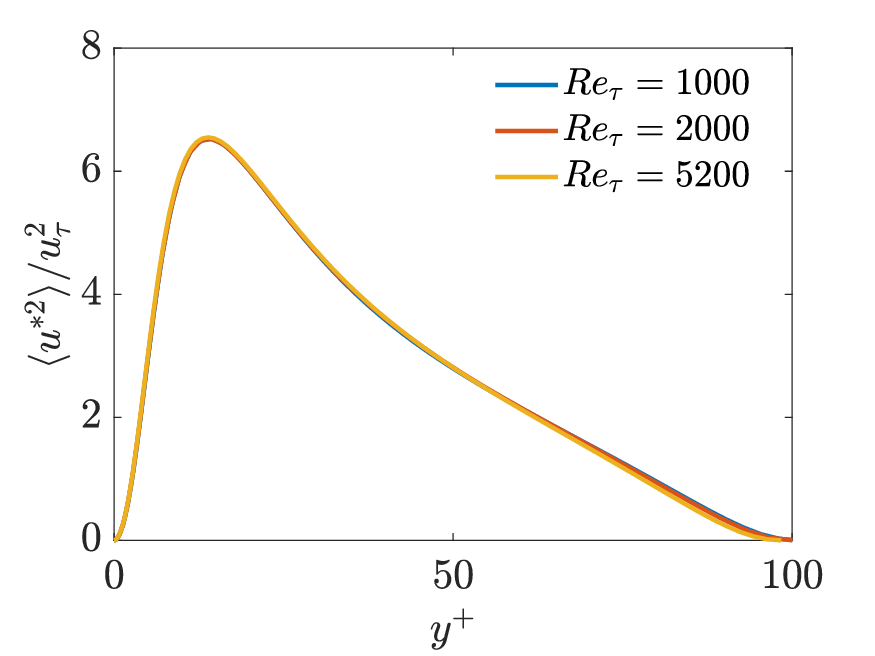}
            };
            \node[anchor=north west] at (image.north west) {{\rmfamily\fontsize{12}{13}\fontseries{l}\selectfont(a)}};
        \end{tikzpicture}
        \label{figtinusa}}
    \subfigure{    
        \begin{tikzpicture}
            \node[anchor=north west, inner sep=0] (image) at (0,0) {
                \includegraphics[width=0.48\textwidth]{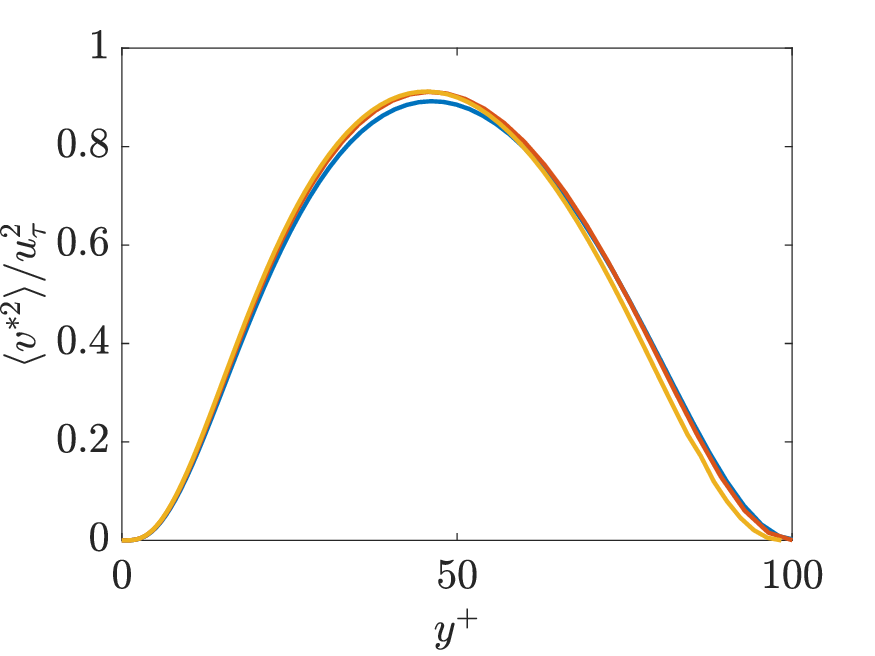}
            };
            \node[anchor=north west] at (image.north west) {{\rmfamily\fontsize{12}{13}\fontseries{l}\selectfont(b)}};
        \end{tikzpicture}
        \label{figtinusb}}
    \subfigure{
        \begin{tikzpicture}
            \node[anchor=north west, inner sep=0] (image) at (0,0) {
                \includegraphics[width=0.48\textwidth]{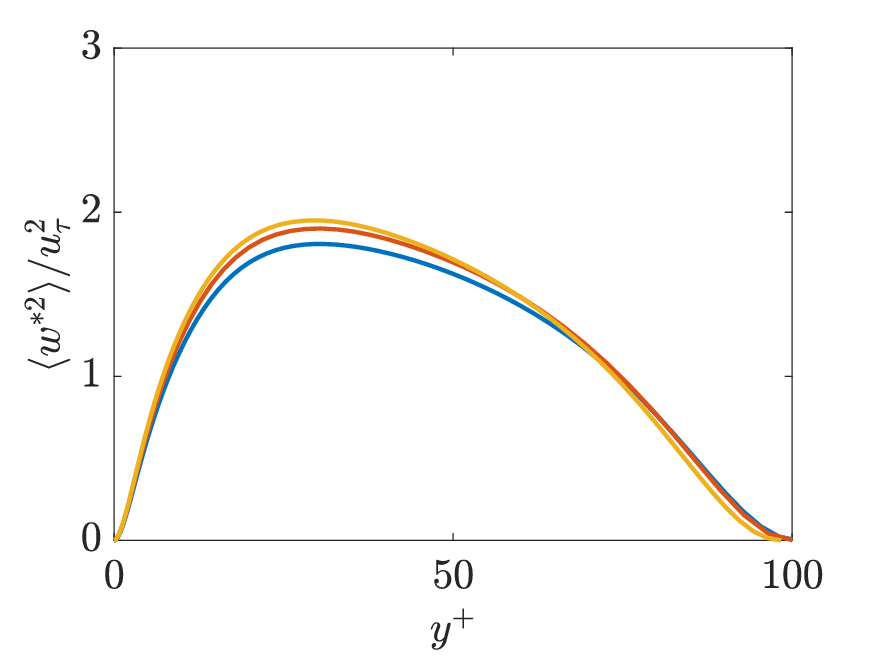}
            };
            \node[anchor=north west] at (image.north west) {{\rmfamily\fontsize{12}{13}\fontseries{l}\selectfont(c)}};
        \end{tikzpicture}
        \label{figtinusc}}
    \subfigure{
        \begin{tikzpicture}
            \node[anchor=north west, inner sep=0] (image) at (0,0) {
                \includegraphics[width=0.48\textwidth]{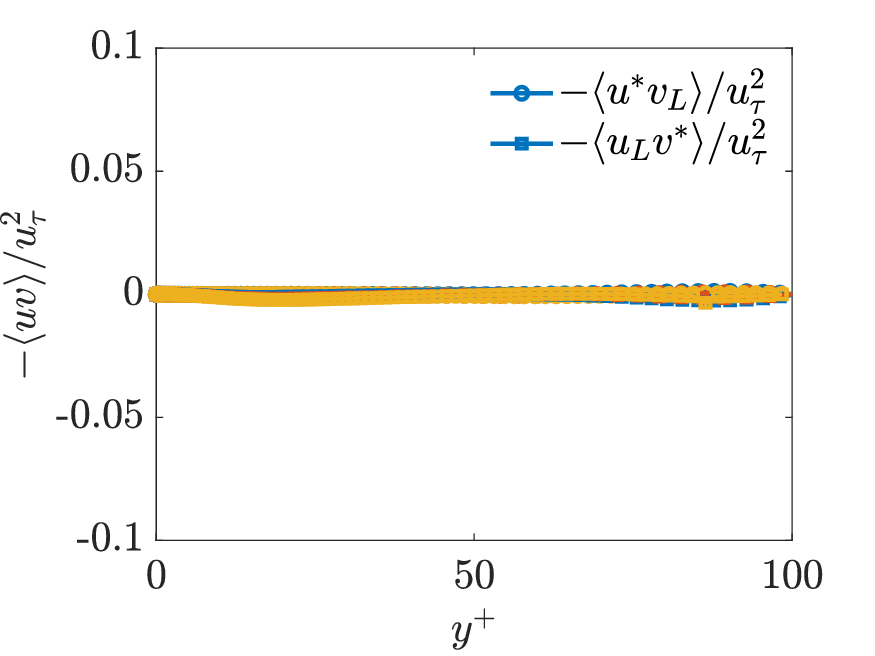}
            };
            \node[anchor=north west] at (image.north west) {{\rmfamily\fontsize{12}{13}\fontseries{l}\selectfont(d)}};
        \end{tikzpicture}
        \label{figtinusd}}
    \caption{Velocity fluctuation intensities of the demodulated inner motions at friction Reynolds numbers $Re_\tau=1000,2000$, and $5200$. (a) Streamwise velocity fluctuations, (b) wall-normal velocity fluctuations, (c) spanwise velocity fluctuations, (d) cross components of the decomposed Reynolds shear stress. The line colours correspond to the cases listed in table~\ref{detailinfor}. }
    \label{figtinus}
\end{figure}

Figure~\ref{figtinus} (a-c) shows that the intensities of the three velocity fluctuations of the inner motions by the improved method remain Reynolds number independent across the three Reynolds numbers. 
Figure~\ref{figtinus} (d) illustrates the cross components of the decomposed Reynolds shear stress, which are close to zero, indicating that there is no correlation between the cross velocities of the inner and outer motions. 
Furthermore, we note that the kernel functions defined in (\ref{kfacu}), (\ref{kfacv}), (\ref{kffacu}), and (\ref{kffacv}) also exhibit Reynolds number independence, similar to those in \citet{Wang2021A}.

For consistency, we maintain the same notation for inner and outer motions obtained through the newly improved decomposition method. All subsequent analyses exclusively use data derived from this new approach. 
{It should be noted, however, that the decomposition of turbulent motions into inner and outer components may not be unique. Different decomposition strategies (e.g., Fourier filtering, proper orthogonal decomposition, or wavelet-based scale separation) may yield different partitions of the velocity field. Consequently, the causal findings reported in the following sections may be specific to the present decomposition framework.}

\section{Causal analysis of inner and outer motions}\label{section5}

In this section, we employ the three causal inference methods, \emph{i.e.} transfer entropy, information flow and SURD, to analyse causal relationships between the inner and outer motions in near-wall turbulence. We aim to elucidate the generation and interaction mechanisms of the inner and outer motions through the causal inference methods. 
{We compute the causal measures over a range of time lag values and select the time lag that maximises the total causal inference for transfer entropy or information flow, with the self causal links set to zero. For SURD, the time lag corresponds to the time lag of the maximum cross-induced unique causality. The selection process based on the time lag dependence is further detailed in the supplementary material. }

\subsection{Dataset}
The dataset used in the present analysis is from a time-resolved DNS of turbulent channel flow at $Re_\tau=1000$. The numerical scheme combines fourth-order compact differences in the homogeneous directions and second-order central differences in the wall-normal direction for solving the incompressible Navier–Stokes equations on a staggered grid \citep{Hu2018application}. 
{The simulation was conducted in a computational domain of $L_x/\delta=4\pi$, $L_y/\delta=2$, and $L_z/\delta=2\pi$. Uniform grids were employed in the streamwise ($x$) and spanwise ($z$) directions with grid numbers of $N_x=N_z=1152$, resulting in grid resolutions of $\Delta x^+=10.9$ and $\Delta z^+=5.5$ in wall units. In the wall-normal direction, a non-uniform grid with $N_y=384$ was used, resulting in a grid spacing distribution of $\Delta y_w^+=0.4$ at the wall and $\Delta y_c^+=13.2$ at the channel centre.}
The dataset contains a flow duration of 19.8$\delta/u_\tau$, with instantaneous flow fields stored in a time interval of $\Delta t^+=0.9$, which yields 20,000 snapshots in total. 
The velocity and pressure fields at four wall-normal locations ($y^+=15,\ 70,\ 100,\ 200$) are stored. As demonstrated in Appendix~\ref{datav}, the present DNS dataset shows excellent agreement with the reference DNS data of \citet{Lee2015} at the same Reynolds number, confirming the accuracy and reliability of the present data. The statistical convergence of causal inference is assessed in Appendix~\ref{appB}.

\subsection{One-point causal anlaysis}

\begin{figure}
    \centering
    \subfigure{
        \begin{tikzpicture}
            \node[anchor=north west, inner sep=0] (image) at (0,0) {
                \includegraphics[width=0.48\textwidth]{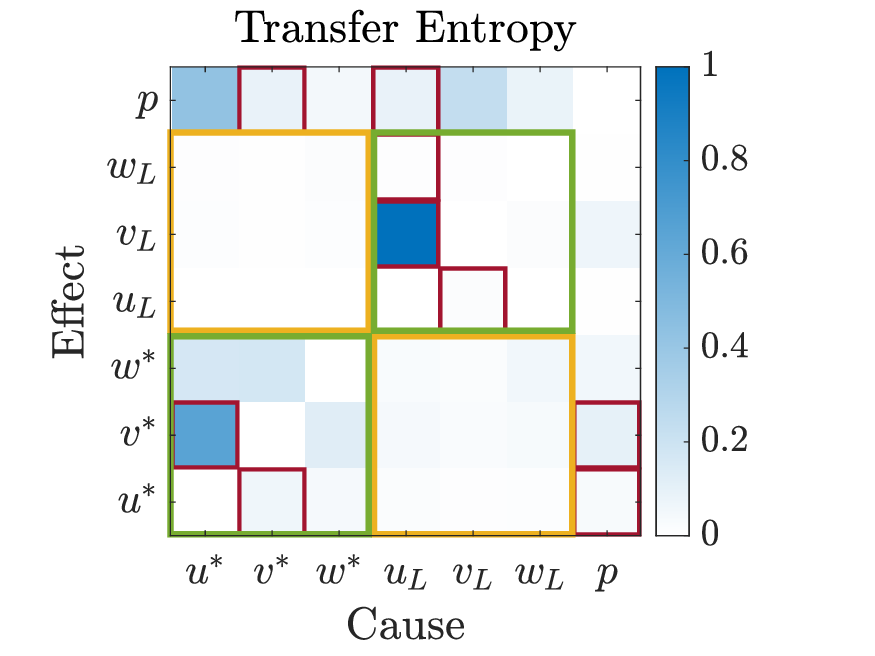}
            };
            \node[anchor=north west] at (image.north west) {{\rmfamily\fontsize{12}{13}\fontseries{l}\selectfont(a)}};
        \end{tikzpicture}
        \label{fig15stara}  
        \begin{tikzpicture}
            \node[anchor=north west, inner sep=0] (image) at (0,0) {
                \includegraphics[width=0.48\textwidth]{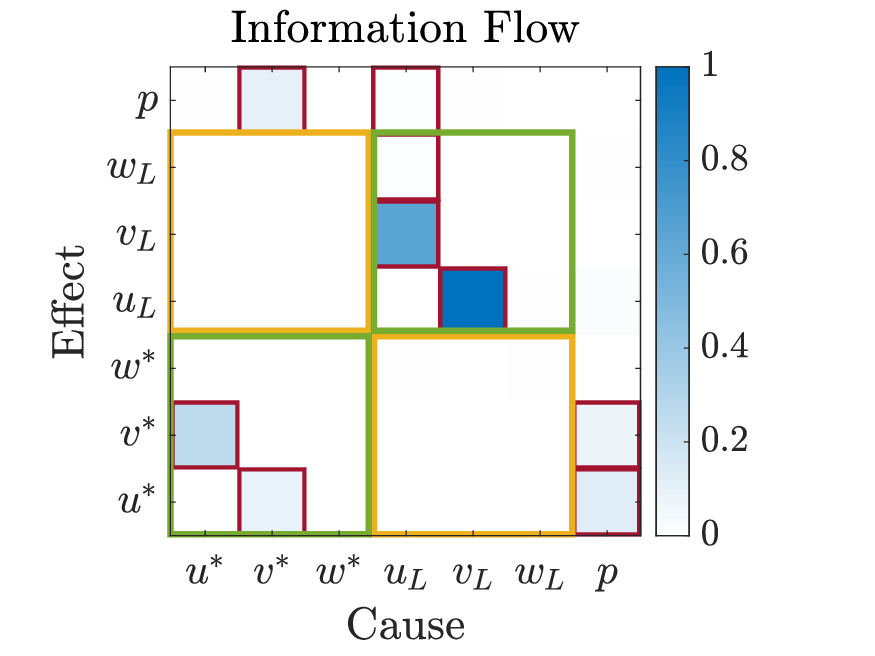}
            };
            \node[anchor=north west] at (image.north west) {{\rmfamily\fontsize{12}{13}\fontseries{l}\selectfont(b)}};
        \end{tikzpicture}
        \label{fig15starb}}
    \subfigure{
        \begin{tikzpicture}
            \node[anchor=north west, inner sep=0] (image) at (0,0) {
                \includegraphics[width=0.85\textwidth]{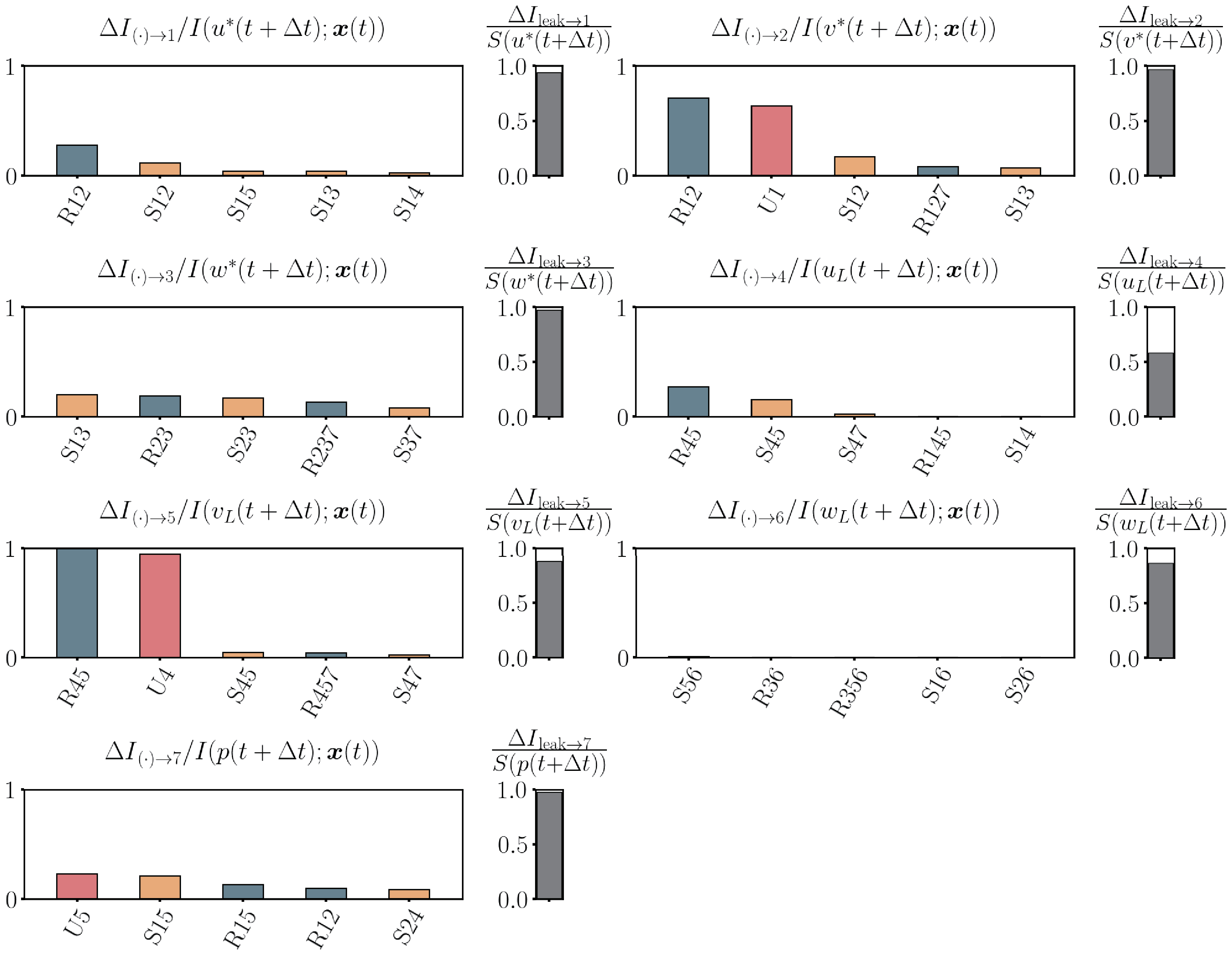}
            };
            \node[anchor=north west] at (image.north west) {{\rmfamily\fontsize{12}{13}\fontseries{l}\selectfont(c)}};
        \end{tikzpicture}
        \label{fig15starc}}
    \caption{{Causal maps for the demodulated inner and outer motions at $y^+=15$. (a) Transfer entropy, (b) information flow, and (c) SURD. {Yellow boxes}: inter-scale causalities. Cyan boxes: intra-scale causalities. Red boxes: causalities identified by both transfer entropy and information flow.} }
    \label{fig15star}
\end{figure}

\begin{figure}
    \centering
    \subfigure{
        \begin{tikzpicture}
            \node[anchor=north west, inner sep=0] (image) at (0,0) {
                \includegraphics[width=0.48\textwidth]{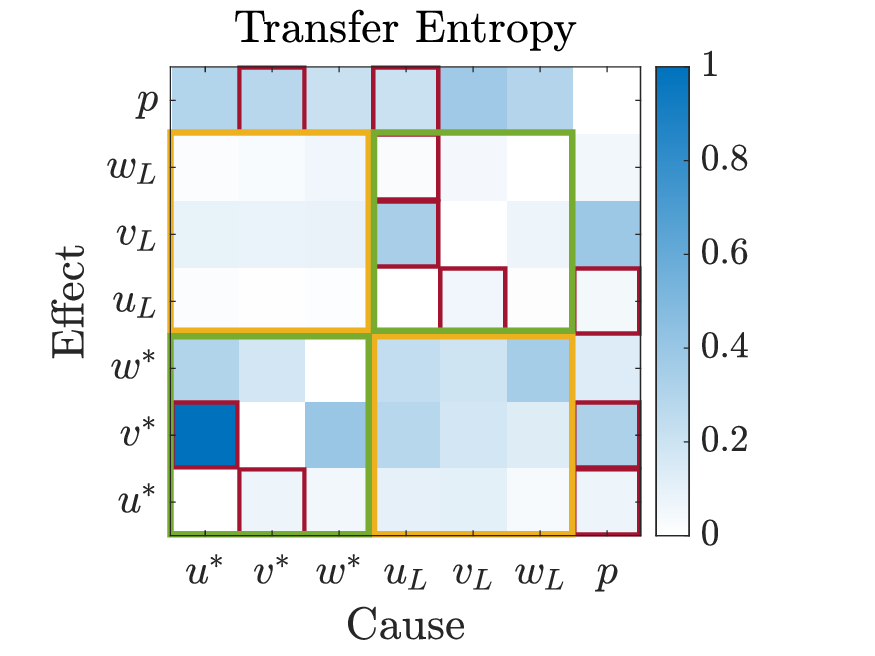}
            };
            \node[anchor=north west] at (image.north west) {{\rmfamily\fontsize{12}{13}\fontseries{l}\selectfont(a)}};
        \end{tikzpicture}
        \label{fig70stara}   
        \begin{tikzpicture}
            \node[anchor=north west, inner sep=0] (image) at (0,0) {
                \includegraphics[width=0.48\textwidth]{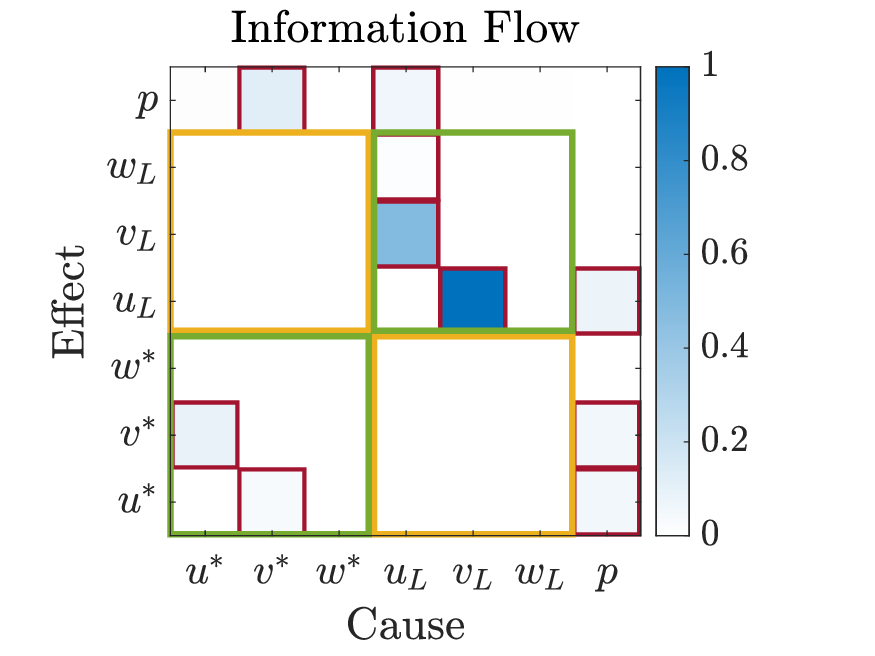}
            };
            \node[anchor=north west] at (image.north west) {{\rmfamily\fontsize{12}{13}\fontseries{l}\selectfont(b)}};
        \end{tikzpicture}
        \label{fig70starb}}
    \subfigure{
        \begin{tikzpicture}
            \node[anchor=north west, inner sep=0] (image) at (0,0) {
                \includegraphics[width=0.85\textwidth]{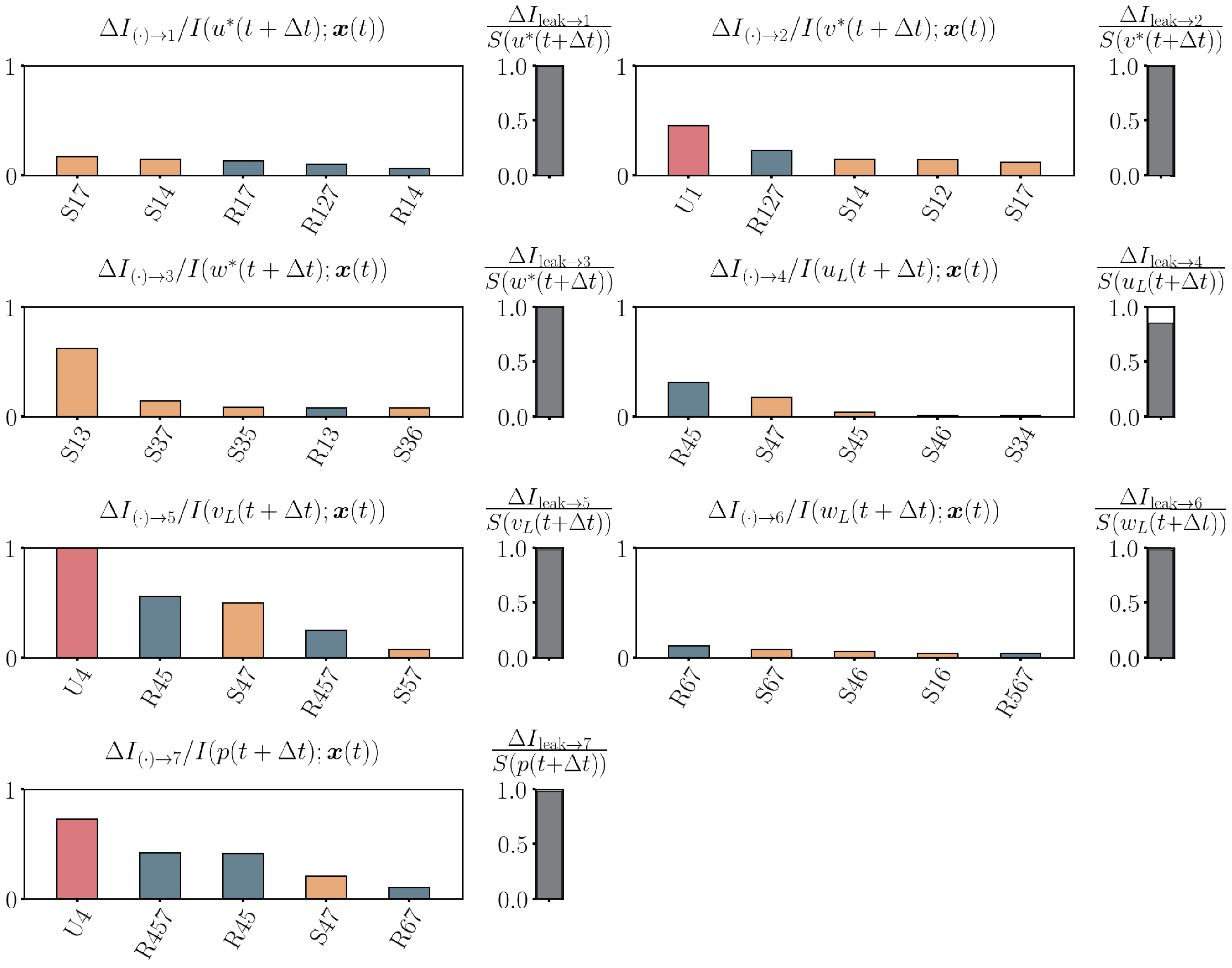}
            };
            \node[anchor=north west] at (image.north west) {{\rmfamily\fontsize{12}{13}\fontseries{l}\selectfont(c)}};
        \end{tikzpicture}
        \label{fig70starc}}
    \caption{{Causal maps for the demodulated inner and outer motions at $y^+=70$. (a) Transfer entropy, (b) information flow, and (c) SURD. {Yellow boxes}: inter-scale causalities. Cyan boxes: intra-scale causalities. Red boxes: causalities identified by both transfer entropy and information flow.} }
    \label{fig70star}
\end{figure}

We first examine the causalities of the {demodulated} inner and outer motions at the same location in a wall-parallel plane, \emph{i.e.} one-point causality. 
Time series for the variable vector $\mathbf{V}=\left(u^*,v^*,w^*,u_L,v_L,w_L,p\right)$ at 10,000 points uniformly distributed in a horizontal plane are adopted for the causal analysis. 

Figure~\ref{fig15star} displays the causal maps at $y^+=15$ where the peak of streamwise turbulence intensity is located approximately. 
From figures~\ref{fig15star} (a) and (b), one can see that transfer entropy can identify more causal links than information flow, which is similar to the results of the low-dimensional model of near-wall turbulence shown in figure~\ref{figlowordert}. 
The prominent causal links identified by both transfer entropy and information flow are highlighted via the red boxes in figure~\ref{fig15star} and the following analysis.
For example, one can find the intra-scale causal links of $u^* \leftrightarrow v^*$, which represent the self-sustaining streak-vortex cycle of the inner motions. In contrast, we can hardly identify any evident inter-scale causal links between the inner motions and the outer footprints at $y^+=15$.  
Second, it is found that pressure fluctuation is an active causal component. The notable causal links of $p \to u^*$ and $p \leftrightarrow v^*$ can be observed, which implies that pressure is actively involved in the self-sustaining near-wall turbulence cycle. This is understandable, since the vortex centre is usually located in a low-pressure region, and a stagnation is in a high-pressure region, both of which are typical flow topologies in turbulence. 

In figure~\ref{fig15star} (c), we can see that the causal inference result by SURD shows similar characteristics to the primary causalities identified by information flow, where the variables are indexed $(1,2,3,4,5,6,7) \to (u^*,v^*,w^*,u_L,v_L,w_L,p)$. 
{For example, the causal links of the inner streak-vortex generation cycle $u^* \leftrightarrow v^*$ are manifested as the redundant causality $\Delta I_{12 \to 1}^R$ and the unique causality $\Delta I_{1 \to 2}^U$ in the SURD result. The causal links of the outer footprint cycle $u_L \leftrightarrow v_L$ are identified by the redundant causality $\Delta I_{45 \to 4}^R$ and the unique causality $\Delta I_{4 \to 5}^U$. The pressure-related causal links $p \leftrightarrow v^*$ are reflected by the redundant causalities $\Delta I_{127 \to 2}^R$ and $\Delta I_{12 \to 7}^R$. }


Figure~\ref{fig70star} shows the causal maps at $y^+=70$, which is close to the upper bound of the buffer layer and more susceptible to the influence of outer motions.
In general, the causal features at this higher wall-normal location are quite similar to those at $y^+=15$, demonstrating some universality. 
Specifically, intra-scale causal links $u^* \leftrightarrow v^*$ and $u_L \leftrightarrow v_L$ can be observed, indicating that the self-sustaining cycles of both the demodulated inner motions and outer footprints exist at this height. This evidence supports the self-sustaining mechanism of turbulent motions at all scales \citep{hwang2010,hwang2011a,Hwang2015b,Rawat2015a,hwang2016a,Cossu2017Selfsustaininga,bae2021}. 
{There are also no evident inter-scale causal links between the inner motions and outer footprints at $y^+=70$. } 
Moreover, pressure fluctuations persist as an active component in the causal relationships at $y^+=70$, where the causal links of $p \to u^*$ and $p \leftrightarrow v^*$ remain discernible. 
This confirms the critical role of pressure in the sustainment of near-wall turbulence. 
{In figure~\ref{fig70star} (c), the causal links $u^* \leftrightarrow v^*$ correspond to the redundant causality $\Delta I_{127 \to 1}^R$ and the unique causality $\Delta I_{1 \to 2}^U$ in the SURD result. The causal links $u_L \leftrightarrow v_L$ are manifested as the redundant causalities $\Delta I_{45 \to 4}^R$ and the unique causality $\Delta I_{4 \to 5}^U$. And the causal links $p \to u^*$ and $p \to v^*$ are reflected by the synergistic causality $\Delta I_{17 \to 1}^S$ and the redundant causality $\Delta I_{127 \to 2}^R$. }



In summary, the one-point causal analysis has yielded the following insights. First, the demodulated inner motions and the outer footprints are demonstrated to be self-sustained and independent of each other, {the latter of which is further demonstrated with other statistics in the supplementary material}. 
This supports the idea that turbulent motions are self-sustained at all scales. In addition, the present causal inference results imply that the outer footprints are not passive with respect to their origin in the logarithmic layer but are active with local generation and evolution. Second, pressure plays a critical role in the self-sustaining cycles and may be regarded as a bridge connecting the inner and outer motions in the local sense.

\subsection{Two-point causal analysis}

\begin{figure}
    \centering
    \subfigure{
        \begin{tikzpicture}
            \node[anchor=north west, inner sep=0] (image) at (0,0) {
                \includegraphics[width=0.48\textwidth]{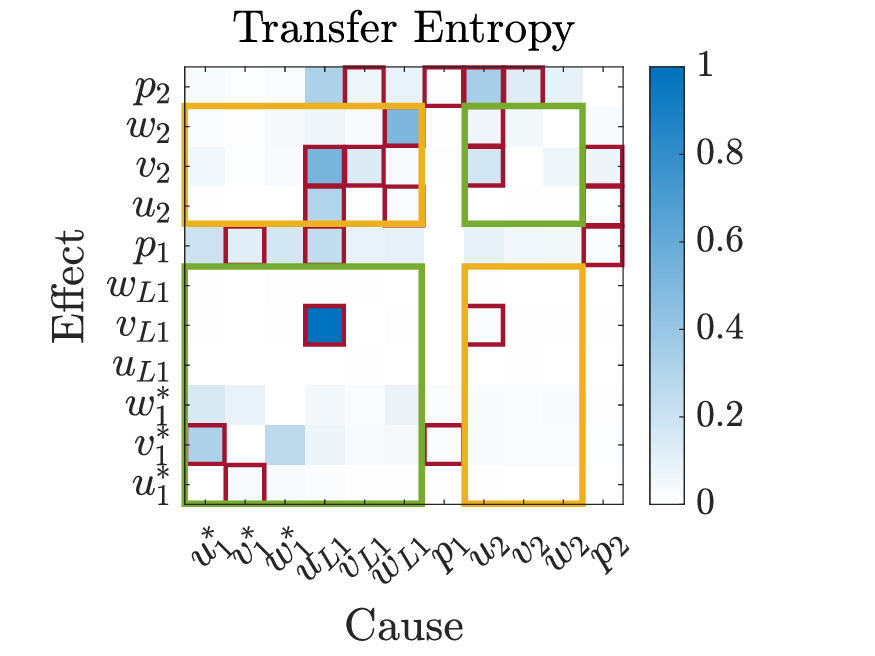}
            };
            \node[anchor=north west] at (image.north west) {{\rmfamily\fontsize{12}{13}\fontseries{l}\selectfont(a)}};
        \end{tikzpicture}
        \label{fig15200a}}
    \subfigure{    
        \begin{tikzpicture}
            \node[anchor=north west, inner sep=0] (image) at (0,0) {
                \includegraphics[width=0.48\textwidth]{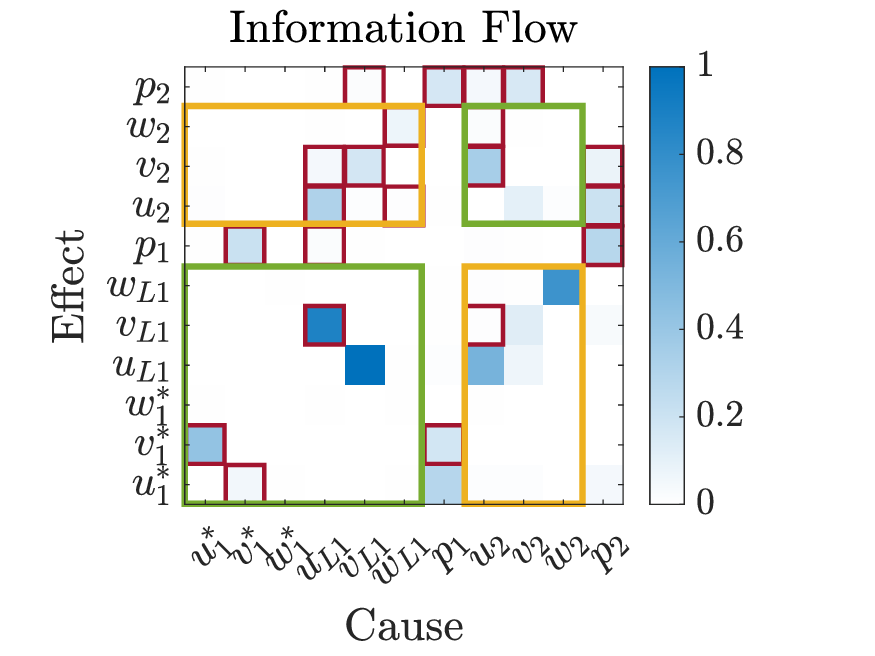}
            };
            \node[anchor=north west] at (image.north west) {{\rmfamily\fontsize{12}{13}\fontseries{l}\selectfont(b)}};
        \end{tikzpicture}
        \label{fig15200b}}
    \subfigure{
        \begin{tikzpicture}
            \node[anchor=north west, inner sep=0] (image) at (0,0) {
                \includegraphics[width=0.85\textwidth]{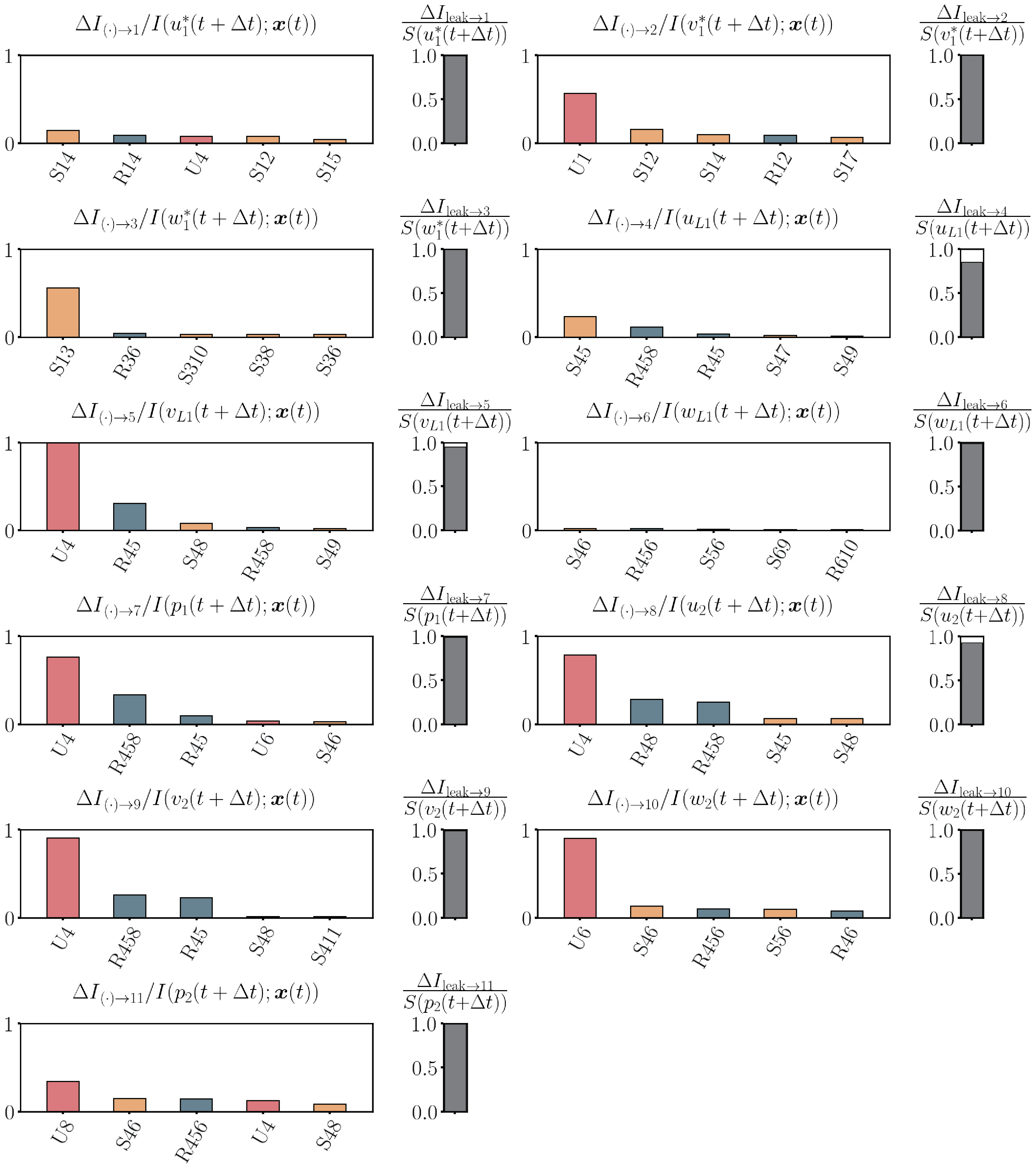}
            };
            \node[anchor=north west] at (image.north west) {{\rmfamily\fontsize{12}{13}\fontseries{l}\selectfont(c)}};
        \end{tikzpicture}
        \label{fig15200c}}
    \caption{{Two-point causal maps for the inner and outer motions at $y^+=15$ and $y^+=200$. (a) Transfer entropy, (b) information flow, and (c) SURD. {Yellow boxes}: causalities between the two heights. Green boxes: causalities at the same height.} }
    \label{fig15200}
\end{figure}

\begin{figure}
    \centering
    \subfigure{
        \begin{tikzpicture}
            \node[anchor=north west, inner sep=0] (image) at (0,0) {
                \includegraphics[width=0.48\textwidth]{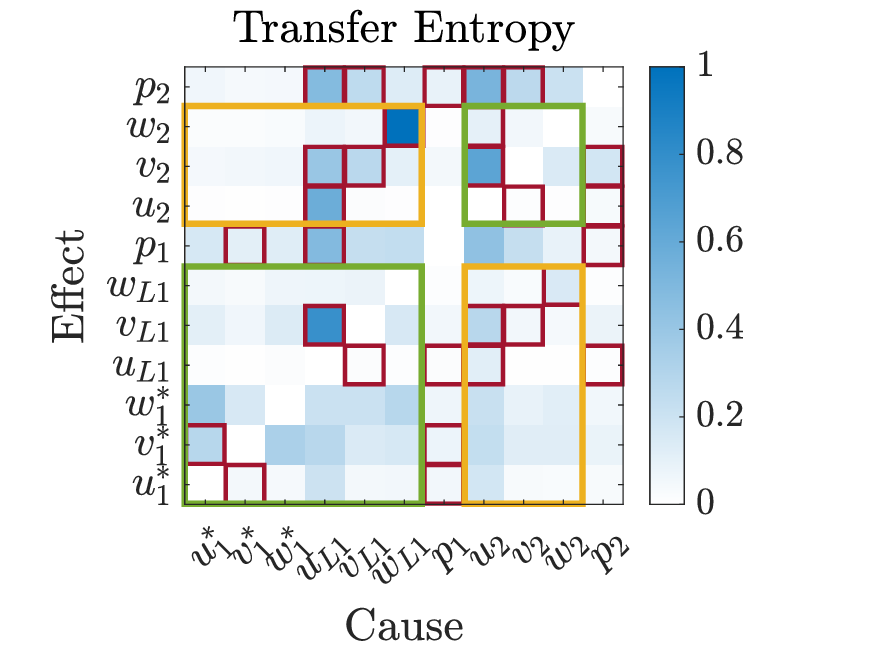}
            };
            \node[anchor=north west] at (image.north west) {{\rmfamily\fontsize{12}{13}\fontseries{l}\selectfont(a)}};
        \end{tikzpicture}
        \label{fig70200a}}
    \subfigure{    
        \begin{tikzpicture}
            \node[anchor=north west, inner sep=0] (image) at (0,0) {
                \includegraphics[width=0.48\textwidth]{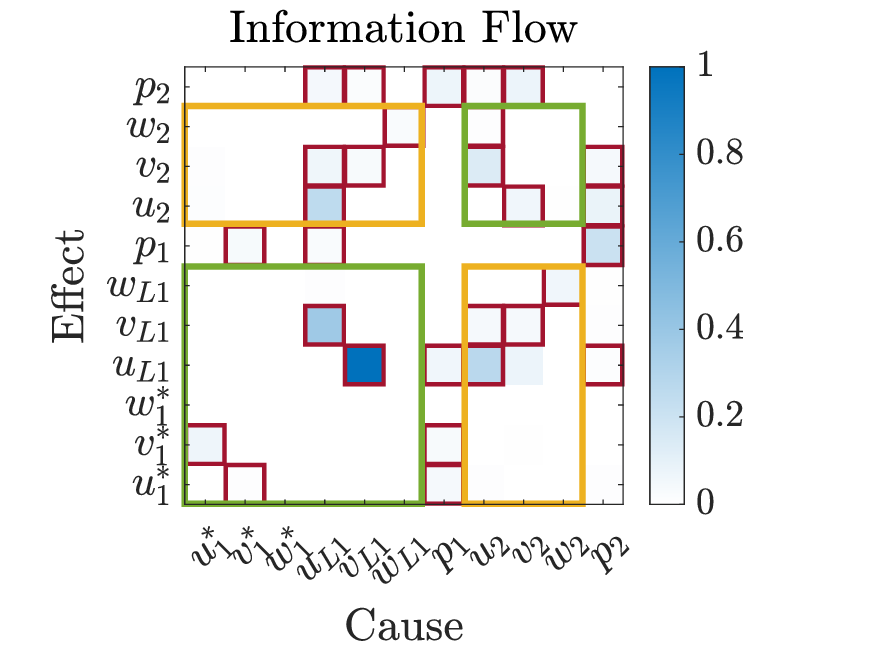}
            };
            \node[anchor=north west] at (image.north west) {{\rmfamily\fontsize{12}{13}\fontseries{l}\selectfont(b)}};
        \end{tikzpicture}
        \label{fig70200b}}
    \subfigure{
        \begin{tikzpicture}
            \node[anchor=north west, inner sep=0] (image) at (0,0) {
                \includegraphics[width=0.85\textwidth]{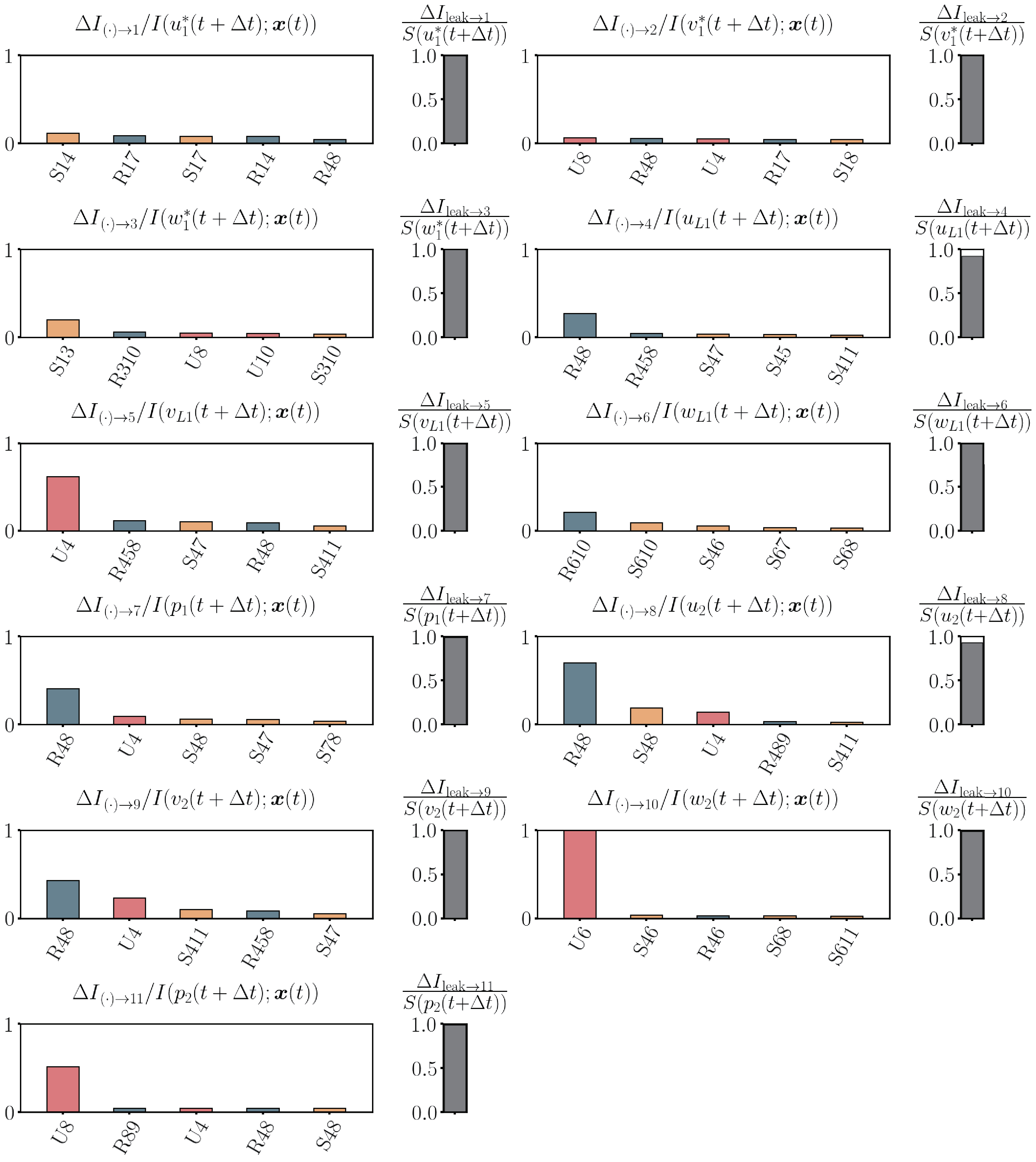}
            };
            \node[anchor=north west] at (image.north west) {{\rmfamily\fontsize{12}{13}\fontseries{l}\selectfont(c)}};
        \end{tikzpicture}
        \label{fig70200c}}
    \caption{{Two-point causal maps for the inner and outer motions at $y^+=70$ and $y^+=200$. (a) Transfer entropy, (b) information flow, and (c) SURD. {Yellow boxes}: causalities between the two heights. Green boxes: causalities at the same height.} }
    \label{fig70200}
\end{figure}

Next, we investigate the causal relationship between velocities and pressure at two wall-normal locations, \emph{i.e.,} one in the near-wall region ($y^+<100$) and the other in the outer region ($y^+>100$). The flow data at 10,000 points uniformly distributed in the horizontal plane of each height are utilised for analysis, and the streamwise and spanwise coordinates of the two paired points are the same.  
The variables in the SURD result are indexed as $(1,2,3,4,5,6,7,8,9,10,11) \to (u_{1}^*,v_{1}^*,w_{1}^*,u_{L1},v_{L1},w_{L1},p_1,u_2,v_2,w_2,p_2)$, in which the subscripts `1' and `2' indicate the inner and outer locations, respectively.

Figure~\ref{fig15200} (a) and (b) display the two-point causal maps of $y_1^+=15$ and $y_2^+=200$ by transfer entropy and information flow. The result is found to be almost the same as the one-point result shown in figure~\ref{fig15star}. The intra-scale self-generation streak-vortex cycle of inner motions ($u_{1}^* \leftrightarrow v_{1}^*$) is still evident. 
It suggests that the inclusion of the outer point in causal inference leaves the near-wall causal links essentially unaffected, confirming the stability of the autonomous regeneration mechanism of the inner motions.
In addition, a similar causal cycle of the outer motions at $y_2^+=200$ is also observed, namely {$u_2 \to v_2$}
, which further supports the self-sustaining hypothesis of turbulent motions at all scales \citep{hwang2010,hwang2011a,Hwang2015b,Rawat2015a,hwang2016a,Cossu2017Selfsustaininga,bae2021}. 

{The causal relationships between the two wall-normal heights $y_1^+=15$ and $y_2^+=200$ are shown in the yellow boxes of figure~\ref{fig15200} (a) and (b). First, the information flow can identify evident causal links of $u_{L1} \leftrightarrow u_{2}$, $v_{L1} \leftrightarrow v_{2}$ and $w_{L1} \leftrightarrow w_{2}$, demonstrating direct causal relationships between the outer motions at $y_2^+=200$ and their near-wall footprints at $y_1^+=15$, which reflect the dynamical coherence of them in the forward-inclined large-scale motions \citep{brownLargeStructureTurbulent1977a,chauhan2013structure,liu2017variation,liScaledependentInclinationAngle2022}. However, the transfer entropy can only detect ``bottom-up'' causalities of $u_{L1} \to u_{2}$, $v_{L1} \to v_{2}$ and $w_{L1} \to w_{2}$. In addition, the causal link $u_{L1} \rightarrow v_{2}$ is also observed, indicating a statistical causality among cross-velocity components. It is important to note that these links may not imply a dynamic ``bottom-up'' generation mechanism. Instead, these are signatures of the same large-scale coherent motions at different wall-normal locations. Therefore, these causal links quantitatively confirm the spatial convection and structural coherence of the LSMs between the outer layer and the near-wall region. Second, the absence of causal links of $(u_2, v_2, w_2) \leftrightarrow (u_{1}^*, v_{1}^*, w_{1}^*)$ implies that the outer motions at $y_2^+=200$ are causally uncorrelated with the near-wall demodulated inner motions at $y_1^+=15$. This also supports the hypothesis of the self-sustaining mechanism of turbulent motions at different scales.}


The pressure fluctuations at the two heights are also active in causal relations. The two-point causal inference results of $p_1$ at $y_1^+=15$ and $p_2$ at $y_2^+=200$ demonstrate that the pressure-related causal links remain consistent with the one-point result at the same height. 
{Taking into account two-point causalities between velocity and pressure, only a weak link $v_{L1} \to p_{2}$ can be identified. } 
{Moreover, the causal link $p_{2} \leftrightarrow p_{1}$ reveals a two-way influence of inner and outer pressures, suggesting that pressure may act as a critical mediator or bridge to coupling the outer and inner motions. }


{From figure~\ref{fig15200} (c), we can see that the SURD result is consistent with the findings of transfer entropy and information flow, providing additional insight into the causal relationships. 
The $u_{1}^*$-targeted causal link $v_{1}^* \to u_{1}^*$ is manifested as the synergistic causality $\Delta I_{12 \to 1}^S$. 
The $v_{1}^*$-targeted causal link $u_{1}^* \to v_{1}^*$ corresponds to the unique causality $\Delta I_{1 \to 2}^U$. 
The $u_{L1}$-targeted causal links $v_{L1} \to u_{L1}$ and $u_{2} \to u_{L1}$ are reflected by the synergistic causality $\Delta I_{45 \to 4}^S$ and the redundant causality $\Delta I_{458 \to 4}^R$. 
The $v_{L1}$-targeted causal link $u_{L1} \to v_{L1}$ is related to the unique causality $\Delta I_{4 \to 5}^U$. 
The $p_1$-targeted causal links $u_{L1} \to p_{1}$ correspond to the unique causality $\Delta I_{4 \to 7}^U$. 
The $u_2$-targeted causal links $u_{L1} \to u_{2}$ and $v_{L1} \to u_{2}$ correspond to the unique causality $\Delta I_{4 \to 8}^U$ and the redundant causality $\Delta I_{458 \to 8}^R$, respectively. 
The $v_2$-targeted causal links $u_{L1} \to v_{2}$, $v_{L1} \to v_{2}$, $u_{2} \to v_{2}$ and $p_{2} \to v_{2}$ are manifested as the unique causality $\Delta I_{4 \to 9}^U$, the redundant causalities $\Delta I_{458 \to 9}^R$ and the synergistic causality $\Delta I_{411 \to 9}^S$, respectively. 
The $p_2$-targeted causal links $u_{2} \to p_{2}$ correspond to the unique causality $\Delta I_{8 \to 11}^U$. 
It is noteworthy that $u_{L1}$ is quite active under the optimal time delay, exhibiting many unique causalities. }

Figure~\ref{fig70200} (a) and (b) show the two-point causal maps of $y_1^+=70$ and $y_2^+=200$ by transfer entropy and information flow.  
It can be seen that the causal relationships at the same height are similar to figure~\ref{fig15200}, in which we can find the streak-vortex links of the inner and outer motions, \emph{i.e.}, 
{$u_{1}^* \leftrightarrow v_{1}^*$, }
$u_{L1} \leftrightarrow v_{L1}$, and $u_2 \leftrightarrow v_2$. 
Compared with figure~\ref{fig15200}, a height dependence similar to the one-point causal inference is observed, \emph{i.e.} the intra-scale causalities of outer footprints at $y_1^+$ are stronger with a higher wall-normal height. 
For the causalities between the two heights, the links $(u_{L1},v_{L1}) \leftrightarrow (u_2,v_2)$ demonstrate bidirectional causal relationships between the outer motions and their footprints at $y^+=70$. 
{These causal links also confirm the spatial convection and inclined structural coherence of the LSMs between the outer layer and the near-wall region.}


The pressure-related causal characteristics illustrated in figure~\ref{fig70200} are basically similar to those in figure~\ref{fig15200}, where most of the links are of the same height. 
{For two-point pressure-velocity causal links, $v_{L1} \to p_2$ and $u_{L1} \leftrightarrow p_2$ can be identified, demonstrating the non-local causal relationships between outer pressure and near-wall velocity footprints. The causal links $p_2 \leftrightarrow p_1$ indicate a two-way coupling between the inner and outer pressures. This result strengthens the role of pressure in bridging the inner and outer turbulent motions.  }



{From figure~\ref{fig70200} (c), the $u_{1}^*$-targeted causal links $u_{L1} \to u_{1}^*$ and $p_1 \to u_{1}^*$ are manifested as the redundant causality $\Delta I_{14 \to 1}^R$ and $\Delta I_{17 \to 1}^R$, the synergistic causalities $\Delta I_{14 \to 1}^S$ and $\Delta I_{17 \to 1}^S$, respectively. 
The $v_{1}^*$-targeted causal link $p_{1} \to v_{1}^*$ corresponds to the redundant causality $\Delta I_{17 \to 2}^R$. 
The causal link $u_{L1} \to v_{1}^*$ corresponds to the unique causality $\Delta I_{4 \to 2}^U$. 
The $u_{L1}$-targeted causal links $v_{L1} \to u_{L1}$ and $u_{2} \to u_{L1}$ are reflected by the synergistic causalities $\Delta I_{45 \to 4}^S$ and the redundant causalities $\Delta I_{48 \to 4}^R$, $\Delta I_{458 \to 4}^R$, respectively. 
The causal link $p_{1} \to u_{L1}$ is reflected by the synergistic causality $\Delta I_{47 \to 4}^S$, which further illustrates the important role of pressure in the self-sustaining cycle. 
The $v_{L1}$-targeted causal link $u_{L1} \to v_{L1}$ is related to the unique causality $\Delta I_{4 \to 5}^R$. 
The causal links $p_{1} \to v_{L1}$ and $u_{2} \to v_{L1}$ are related to the redundant causality $\Delta I_{458 \to 5}^R$ and the synergistic causality $\Delta I_{47 \to 5}^S$, respectively, which reveal the interactions between the outer motions and their footprints. 
The $p_1$-targeted causal links $u_{L1} \to p_{1}$ and $u_{2} \to p_{1}$ correspond to the unique causality $\Delta I_{4 \to 7}^U$ and the redundant causality $\Delta I_{48 \to 7}^R$, respectively. 
The $u_2$-targeted causal links $u_{L1} \to u_{2}$, $v_{2} \to u_{2}$ and $p_{2} \to u_{2}$ correspond to the redundant causalities $\Delta I_{48 \to 8}^R$, $\Delta I_{489 \to 8}^R$, the synergistic causality $\Delta I_{411 \to 8}^S$ and the unique causality $\Delta I_{4 \to 8}^U$ respectively. 
The $v_2$-targeted causal links $u_{L1} \to v_{2}$, $v_{L1} \to v_{2}$, $u_{2} \to v_{2}$ and $p_{2} \to v_{2}$ are manifested as the redundant causalities $\Delta I_{48 \to 9}^R$, $\Delta I_{458 \to 9}^R$, the synergistic causality $\Delta I_{411 \to 9}^S$ and the unique causality $\Delta I_{4 \to 9}^U$, respectively. 
The $p_2$-targeted causal links $u_{L1} \to p_2$ and $u_{2} \to p_{2}$ correspond to the redundant causalities $\Delta I_{89 \to 11}^R$, $\Delta I_{48 \to 11}^R$, the synergistic causality $\Delta I_{48 \to 11}^S$ and the unique causalities $\Delta I_{8 \to 11}^U$ and $\Delta I_{4 \to 11}^U$, respectively. }

In summary, the two-point causal analysis has provided the following insights. First, the causally identified intra-scale streak-vortex self-sustaining cycles of the inner motions and outer footprints are the same as the one-point result. A similar causal cycle in the outer region is also found, which further supports the self-sustaining mechanism of turbulent motions at all scales. 
However, direct causal links between the inner and outer velocities (or footprints) are absent.
{Second, there exist two-way causal couplings between outer velocity components and their near-wall footprints as a result of the inclined LSMs. Moreover, there also exist two-way causal couplings between their cross-velocity components.}
{We have also verified that the identified causal structures consistently persist at $Re_{\tau}=2000$ and with a different outer location ($y_2^+=100$) as in the supplementary material, indicating the robustness of the present findings. }

\begin{figure}
    \centering
    \includegraphics[width=1\textwidth]{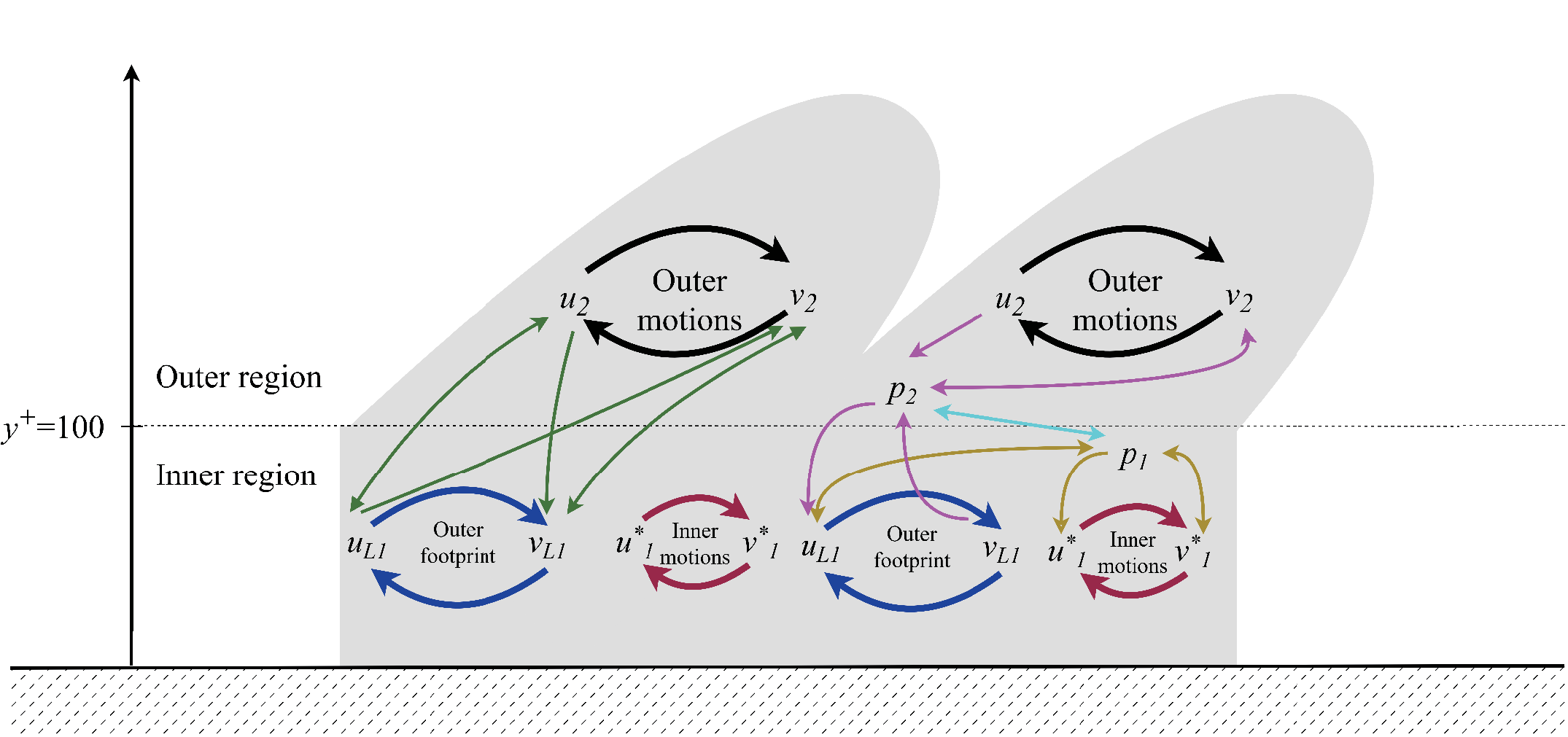}
    \caption{Schematic diagram of the causal links between inner and outer motions. Black arrows: the streak-vortex self-sustaining cycle of outer motions; blue arrows: the streak-vortex self-sustaining cycle of outer footprints; red arrows: the streak-vortex self-sustaining cycle of inner motions; green arrows: the bidirectional causalities of the outer motions and their near-wall footprints; brown arrows: the inner-pressure-related causal links; magenta arrows: the outer-pressure-related arrows; cyan arrow: the causal links between the inner and outer pressure.}
    \label{fig:causal_diag}
\end{figure}

\section{Conclusions}\label{section6}

In this work, we have applied three typical causal inference methods,  \emph{i.e.} transfer entropy, information flow, and SURD, to study the causal relationship of the inner and outer turbulent motions. Initially, several canonical problems, from simple linear and nonlinear dynamical systems to low-order dynamics of near-wall turbulence, were examined. The three methods can predict the same causal relationships for the linear problem, but some differences exist in the nonlinear problem. 
Richer causalities can be identified by transfer entropy and SURD, while more pronounced causal values can be revealed by information flow. For the low-dimensional model of near-wall turbulence, all three methods can capture the prominent self-sustaining mechanism of near-wall turbulence, \emph{i.e.} the streak-vortex regeneration cycle.

By combining the three methods, we conducted causal analysis using a newly generated time-resolved DNS dataset of turbulent channel flow at $Re_\tau = 1000$. An improved inner-outer decomposition method is proposed and employed.
From the one-point analysis, it was found that the inner motions and the outer footprints are self-sustained. From the two-point analysis, it was further revealed that both inner and outer motions are self-sustained, which supports the self-sustaining mechanism of turbulent motions at all scales \citep{hwang2010,hwang2011a,Hwang2015b,Rawat2015a,hwang2016a,Cossu2017Selfsustaininga,bae2021}. In particular, the outer footprints in the near-wall region can also be self-sustained, implying that they are not inactive in the causal cycle.  
{Moreover, there exist two-way causal links between the outer velocities and their near-wall footprints (their own or cross velocities) owing to the dynamical coherence of inclined large-scale turbulent motions.} 

Figure~\ref{fig:causal_diag} illustrates the schematics of the causal links between the inner and outer motions identified by the present study. 
Regarding inner-outer interactions, an interesting finding is that although there is no direct causal connection between the inner and outer motions, pressure may act as a bridge in linking them, both locally (one point) and non-locally (two points). In the near-wall region, local pressure is involved in the causal links with the inner velocities and outer footprint velocities, as shown by the brown arrows in figure~\ref{fig:causal_diag}. The outer pressure is causally connected to the outer velocities and their near-wall footprint velocities, as shown by the magenta arrows in figure~\ref{fig:causal_diag}. Moreover, there exist bidirectional causal links between the inner and outer pressures, as indicated by the cyan arrows in figure~\ref{fig:causal_diag}.
Therefore, a path of inner-outer causality (bidirectional in both top-down and bottom-up) is: outer velocities $\leftrightarrow$ outer pressure $\leftrightarrow$ inner pressure $\leftrightarrow$ inner velocities.
From figure~\ref{fig:causal_diag}, we can find another path of inner-outer causality (unidirectional in top-down) as: outer velocities $\rightarrow$ outer footprint velocities $\rightarrow$ inner pressure $\rightarrow$ inner velocities. The two identified pathways can provide further insight into the inner-outer interactions in wall-bounded turbulent flows. 

{Finally, we emphasise that the causal relationships identified in this study are contingent upon the specific inner–outer decomposition method employed. The present findings should therefore be regarded as decomposition-specific rather than universal. Future investigations using alternative scale-separation techniques are needed to assess the universality of these causal relationships.}

\section*{Acknowledgement}
The authors thank Profs. X. San Liang, R. Vinuesa and A. Lozano-Dur\'an for helpful discussions. The financial support from the National Natural Science Foundation of China (12388101 and 12472221), the Fundamental Research Funds for the Central Universities (lzujbky-2024-oy10) and the Natural Science Foundation of Gansu Province (25JRRA636) is acknowledged.

\section*{Declaration of interests}
The authors report no conflict of interest.

\section*{Data Availability Statement}
The data that support the findings of this study are available from the corresponding author upon reasonable request.

\appendix

\section{Formulation of the low-dimensional near-wall turbulence model}\label{appA}
{
The low-dimensional model of near-wall turbulence is governed by the following system of nine ordinary differential equations \citep{moehlis2004low}: 
\begin{equation}
    \frac{\mathrm{d}a_1}{\mathrm{d}t}=\frac{\beta^2}{Re}-\frac{\beta^2}{Re}a_1-\sqrt{\frac{3}{2}}\frac{\beta\gamma}{\kappa_{\alpha\beta\gamma}}a_6a_8+\sqrt{\frac{3}{2}}\frac{\beta\gamma}{\kappa_{\beta\gamma}}a_2a_3, 
    \label{eqn:a1}
\end{equation}
\begin{equation}
    \begin{aligned}
    \frac{\mathrm{d}a_2}{\mathrm{d}t} =& -\big(\frac{4\beta^2}{3}+\gamma^2\big)\frac{a_2}{Re} + \frac{5\sqrt{2}}{3\sqrt{3}}\frac{\gamma^2}{\kappa_{\alpha\gamma}} a_4 a_6 - \frac{\gamma^2}{\sqrt{6}\,\kappa_{\alpha\gamma}} a_5 a_7 \\
    &- \frac{\alpha\beta\gamma}{\sqrt{6}\,\kappa_{\alpha\gamma}\kappa_{\alpha\beta\gamma}} a_5 a_8 - \sqrt{\frac{3}{2}}\frac{\beta\gamma}{\kappa_{\beta\gamma}} a_1 a_3 - \sqrt{\frac{3}{2}}\frac{\beta\gamma}{\kappa_{\beta\gamma}} a_3 a_9, 
    \end{aligned}
\end{equation}
\begin{equation}
    \begin{aligned}
    \frac{\mathrm{d}a_3}{\mathrm{d}t} =& -\frac{\beta^2+\gamma^2}{Re} a_3 + \frac{2}{\sqrt{6}}\frac{\alpha\beta\gamma}{\kappa_{\alpha\gamma}\kappa_{\beta\gamma}} (a_4 a_7 + a_5 a_6) \\
    &+ \frac{\beta^2(3\alpha^2+\gamma^2)-3\gamma^2(\alpha^2+\gamma^2)}{\sqrt{6}\,\kappa_{\alpha\gamma}\kappa_{\beta\gamma}\kappa_{\alpha\beta\gamma}} a_4 a_8, 
    \end{aligned}
    \label{eqn:a3}
\end{equation}
\begin{equation}
    \begin{aligned}
    \frac{\mathrm{d}a_4}{\mathrm{d}t} =& -\frac{3\alpha^2+4\beta^2}{3Re} a_4 - \frac{10}{3\sqrt{6}}\frac{\alpha^2}{\kappa_{\alpha\gamma}} a_2 a_6 - \sqrt{\frac{3}{2}}\frac{\alpha\beta\gamma}{\kappa_{\alpha\gamma}\kappa_{\beta\gamma}} a_3 a_7 \\
    &- \sqrt{\frac{3}{2}}\frac{\alpha^2\beta^2}{\kappa_{\alpha\gamma}\kappa_{\beta\gamma}\kappa_{\alpha\beta\gamma}} a_3 a_8 - \frac{\alpha}{\sqrt{6}} a_1 a_5 - \frac{\alpha}{\sqrt{6}} a_5 a_9, 
    \end{aligned}
\end{equation}
\begin{equation}
    \begin{aligned}
    \frac{\mathrm{d}a_5}{\mathrm{d}t} =& -\frac{\alpha^2+\beta^2}{Re} a_5 + \frac{\alpha}{\sqrt{6}} a_1 a_4 + \frac{\alpha^2}{\sqrt{6}\,\kappa_{\alpha\gamma}} a_2 a_7 \\
    &- \frac{\alpha\beta\gamma}{\sqrt{6}\,\kappa_{\alpha\gamma}\kappa_{\alpha\beta\gamma}} a_2 a_8 + \frac{\alpha}{\sqrt{6}} a_4 a_9 + \frac{2}{\sqrt{6}}\frac{\alpha\beta\gamma}{\kappa_{\alpha\gamma}\kappa_{\beta\gamma}} a_3 a_6, 
    \end{aligned}
\end{equation}
\begin{equation}
    \begin{aligned}
    \frac{\mathrm{d}a_6}{\mathrm{d}t} =& -\frac{3\alpha^2+4\beta^2+3\gamma^2}{3Re} a_6 + \frac{\alpha}{\sqrt{6}} a_1 a_7 + \sqrt{\frac{3}{2}}\frac{\beta\gamma}{\kappa_{\alpha\beta\gamma}} a_1 a_8 \\
    &+ \frac{10}{3\sqrt{6}}\frac{\alpha^2-\gamma^2}{\kappa_{\alpha\gamma}} a_2 a_4 - 2\sqrt{\frac{2}{3}}\frac{\alpha\beta\gamma}{\kappa_{\alpha\gamma}\kappa_{\beta\gamma}} a_3 a_5 \\
    &+ \frac{\alpha}{\sqrt{6}} a_7 a_9 + \sqrt{\frac{3}{2}}\frac{\beta\gamma}{\kappa_{\alpha\beta\gamma}} a_8 a_9, 
    \end{aligned}
\end{equation}
\begin{equation}
    \begin{aligned}
    \frac{\mathrm{d}a_7}{\mathrm{d}t} =& -\frac{\alpha^2+\beta^2+\gamma^2}{Re} a_7 - \frac{\alpha}{\sqrt{6}}(a_1 a_6 + a_6 a_9) \\
    &+ \frac{1}{\sqrt{6}}\frac{\gamma^2-\alpha^2}{\kappa_{\alpha\gamma}} a_2 a_5 + \frac{1}{\sqrt{6}}\frac{\alpha\beta\gamma}{\kappa_{\alpha\gamma}\kappa_{\beta\gamma}} a_3 a_4, 
    \end{aligned}
\end{equation}
\begin{equation}
    \frac{\mathrm{d}a_8}{\mathrm{d}t} = -\frac{\alpha^2+\beta^2+\gamma^2}{Re} a_8 + \frac{2}{\sqrt{6}}\frac{\alpha\beta\gamma}{\kappa_{\alpha\gamma}\kappa_{\alpha\beta\gamma}} a_2 a_5 + \frac{\gamma^2(3\alpha^2-\beta^2+3\gamma^2)}{\sqrt{6}\,\kappa_{\alpha\gamma}\kappa_{\beta\gamma}\kappa_{\alpha\beta\gamma}} a_3 a_4, 
\end{equation}
\begin{equation}
    \frac{\mathrm{d}a_9}{\mathrm{d}t} = -\frac{9\beta^2}{Re} a_9 + \sqrt{\frac{3}{2}}\frac{\beta\gamma}{\kappa_{\beta\gamma}} a_2 a_3 - \sqrt{\frac{3}{2}}\frac{\beta\gamma}{\kappa_{\alpha\beta\gamma}} a_6 a_8, 
    \label{eqn:a9}
\end{equation}
where $Re$ denotes the Reynolds number, and the constants are defined as 
\begin{equation}
    \begin{gathered}
    \alpha = {2\pi}/{L_x},\quad \beta = {\pi}/{2},\quad \gamma = {2\pi}/{L_z}, \\
    \kappa_{\alpha\gamma} = \sqrt{\alpha^2 + \gamma^2},\quad \kappa_{\beta\gamma} = \sqrt{\beta^2 + \gamma^2},\quad \kappa_{\alpha\beta\gamma} = \sqrt{\alpha^2 + \beta^2 + \gamma^2},
    \end{gathered}
\end{equation}
in which, $L_x$ and $L_z$ are streamwise and spanwise domain lengths, respectively. 

Subtracting equation~(\ref{eqn:a9}) from (\ref{eqn:a1}) yields a linear ordinary differential equation that contains only $a_1$ and $a_9$ (and their derivatives), while the nonlinear terms from all other modes cancel out completely: 
\begin{equation}
    \frac{\mathrm{d}(a_1 - a_9)}{\mathrm{d}t} = \frac{\beta^2}{Re} - \frac{\beta^2}{Re} a_1 + \frac{9\beta^2}{Re} a_9. 
    \label{eqn:a1_a9}
\end{equation}

Multiplying the equations of each mode by its respective coefficient, one can obtain the following equations: 
\begin{equation}
    \frac{1}{2}\frac{\mathrm{d}a_1^2}{\mathrm{d}t} = \frac{\beta^2}{Re}a_1 - \frac{\beta^2}{Re}a_1^2 - \sqrt{\frac{3}{2}}\frac{\beta\gamma}{\kappa_{\alpha\beta\gamma}} a_1 a_6 a_8 + \sqrt{\frac{3}{2}}\frac{\beta\gamma}{\kappa_{\beta\gamma}} a_1 a_2 a_3, 
    \label{eqn:a12}
\end{equation}
\begin{equation}
    \begin{aligned}
    \frac{1}{2}\frac{\mathrm{d}a_2^2}{\mathrm{d}t} =& -\big(\frac{4\beta^2}{3}+\gamma^2\big)\frac{a_2^2}{Re} + \frac{5\sqrt{2}}{3\sqrt{3}}\frac{\gamma^2}{\kappa_{\alpha\gamma}} a_2 a_4 a_6 - \frac{\gamma^2}{\sqrt{6}\,\kappa_{\alpha\gamma}} a_2 a_5 a_7 \\
    &- \frac{\alpha\beta\gamma}{\sqrt{6}\,\kappa_{\alpha\gamma}\kappa_{\alpha\beta\gamma}} a_2 a_5 a_8 - \sqrt{\frac{3}{2}}\frac{\beta\gamma}{\kappa_{\beta\gamma}} a_1 a_2 a_3 - \sqrt{\frac{3}{2}}\frac{\beta\gamma}{\kappa_{\beta\gamma}} a_2 a_3 a_9, 
    \end{aligned}
\end{equation}
\begin{equation}
    \begin{aligned}
    \frac{1}{2}\frac{\mathrm{d}a_3^2}{\mathrm{d}t} =& -\frac{\beta^2+\gamma^2}{Re} a_3^2 + \frac{2}{\sqrt{6}}\frac{\alpha\beta\gamma}{\kappa_{\alpha\gamma}\kappa_{\beta\gamma}} a_3 a_4 a_7 + \frac{2}{\sqrt{6}}\frac{\alpha\beta\gamma}{\kappa_{\alpha\gamma}\kappa_{\beta\gamma}} a_3 a_5 a_6 \\
    &+ \frac{\beta^2(3\alpha^2+\gamma^2)-3\gamma^2(\alpha^2+\gamma^2)}{\sqrt{6}\,\kappa_{\alpha\gamma}\kappa_{\beta\gamma}\kappa_{\alpha\beta\gamma}} a_3 a_4 a_8, 
    \end{aligned}
\end{equation}
\begin{equation}
    \begin{aligned}
    \frac{1}{2}\frac{\mathrm{d}a_4^2}{\mathrm{d}t} =& -\frac{3\alpha^2+4\beta^2}{3Re} a_4^2 - \frac{10}{3\sqrt{6}}\frac{\alpha^2}{\kappa_{\alpha\gamma}} a_2 a_4 a_6 - \sqrt{\frac{3}{2}}\frac{\alpha\beta\gamma}{\kappa_{\alpha\gamma}\kappa_{\beta\gamma}} a_3 a_4 a_7 \\
    &- \sqrt{\frac{3}{2}}\frac{\alpha^2\beta^2}{\kappa_{\alpha\gamma}\kappa_{\beta\gamma}\kappa_{\alpha\beta\gamma}} a_3 a_4 a_8 - \frac{\alpha}{\sqrt{6}} a_1 a_4 a_5 - \frac{\alpha}{\sqrt{6}} a_4 a_5 a_9, 
    \end{aligned}
\end{equation}
\begin{equation}
    \begin{aligned}
    \frac{1}{2}\frac{\mathrm{d}a_5^2}{\mathrm{d}t} =& -\frac{\alpha^2+\beta^2}{Re} a_5^2 + \frac{\alpha}{\sqrt{6}} a_1 a_4 a_5 + \frac{\alpha^2}{\sqrt{6}\,\kappa_{\alpha\gamma}} a_2 a_5 a_7 \\
    &- \frac{\alpha\beta\gamma}{\sqrt{6}\,\kappa_{\alpha\gamma}\kappa_{\alpha\beta\gamma}} a_2 a_5 a_8 + \frac{\alpha}{\sqrt{6}} a_4 a_5 a_9 + \frac{2}{\sqrt{6}}\frac{\alpha\beta\gamma}{\kappa_{\alpha\gamma}\kappa_{\beta\gamma}} a_3 a_5 a_6, 
    \end{aligned}
\end{equation}
\begin{equation}
    \begin{aligned}
    \frac{1}{2}\frac{\mathrm{d}a_6^2}{\mathrm{d}t} =& -\frac{3\alpha^2+4\beta^2+3\gamma^2}{3Re} a_6^2 + \frac{\alpha}{\sqrt{6}} a_1 a_6 a_7 + \sqrt{\frac{3}{2}}\frac{\beta\gamma}{\kappa_{\alpha\beta\gamma}} a_1 a_6 a_8 \\
    &+ \frac{10}{3\sqrt{6}}\frac{\alpha^2-\gamma^2}{\kappa_{\alpha\gamma}} a_2 a_4 a_6 - 2\sqrt{\frac{2}{3}}\frac{\alpha\beta\gamma}{\kappa_{\alpha\gamma}\kappa_{\beta\gamma}} a_3 a_5 a_6 \\
    &+ \frac{\alpha}{\sqrt{6}} a_6 a_7 a_9 + \sqrt{\frac{3}{2}}\frac{\beta\gamma}{\kappa_{\alpha\beta\gamma}} a_6 a_8 a_9, 
    \end{aligned}
\end{equation}
\begin{equation}
    \begin{aligned}
    \frac{1}{2}\frac{\mathrm{d}a_7^2}{\mathrm{d}t} =& -\frac{\alpha^2+\beta^2+\gamma^2}{Re} a_7^2 - \frac{\alpha}{\sqrt{6}} a_1 a_6 a_7 - \frac{\alpha}{\sqrt{6}} a_6 a_7 a_9 \\
    &+ \frac{1}{\sqrt{6}}\frac{\gamma^2-\alpha^2}{\kappa_{\alpha\gamma}} a_2 a_5 a_7 + \frac{1}{\sqrt{6}}\frac{\alpha\beta\gamma}{\kappa_{\alpha\gamma}\kappa_{\beta\gamma}} a_3 a_4 a_7, 
    \end{aligned}
\end{equation}
\begin{equation}
    \begin{aligned}
    \frac{1}{2}\frac{\mathrm{d}a_8^2}{\mathrm{d}t} =& -\frac{\alpha^2+\beta^2+\gamma^2}{Re} a_8^2 + \frac{2}{\sqrt{6}}\frac{\alpha\beta\gamma}{\kappa_{\alpha\gamma}\kappa_{\alpha\beta\gamma}} a_2 a_5 a_8 \\
    &+ \frac{\gamma^2(3\alpha^2-\beta^2+3\gamma^2)}{\sqrt{6}\,\kappa_{\alpha\gamma}\kappa_{\beta\gamma}\kappa_{\alpha\beta\gamma}} a_3 a_4 a_8, 
    \end{aligned}
\end{equation}
\begin{equation}
    \frac{1}{2}\frac{\mathrm{d}a_9^2}{\mathrm{d}t} = -\frac{9\beta^2}{Re} a_9^2 + \sqrt{\frac{3}{2}}\frac{\beta\gamma}{\kappa_{\beta\gamma}} a_2 a_3 a_9 - \sqrt{\frac{3}{2}}\frac{\beta\gamma}{\kappa_{\alpha\beta\gamma}} a_6 a_8 a_9. 
    \label{eqn:a92}
\end{equation}

Summing the equations (\ref{eqn:a12}) to (\ref{eqn:a92}) eliminates all crossing terms completely, yielding the following equation: 
\begin{equation}
    \begin{aligned}
    \frac{1}{2}\frac{\mathrm{d}}{\mathrm{d}t}\sum_{i=1}^{9} a_i^2 =& \frac{\beta^2}{Re}a_1 - \frac{1}{Re}\big[
    \beta^2 a_1^2 + \big(\frac{4\beta^2}{3}+\gamma^2\big)a_2^2 + (\beta^2+\gamma^2)a_3^2 \\
    &+ \frac{3\alpha^2+4\beta^2}{3}a_4^2 + (\alpha^2+\beta^2)a_5^2 + \frac{3\alpha^2+4\beta^2+3\gamma^2}{3}a_6^2 \\
    &+ (\alpha^2+\beta^2+\gamma^2)(a_7^2+a_8^2) + 9\beta^2 a_9^2 \big]. 
    \end{aligned}
    \label{eqn:sum_ai}
\end{equation}
}

\section{DNS validation}\label{datav}

\begin{figure}
    \centering
    \subfigure{
        \begin{tikzpicture}
            \node[anchor=north west, inner sep=0] (image) at (0,0) {
                \includegraphics[width=0.45\textwidth]{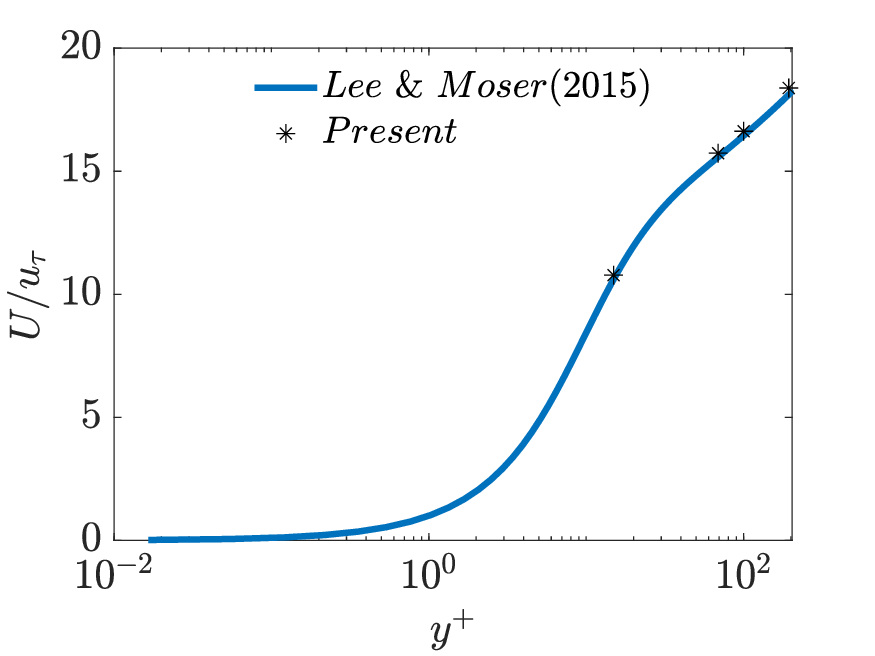}
            };
            \node[anchor=north west, xshift=-1.5ex] at (image.north west) {{\rmfamily\fontsize{12}{13}\fontseries{l}\selectfont(a)}};
        \end{tikzpicture}
        \label{figdatava}}
    \subfigure{    
        \begin{tikzpicture}
            \node[anchor=north west, inner sep=0] (image) at (0,0) {
                \includegraphics[width=0.45\textwidth]{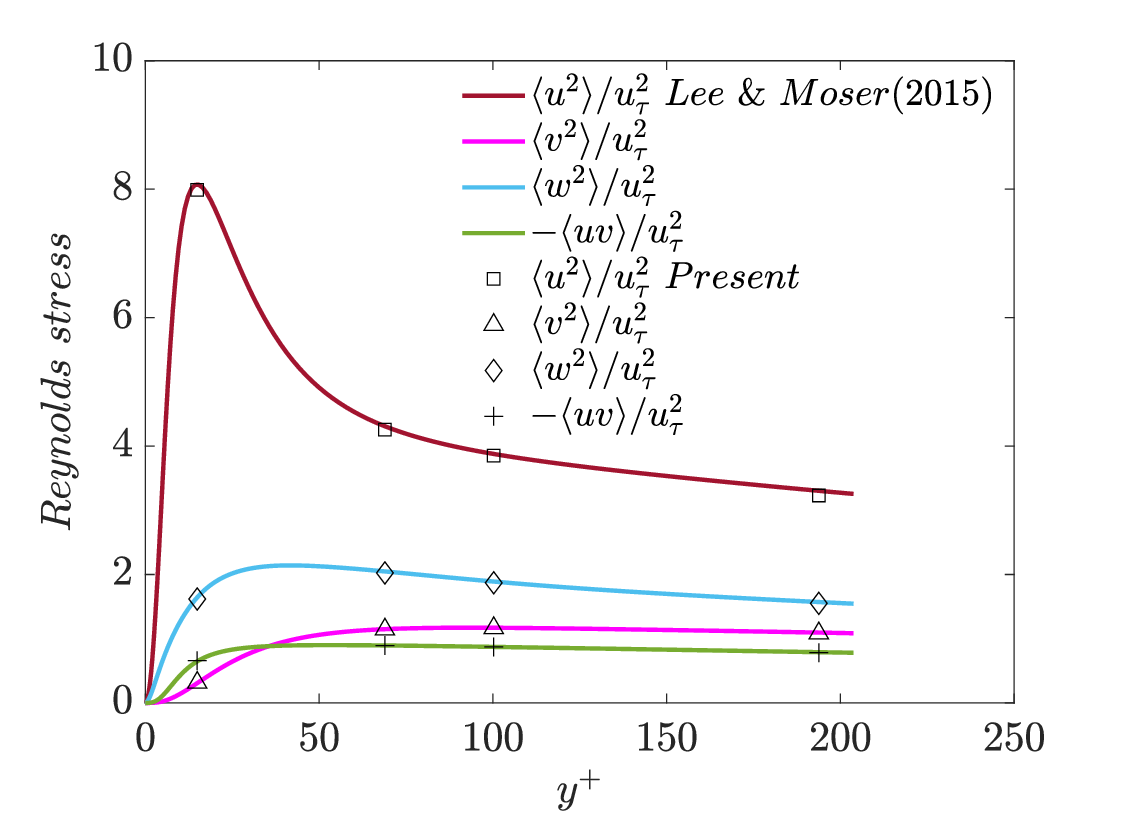}
            };
            \node[anchor=north west, xshift=-1.5ex] at (image.north west) {{\rmfamily\fontsize{12}{13}\fontseries{l}\selectfont(b)}};
        \end{tikzpicture}
        \label{figdatavb}}
    \caption{Comparison of turbulence statistics of the present DNS with \citet{Lee2015}. (a) Mean streamwise flow velocity, (b) Reynolds normal and shear stresses. }
    \label{figdatav}
\end{figure}

In this study, we utilised a newly generated DNS dataset of turbulent channel flow at $Re_\tau = 1000$. The accuracy of the DNS code at low-$Re_\tau$ channel flows has been validated in \citet{Hu2018application} and \citet{Hu2018energy}. Here we present a validation of the DNS data at $Re_\tau = 1000$ with that of \citet{Lee2015}. 
Figure~\ref{figdatav} displays the comparison of the profiles of the mean streamwise flow velocity and Reynolds stresses. It can be seen that the present DNS data at $y^+=15$, 70, 100 and 200 perfectly match the reference. 

\begin{figure}
    \centering
    \subfigure{
        \begin{tikzpicture}
            \node[anchor=north west, inner sep=0] (image) at (0,0) {
                \includegraphics[width=0.48\textwidth]{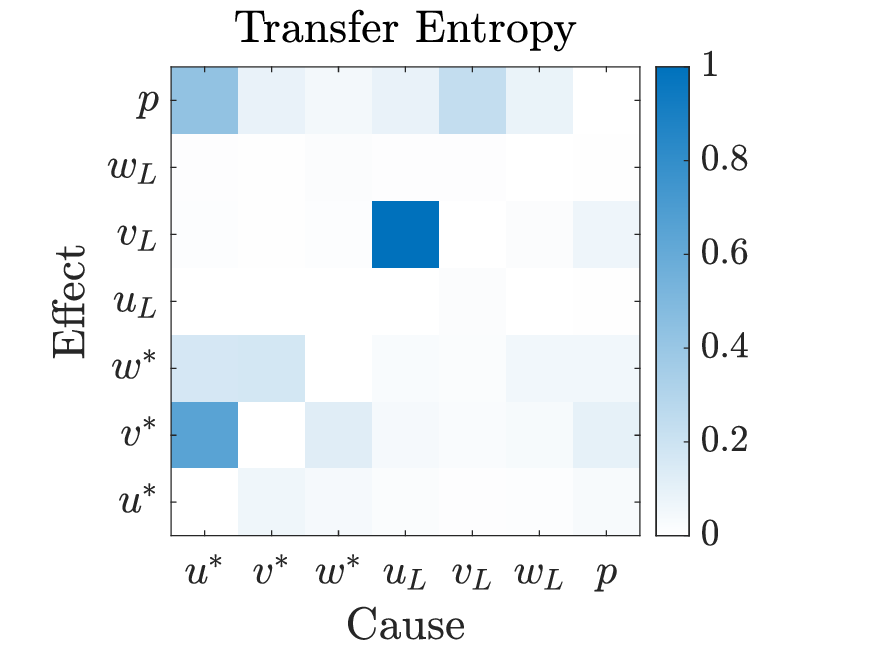}
            };
            \node[anchor=north west] at (image.north west) {{\rmfamily\fontsize{12}{13}\fontseries{l}\selectfont(a)}};
        \end{tikzpicture}
        \label{figtec}}
    \subfigure{    
        \begin{tikzpicture}
            \node[anchor=north west, inner sep=0] (image) at (0,0) {
                \includegraphics[width=0.48\textwidth]{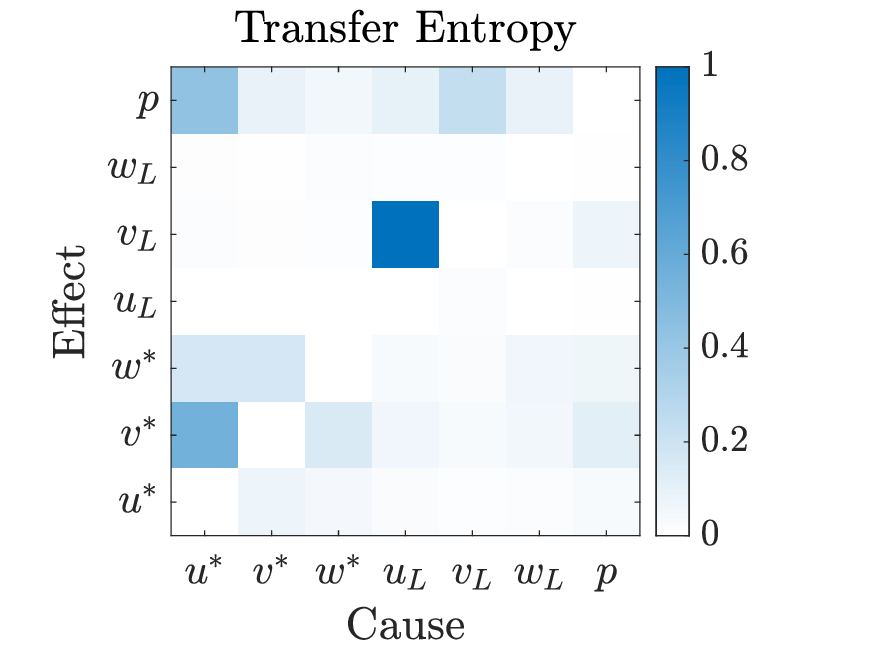}
            };
            \node[anchor=north west] at (image.north west) {{\rmfamily\fontsize{12}{13}\fontseries{l}\selectfont(b)}};
        \end{tikzpicture}
        \label{figteh}}
    \subfigure{
        \begin{tikzpicture}
            \node[anchor=north west, inner sep=0] (image) at (0,0) {
                \includegraphics[width=0.48\textwidth]{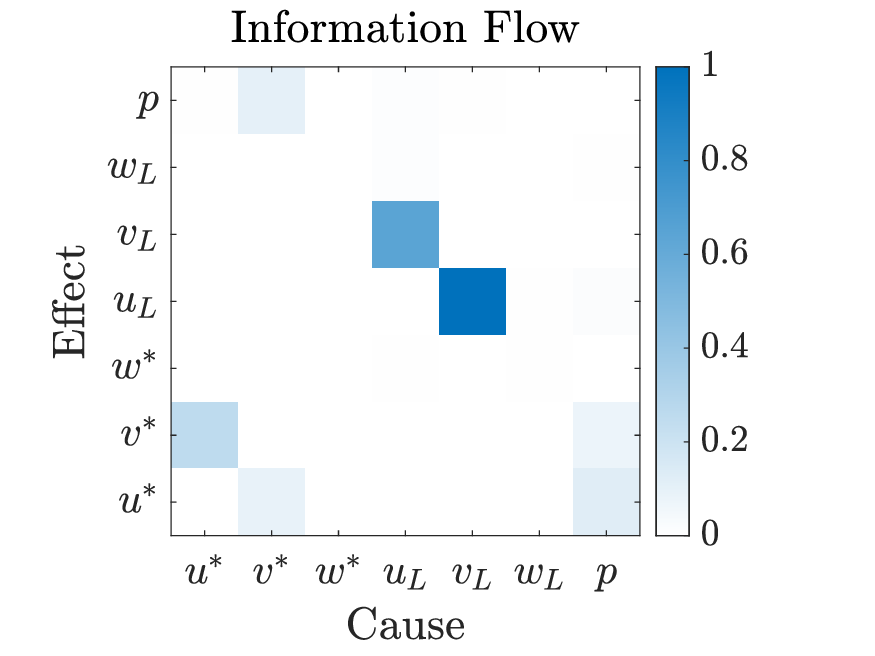}
            };
            \node[anchor=north west] at (image.north west) {{\rmfamily\fontsize{12}{13}\fontseries{l}\selectfont(c)}};
        \end{tikzpicture}
        \label{figifc}}
    \subfigure{    
        \begin{tikzpicture}
            \node[anchor=north west, inner sep=0] (image) at (0,0) {
                \includegraphics[width=0.48\textwidth]{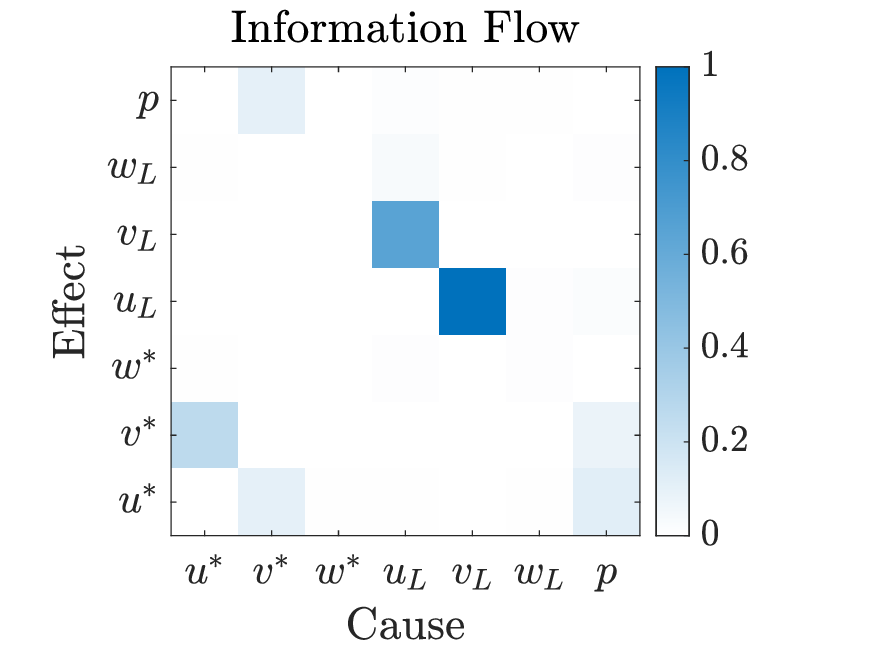}
            };
            \node[anchor=north west] at (image.north west) {{\rmfamily\fontsize{12}{13}\fontseries{l}\selectfont(d)}};
        \end{tikzpicture}
        \label{figifh}}
    \caption{Causal maps computed using $(a,c)$ the complete dataset, and $(b,d)$ half of the of the dataset in temporal length. }
    \label{figch}
\end{figure}

\begin{figure}
    \centering
    \subfigure{
        \begin{tikzpicture}
            \node[anchor=north west, inner sep=0] (image) at (0,0) {
                \includegraphics[width=0.85\textwidth]{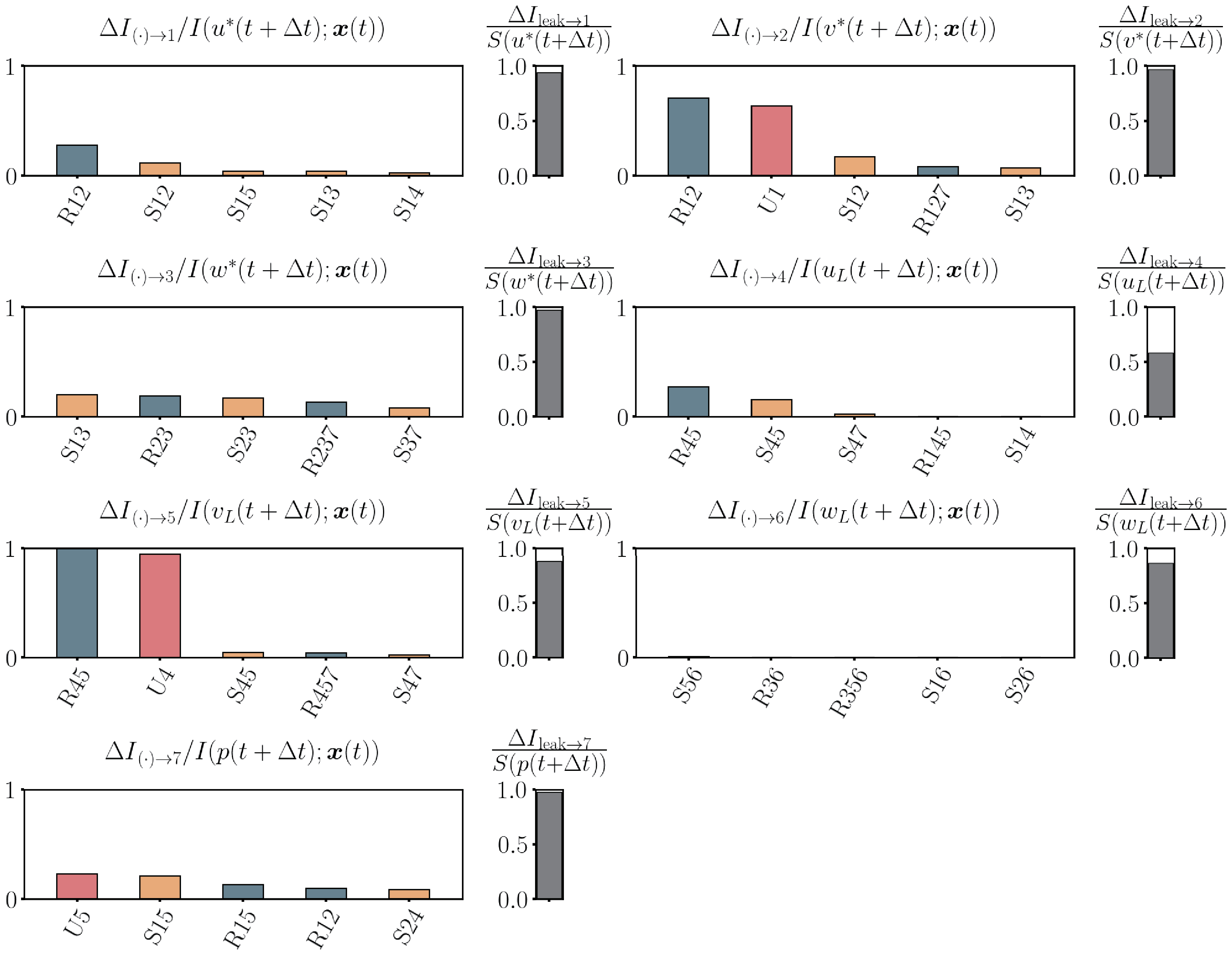}
            };
            \node[anchor=north west] at (image.north west) {{\rmfamily\fontsize{12}{13}\fontseries{l}\selectfont(a)}};
        \end{tikzpicture}
        \label{figsurdc}}
    \subfigure{
        \begin{tikzpicture}
            \node[anchor=north west, inner sep=0] (image) at (0,0) {
                \includegraphics[width=0.85\textwidth]{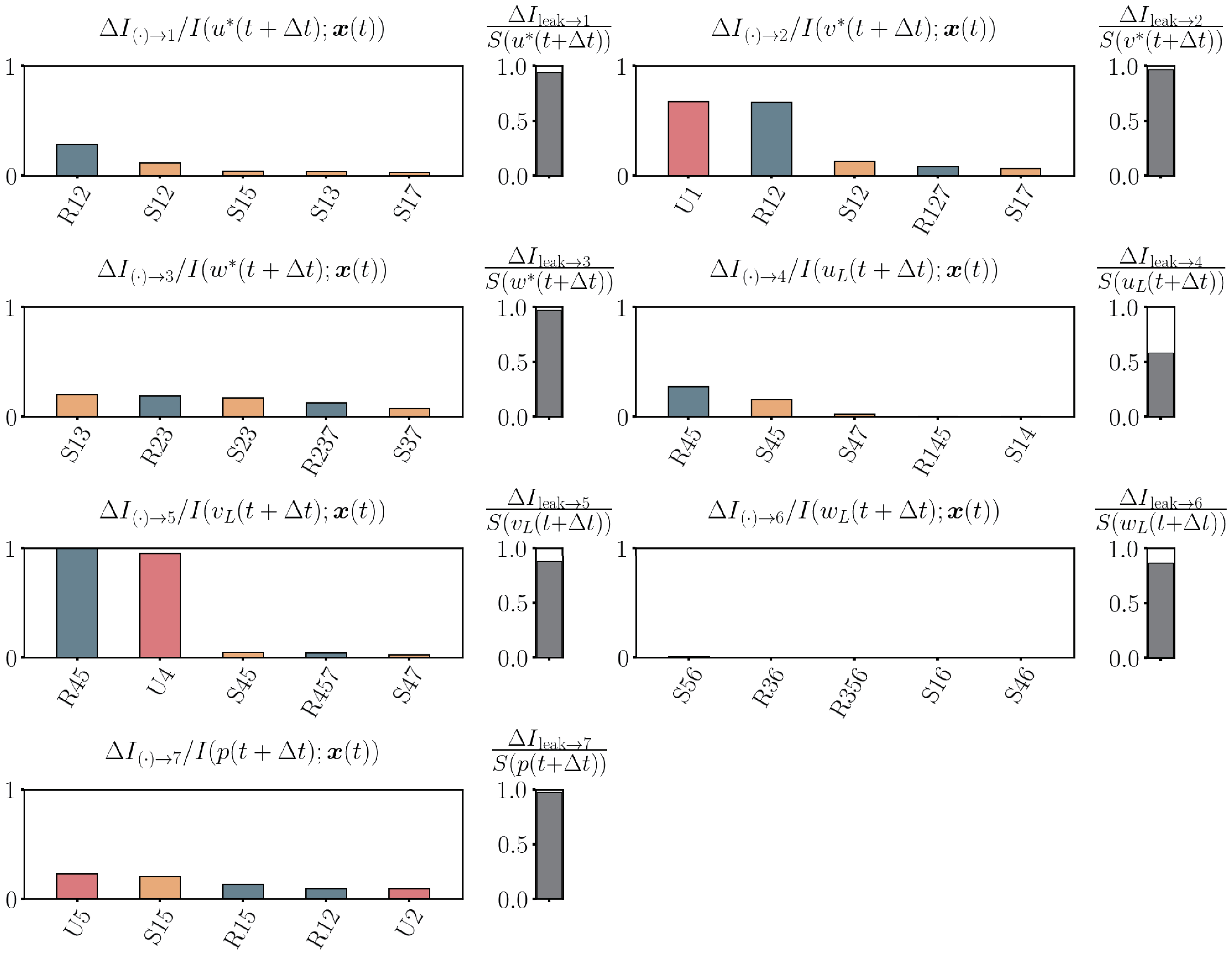}
            };
            \node[anchor=north west] at (image.north west) {{\rmfamily\fontsize{12}{13}\fontseries{l}\selectfont(b)}};
        \end{tikzpicture}
        \label{figsurdh}}
    \caption{SURD computed using $(a)$ the complete dataset, and $(b)$ half of the dataset in temporal length. }
    \label{figsurdch}
\end{figure}

\begin{table}
    \centering
    \begin{tabular}{c  c  c  c  c}
        \diagbox{Methods}{Case} & $y^+=15$ & $y^+=70$ & $y^+=15 \leftrightarrow 200$ & $y^+=70 \leftrightarrow 200$ \\
        \hline
            TE                 &  0.0069  &  0.0353  &            0.0123            &              0.0315           \\
        \hline
            IF                 &  0.0022  &  0.0014  &            0.0046            &              0.0015           \\
        \hline
           SURD                &  0.0136  &  0.0131 &             0.0081           &                0.0168
    \end{tabular}
    \caption{The mean absolute errors of the three methods with half of the dataset in temporal length. }
    \label{meanerror}
\end{table}

\section{Statistical convergence assessment}\label{appB}

The result of causal inference heavily relies on the data. To verify statistical convergence, we compare the causal inference results using the complete dataset and a reduced one that only takes into account half of the data in temporal length. 

Figures~\ref{figch} and \ref{figsurdch} show the causal maps of the inner and outer motions of the near-wall turbulence at $y^+ = 15$. It can be seen that the identified causal links are the same using half and the complete dataset, regardless of the inference methods (transfer entropy, information flow, or SURD) used. 
We also present the mean absolute errors (defined by the absolute difference between the results using two data lengths) of the three methods in table~\ref{meanerror}, in which `$y^+=15$' and `$y^+=70$' indicate the one-point causalities at each height, respectively, and `$y^+=15 \leftrightarrow 200$' and `$y^+=70 \leftrightarrow 200$' represent the two-point causalities between the inner and outer motions. 
We can see that the mean absolute errors of all three methods are very small (much smaller than one), which confirms the statistical convergence of the present study.

\section{Causal analysis of the modulated inner and outer motions}\label{appD}

\begin{figure}
    \centering
    \subfigure{
        \begin{tikzpicture}
            \node[anchor=north west, inner sep=0] (image) at (0,0) {
                \includegraphics[width=0.48\textwidth]{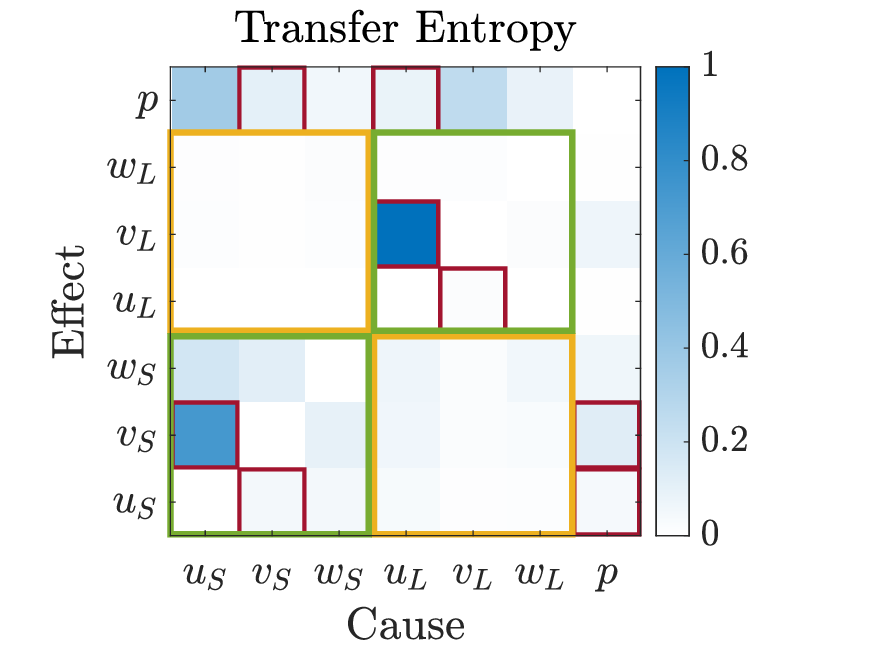}
            };
            \node[anchor=north west] at (image.north west) {{\rmfamily\fontsize{12}{13}\fontseries{l}\selectfont(a)}};
        \end{tikzpicture}
        \label{fig15a}}
    \subfigure{    
        \begin{tikzpicture}
            \node[anchor=north west, inner sep=0] (image) at (0,0) {
                \includegraphics[width=0.48\textwidth]{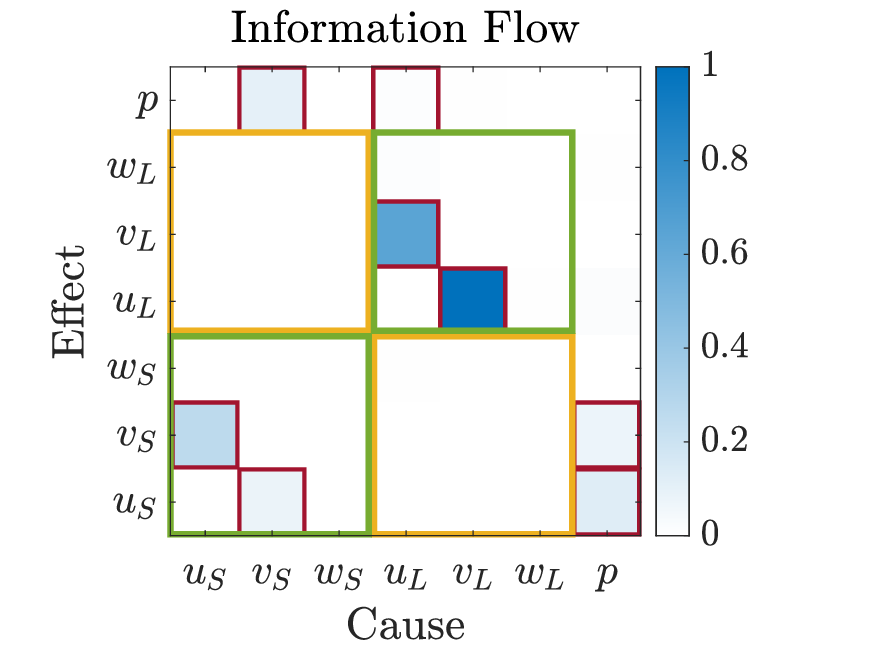}
            };
            \node[anchor=north west] at (image.north west) {{\rmfamily\fontsize{12}{13}\fontseries{l}\selectfont(b)}};
        \end{tikzpicture}
        \label{fig15b}}
    \subfigure{
        \begin{tikzpicture}
            \node[anchor=north west, inner sep=0] (image) at (0,0) {
                \includegraphics[width=0.85\textwidth]{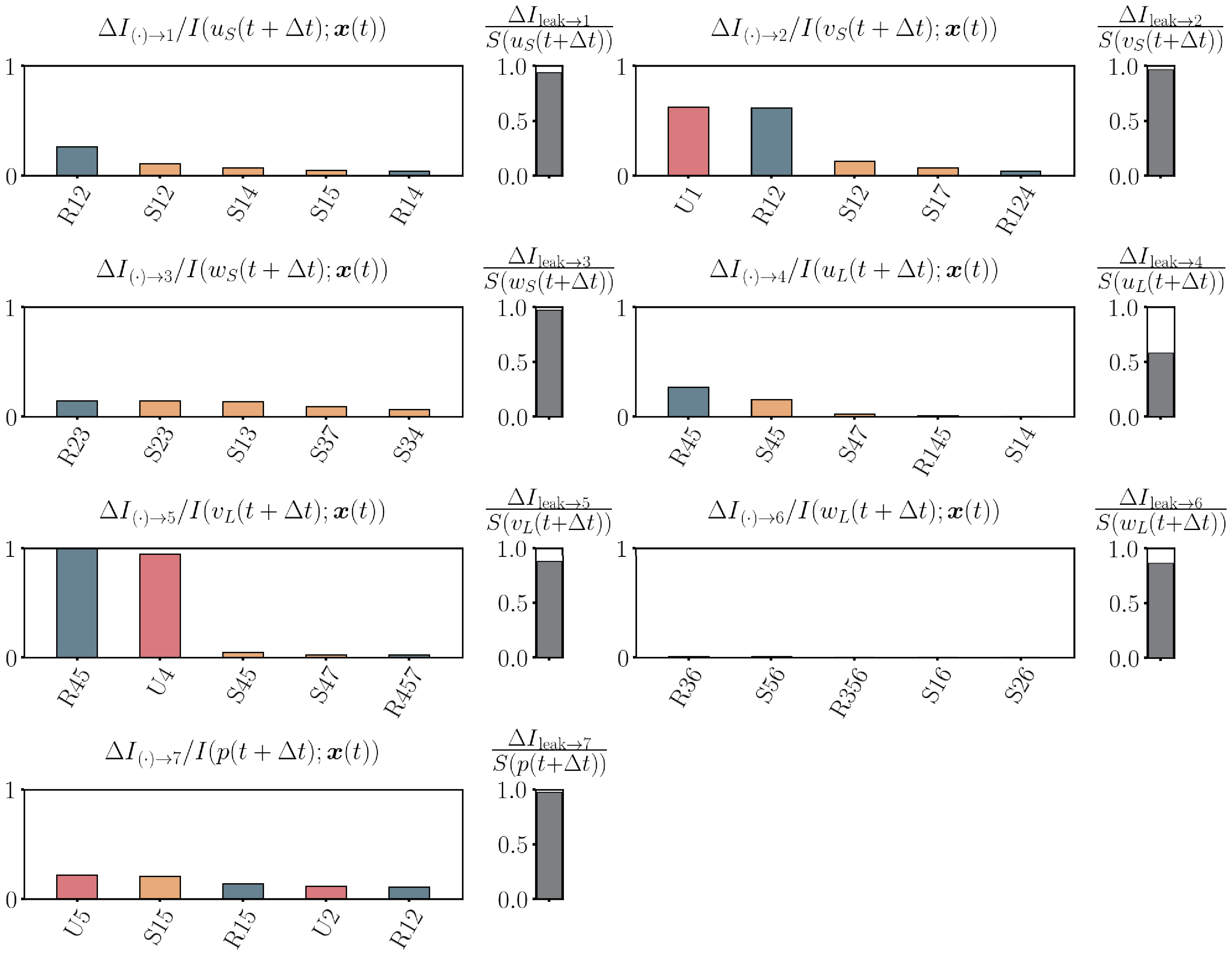}
            };
            \node[anchor=north west] at (image.north west) {{\rmfamily\fontsize{12}{13}\fontseries{l}\selectfont(c)}};
        \end{tikzpicture}
        \label{fig15c}}
    \caption{Causal maps for the inner and outer motions at $y^+=15$. (a) Transfer entropy, (b) information flow, and (c) SURD. {Yellow boxes}: inter-scale causalities. Green boxes: intra-scale causalities. Red boxes: causalities identified by both transfer entropy and information flow. }
    \label{fig15}
\end{figure}

\begin{figure}
    \centering
    \subfigure{
        \begin{tikzpicture}
            \node[anchor=north west, inner sep=0] (image) at (0,0) {
                \includegraphics[width=0.48\textwidth]{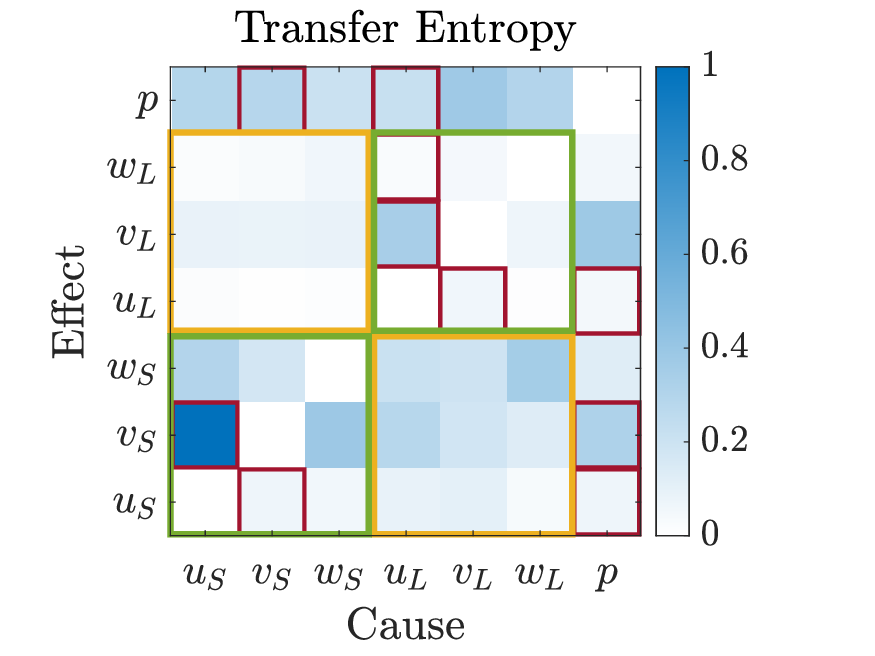}
            };
            \node[anchor=north west] at (image.north west) {{\rmfamily\fontsize{12}{13}\fontseries{l}\selectfont(a)}};
        \end{tikzpicture}
        \label{fig70a}}
    \subfigure{    
        \begin{tikzpicture}
            \node[anchor=north west, inner sep=0] (image) at (0,0) {
                \includegraphics[width=0.48\textwidth]{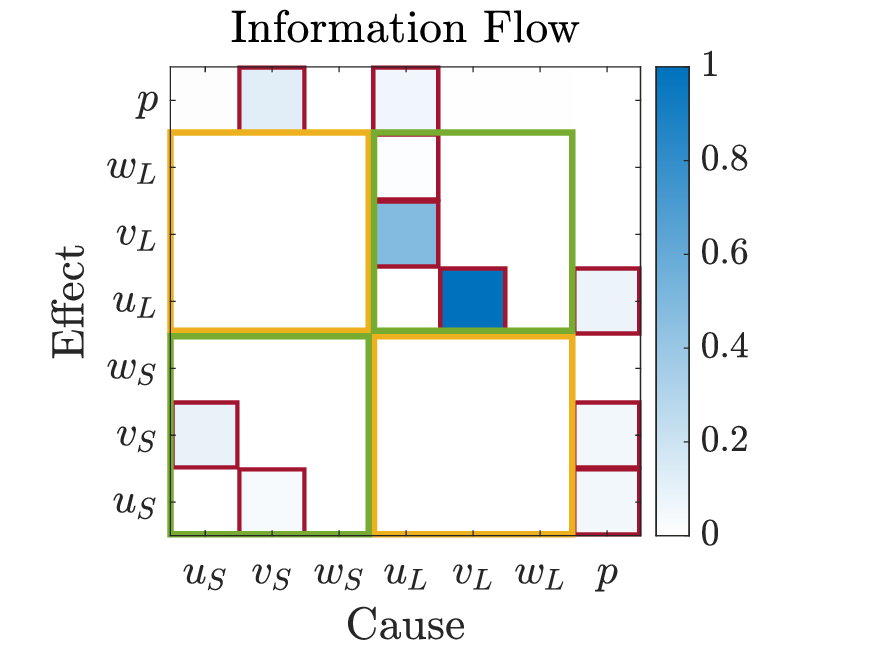}
            };
            \node[anchor=north west] at (image.north west) {{\rmfamily\fontsize{12}{13}\fontseries{l}\selectfont(b)}};
        \end{tikzpicture}
        \label{fig70b}}
    \subfigure{
        \begin{tikzpicture}
            \node[anchor=north west, inner sep=0] (image) at (0,0) {
                \includegraphics[width=0.85\textwidth]{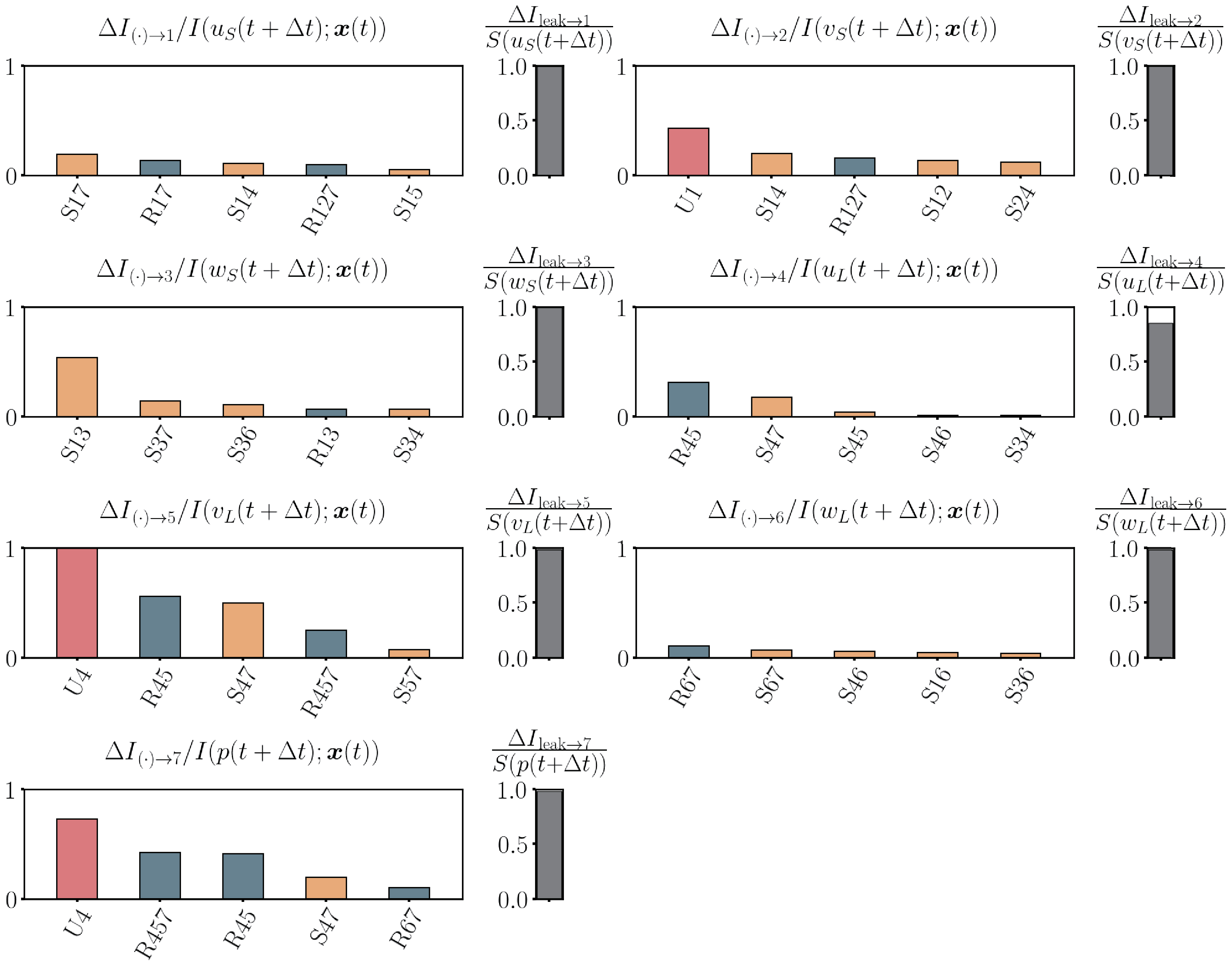}
            };
            \node[anchor=north west] at (image.north west) {{\rmfamily\fontsize{12}{13}\fontseries{l}\selectfont(c)}};
        \end{tikzpicture}
        \label{fig70c}}
    \caption{Causal maps for the inner and outer motions at $y^+=70$. (a) Transfer entropy, (b) information flow, and (c) SURD. {Yellow boxes}: inter-scale causalities. Green boxes: intra-scale causalities. Red boxes: causalities identified by both transfer entropy and information flow. }
    \label{fig70}
\end{figure}

{
All three causal inference methods are also applied using the modulated inner signals ($u_S$, $v_S$, $w_S$) for one-point causal analysis, as shown in figures~\ref{fig15} and \ref{fig70}. 
The causal maps derived from the modulated signals show no significant changes compared to the results using the demodulated signals ($u^*$, $v^*$, $w^*$). {This indicates that the observed lack of causal links between the inner and outer motions is not merely because of the amplitude modulation but the inherent self-sustaining nature.} Even when the modulation effect exists, the causal structure persists. 
On the other hand, it also indicates that the causal inference methods cannot identify the modulation-induced causality between the modulated inner and outer motions using only velocity signals.}

\begin{figure}
    \centering
    \subfigure{
        \begin{tikzpicture}
            \node[anchor=north west, inner sep=0] (image) at (0,0) {
                \includegraphics[width=0.48\textwidth]{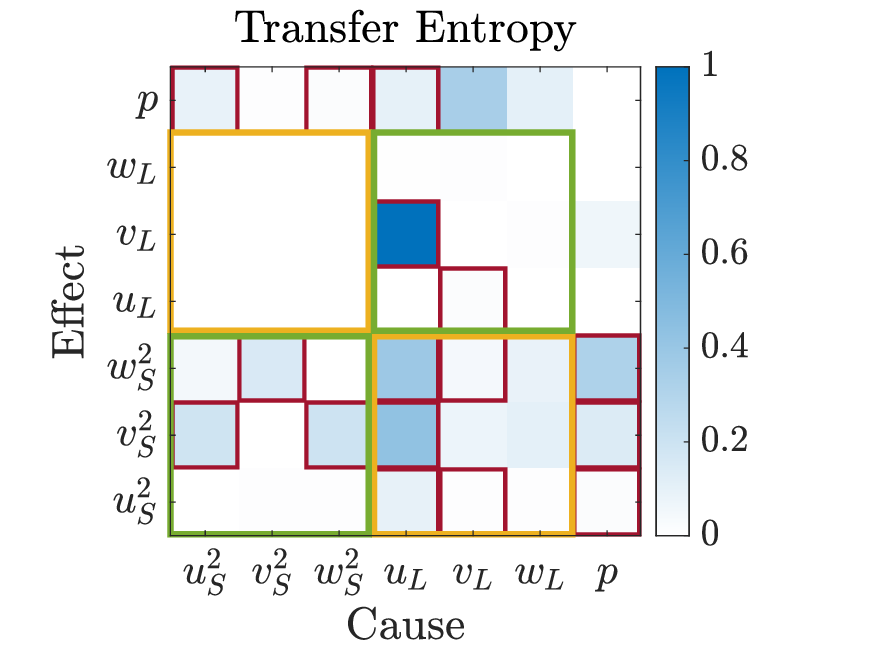}
            };
            \node[anchor=north west] at (image.north west) {{\rmfamily\fontsize{12}{13}\fontseries{l}\selectfont(a)}};
        \end{tikzpicture}
        \label{fig15S2a}  
        \begin{tikzpicture}
            \node[anchor=north west, inner sep=0] (image) at (0,0) {
                \includegraphics[width=0.48\textwidth]{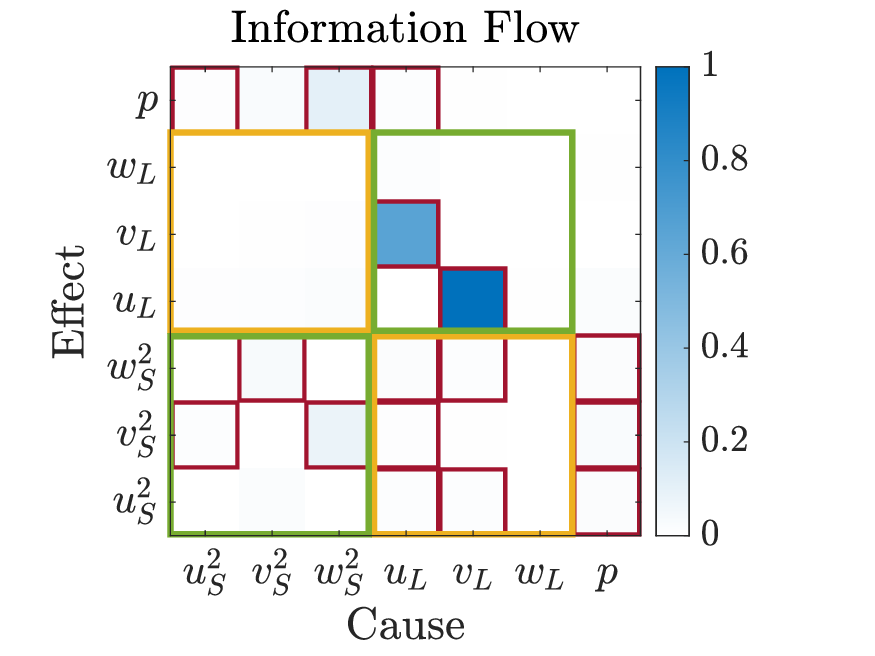}
            };
            \node[anchor=north west] at (image.north west) {{\rmfamily\fontsize{12}{13}\fontseries{l}\selectfont(b)}};
        \end{tikzpicture}
        \label{fig15S2b}}
    \subfigure{
        \begin{tikzpicture}
            \node[anchor=north west, inner sep=0] (image) at (0,0) {
                \includegraphics[width=0.85\textwidth]{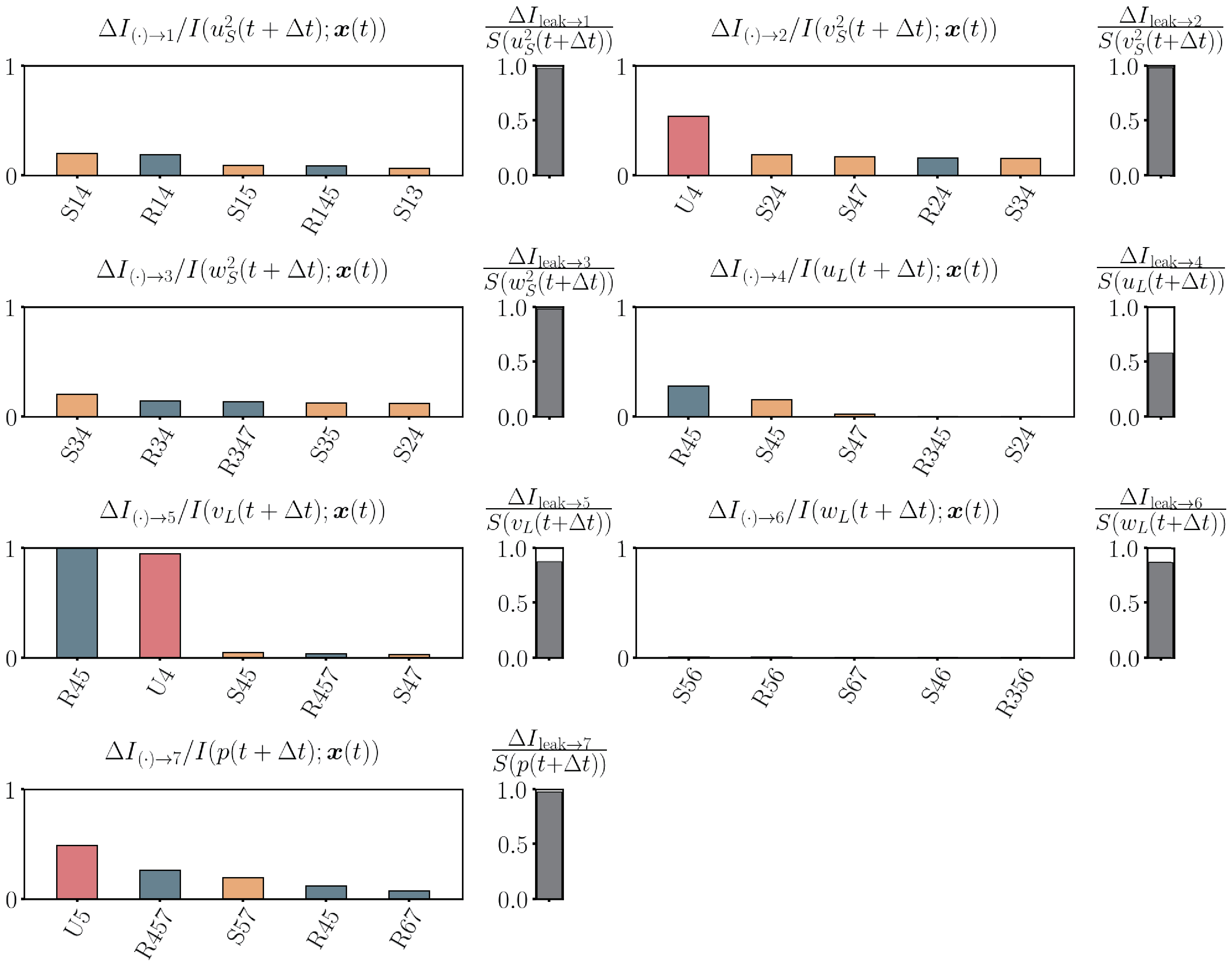}
            };
            \node[anchor=north west] at (image.north west) {{\rmfamily\fontsize{12}{13}\fontseries{l}\selectfont(c)}};
        \end{tikzpicture}
        \label{fig15S2c}}
    \caption{Causal maps for the square of the modulated inner motions and the outer motions at $y^+=15$. (a) Transfer entropy, (b) information flow, and (c) SURD. {Yellow boxes}: inter-scale causalities. Cyan boxes: intra-scale causalities. Red boxes: causalities identified by both transfer entropy and information flow. }
    \label{fig15S2}
\end{figure}

\begin{figure}
    \centering
    \subfigure{
        \begin{tikzpicture}
            \node[anchor=north west, inner sep=0] (image) at (0,0) {
                \includegraphics[width=0.48\textwidth]{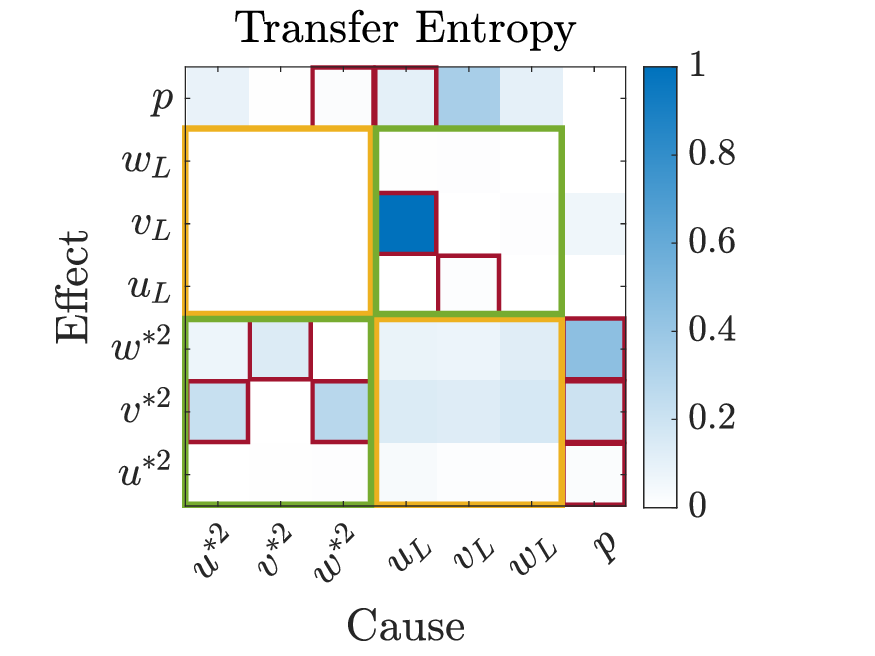}
            };
            \node[anchor=north west] at (image.north west) {{\rmfamily\fontsize{12}{13}\fontseries{l}\selectfont(a)}};
        \end{tikzpicture}
        \label{fig15starS2a}  
        \begin{tikzpicture}
            \node[anchor=north west, inner sep=0] (image) at (0,0) {
                \includegraphics[width=0.48\textwidth]{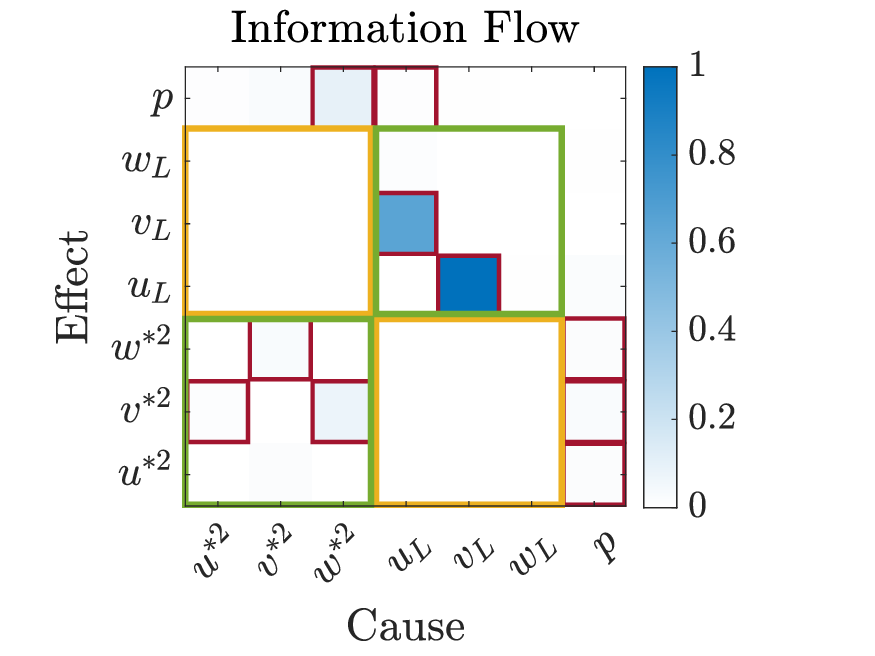}
            };
            \node[anchor=north west] at (image.north west) {{\rmfamily\fontsize{12}{13}\fontseries{l}\selectfont(b)}};
        \end{tikzpicture}
        \label{fig15starS2b}}
    \subfigure{
        \begin{tikzpicture}
            \node[anchor=north west, inner sep=0] (image) at (0,0) {
                \includegraphics[width=0.85\textwidth]{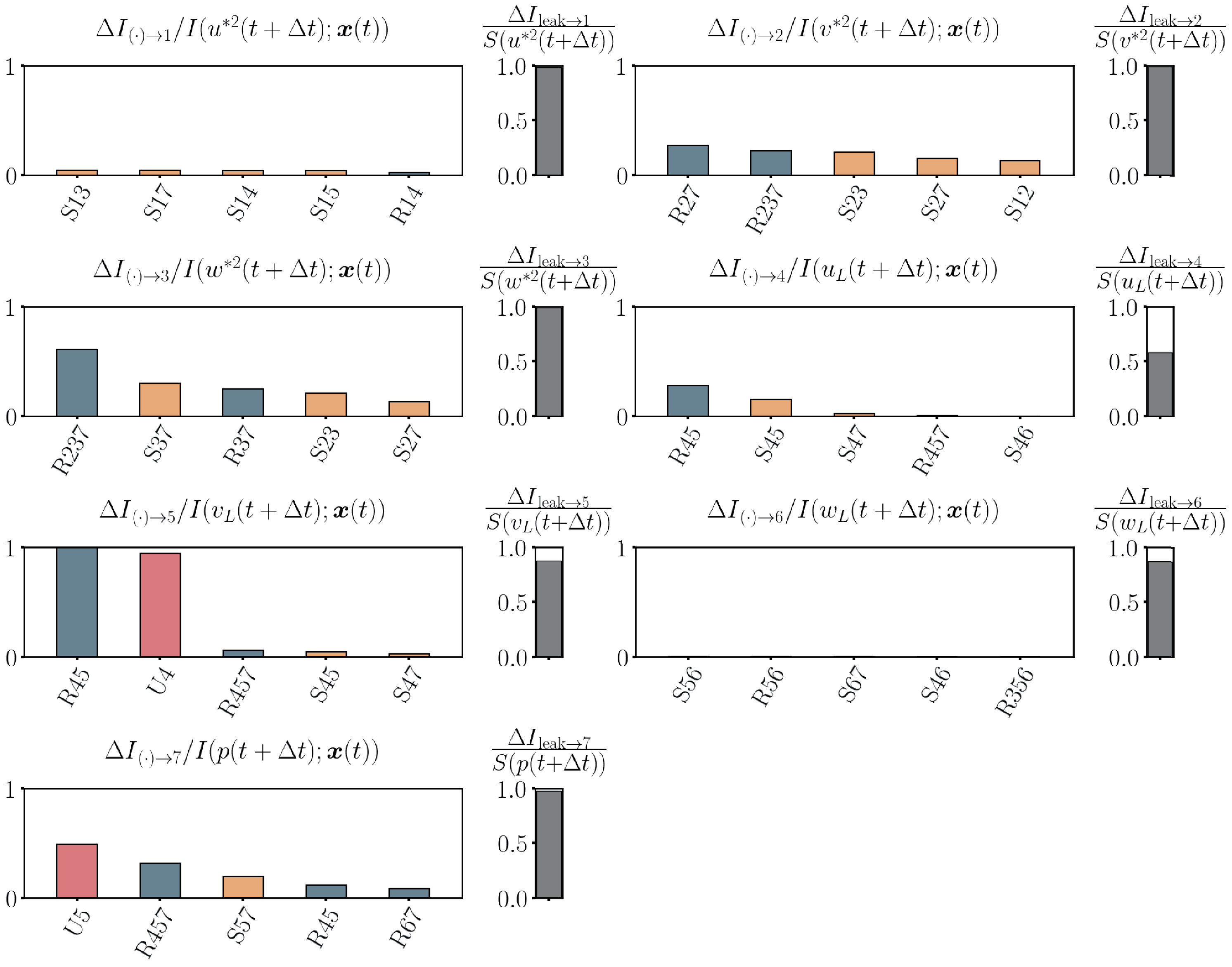}
            };
            \node[anchor=north west] at (image.north west) {{\rmfamily\fontsize{12}{13}\fontseries{l}\selectfont(c)}};
        \end{tikzpicture}
        \label{fig15starS2c}}
    \caption{Causal maps for the square of the demodulated inner motions and the outer motions at $y^+=15$. (a) Transfer entropy, (b) information flow, and (c) SURD. {Yellow boxes}: inter-scale causalities. Cyan boxes: intra-scale causalities. Red boxes: causalities identified by both transfer entropy and information flow. }
    \label{fig15starS2}
\end{figure}

{
Thus, we apply the causal inference methods to the rectified inner motions (velocity square) and the demodulated inner motions as a surrogate of velocity envelope to assess whether the amplitude modulation effect can be identified \citep{bandyopadhyayCouplingScalesShear1984b,chungLargeeddySimulationLargescale2010b,mathisRelationshipVelocitySkewness2011a,duvvuriTriadicScaleInteractions2015}. Figures~\ref{fig15S2} and 
\ref{fig15starS2} present the causal maps obtained using the squares of the modulated and demodulated inner motions at $y^+=15$, respectively. 
One can see from the figure~\ref{fig15S2} (a) and \ref{fig15S2} (b) that causal relationships between $u_L$ and ($u_S^2$, $v_S^2$, $w_S^2$) are detected, which manifest the amplitude modulation effect of the outer streamwise velocity on the streamwise, wall-normal, and spanwise velocity components of the modulated inner motions. 
In figure~\ref{fig15S2c}, the causal links $u_L \to (u_S^2,\ v_S^2,\ w_S^2)$ correspond to the unique causality $\Delta I_{4 \to 2}^U$, the redundant causalities $\Delta I_{14 \to 1}^R$, $\Delta I_{145 \to 1}^R$, $\Delta I_{24 \to 2}^R$, $\Delta I_{34 \to 3}^R$ and $\Delta I_{347 \to 3}^R$, the synergistic causalities $\Delta I_{14 \to 1}^S$, $\Delta I_{24 \to 2}^S$, $\Delta I_{47 \to 2}^S$, $\Delta I_{34 \to 2}^S$, $\Delta I_{34 \to 3}^S$ and $\Delta I_{24 \to 3}^S$ in the SURD result. 
}
{
With the demodulated inner motions, as shown in the figure~\ref{fig15starS2}, the causal links of $u_L \to (u^{*2},\ v^{*2},\ w^{*2})$ and $v_L \to (u_S^2,\ w_S^2)$ are no longer detected, indicating that the effect of amplitude modulation is fully removed. The above results indicate that the causalities due to amplitude modulation can be successfully identified using the rectified inner velocities. 
}

\bibliographystyle{jfm}
\bibliography{jfm}

\end{document}